\documentclass[a4paper,titlepage,11pt]{article}

\usepackage{graphicx}
\usepackage{amsmath, amsfonts, amssymb}
\usepackage{nicefrac}
\usepackage[english]{babel}
\usepackage[sort&compress,merge,numbers]{natbib}
\usepackage[colorlinks=true,urlcolor=blue,citecolor=magenta]{hyperref}

\numberwithin{equation}{section}

\newcommand{\be}{\begin{eqnarray}}
\newcommand{\ee}{\end{eqnarray}}
\newcommand{\beq}{\begin{eqnarray}}
\newcommand{\eeq}{\end{eqnarray}}

\usepackage[a4paper, hmargin={2.4cm,2.4cm}, vmargin={2.5cm,3cm}]{geometry}

\newcommand{\quoting}[1]{``#1''}

\newcommand{\mc}[1]{\mathcal{#1}}

\newcommand{\td}{\text{d}}

\newcommand{\ts}{\thinspace}

\newcommand{\bt}{\textbf{T}^{\mu\nu}}
\newcommand{\jt}{\textbf{J}}
\newcommand{\jmt}{\boldsymbol{\mathcal{J}}}
\newcommand{\btd}{\textbf{T}_{\mu\nu}}

\bibpunct{[}{]}{;}{n}{,}{,}

\def\FF{{\cal F}}

\def\RR{{\cal R}}

\def\WW{{\cal W}}

\def\d{{\partial}}

\def\clock{{\count0=\time
           \divide\count0 60
           \ifnum\count0<10 0\fi\the\count0
           \multiply\count0 -60 \advance\count0 \time
           :\ifnum\count0<10 0\fi \the\count0
         }}
\newcommand{\timestamp}{{\small\vbox{\hbox{\tt\jobname.tex}
\hbox{\the\day/\the\month/\the\year, \clock}}}}

\begin{document}

\hypersetup{pageanchor=false}
\begin{titlepage}
\vskip -1cm
\begin{flushright}
ITCP-IPP-2016-09\\
CCQCN-2016-149\\
CCTP-2016-11
\vskip -1cm
\end{flushright}
\ \ \vskip 0.2cm

\centerline{\LARGE \bf Forced Fluid Dynamics from Blackfolds}
\vskip 0.3cm
\centerline{\LARGE \bf  in General Supergravity Backgrounds}

\vskip 1.cm \centerline{\bf Jay Armas$^{1}$, Jakob Gath$^{2}$, Vasilis Niarchos$^{3}$,}
\vskip 0.2cm
\centerline{\bf Niels A. Obers$^{4}$ and Andreas Vigand Pedersen$^{4}$} \vskip 0.7cm

\begin{center}
\sl $^{1}$ \small\slshape Physique Th\'{e}orique et Math\'{e}matique \\
Universit\'{e} Libre de Bruxelles and International Solvay Institutes \\
ULB-Campus Plaine CP231, B-1050 Brussels, Belgium\\
\vskip 0.2cm
{$^2$ \small\slshape Centre de Physique Th\'{e}orique, \'{E}cole Polytechnique, \\
CNRS UMR 7644, Universit\'{e} Paris-Saclay F-91128 Palaiseau, France.}\\
\vskip 0.2cm
$^{3}$ \small\slshape Crete Center for Theoretical Physics, Institute of Theoretical and Computational Physics\\
Crete Center for Quantum Complexity and Nanotechnology\\
Department of Physics, University of Crete, 71303, Greece\\
\vskip 0.2cm
 $^{4}$ The Niels Bohr Institute, Copenhagen University,\\
  Blegdamsvej 17, DK-2100 Copenhagen \O , Denmark.
\vskip 0.4cm

\end{center}
\vskip 0.3cm

\centerline{\small\tt jarmas@ulb.ac.be, jakob.gath@polytechnique.edu, niarchos@physics.uoc.gr,}
\centerline{\small\tt obers@nbi.ku.dk, andreasvigand@gmail.com}

\vskip 1.cm \centerline{\bf Abstract} \vskip 0.2cm \noindent
We present a general treatment of the leading order dynamics of  the collective modes 
of charged dilatonic $p$-brane solutions of (super)gravity theories in arbitrary backgrounds.
To this end we employ the general strategy of the blackfold approach which is based on a long-wavelength
derivative expansion around an exact or approximate solution of the (super)gravity equations of motion.
The resulting collective mode equations are formulated as forced hydrodynamic equations on dynamically embedded hypersurfaces. We derive them in full generality (including all possible asymptotic fluxes and dilaton profiles) in a far-zone analysis of the (super)gravity equations and in representative examples in a near-zone analysis. An independent treatment based on the study of external couplings in hydrostatic partition functions is also presented. Special emphasis is given to the forced collective mode equations that arise in type IIA/B and eleven-dimensional supergravities, where besides the standard Lorentz force couplings our analysis reveals additional couplings to the background, including terms that arise from Chern-Simons interactions.
We also present a general overview of the blackfold approach  and some of the key conceptual issues that arise when applied to arbitrary backgrounds.

\end{titlepage}
\hypersetup{pageanchor=true}

{
\noindent\rule{\textwidth}{1.2pt}
\vspace{-1cm}
\tableofcontents
\noindent\rule{\textwidth}{1.2pt}
}


\section{Introduction}
\label{intro}

Hydrodynamics has proven to be a powerful universal description of the low energy effective dynamics of 
interacting quantum systems
at finite temperature, valid in the regime where fluctuations have sufficiently long wavelength. Modern developments have revealed a close connection
between hydrodynamics and gravity, in particular black holes, starting with the pioneering work in AdS/CFT \cite{Policastro:2001yc,Kovtun:2004de} and the discovery of fluid/gravity
duality \cite{Bhattacharyya:2008jc,Baier:2007ix} (see \cite{Rangamani:2009xk} for a review). In the latter case the fluid lives on the AdS boundary, but more general fluid dynamical descriptions of black brane dynamics in diverse asymptotic
spacetimes (including flat space) have also been found
in the context of the blackfold approach \cite{Emparan:2009cs,Emparan:2009at,Camps:2010br}. 
The relation between AdS fluid/gravity and flat space blackfolds in the context of D3-branes has been 
considered in \cite{Emparan:2013ila}. 

In hydrodynamics the dynamical equations of a system are captured by the combined set of 
a few conservation equations (typically, the conservation equations of the stress-energy tensor
and some abelian currents) and a set of constitutive relations that reduce the number of independent
degrees of freedom. Important modifications to these equations arise when external forces are applied, or 
when the symmetries underlying the conservation equations are anomalous. Although the standard formulation
of hydrodynamic equations refers to long-wavelength deformations of finite temperature homogeneous 
configurations, general hydrodynamic systems can exhibit a variety of interesting extensions.
These extensions can include, for instance,
the presence of anisotropies, symmetries associated to higher spin currents, or
the propagation of the fluid on dynamical hypersurfaces. In this paper we will encounter 
ideal, non-anomalous, forced hydrodynamic systems with many of these extensions.

The hydrodynamic systems that will be considered in this paper are derived from the classical long-wavelength
dynamics of black holes and branes in general (super)gravity theories in the spirit of the general connection
between fluids and gravity outlined above.
Forced fluids have been discussed in the context of AdS black hole solutions and the fluid-gravity correspondence in several papers in the past, see for instance \cite{Bhattacharyya:2008ji}. The extension beyond AdS along the lines of the blackfold formalism introduces new ingredients and new technical complications. Let us quickly summarize some of these issues. 

\subsubsection*{Forced blackfold equations in (super)gravity}

In blackfold generalizations of the fluid/gravity correspondence the long-wavelength
expansion of the (super)gravity equations around black brane solutions is an affair
that combines the fluid dynamical 
nature of black hole physics \cite{Emparan:2009at,Camps:2010br,Gath:2013qya,Emparan:2013ila,DiDato:2015dia}  with the extrinsic (elastic) dynamics 
\cite{Emparan:2007wm,Caldarelli:2008pz,Camps:2008hb,Emparan:2009at,Emparan:2009vd,Grignani:2010xm,Caldarelli:2010xz,Armas:2010hz,Emparan:2011hg,Armas:2011uf,Camps:2012hw,Niarchos:2012pn,Armas:2012ac,Armas:2013aka}
of hypersurfaces in ambient spacetimes that is characteristic of D-brane physics. Unfortunately, even the
case of black holes in flat space is sufficiently complicated and it has not yet been studied systematically beyond the first order in the derivative expansion. The case of black holes in asymptotic spacetimes with 
arbitrary curved geometry and other non-trivial fluxes, which is the main case of interest in this paper, 
clearly involves an even more demanding technical treatment.
Some of the main conceptual and technical issues that arise in this context
are summarized in the discussion of Sec.~\ref{conclusions}. In this paper we will not attempt to address 
a complete solution of these issues. Our main focus will remain on the leading order 
perturbations in supergravity aiming to isolate and determine  
the generic features of the effective fluid dynamical description that arise at this order.

One of the new ingredients that blackfolds introduce, and whose implications we want to emphasize here, is the simultaneous presence and interplay of different higher spin currents and background abelian gauge fields. 
Black holes and branes can be coupled electrically, magnetically or dyonically with respect to these gauge fields. 
The long-wavelength analysis of the dynamics of such solutions in arbitrary backgrounds leads to  
forced effective fluids with external forces that involve a variety of different couplings between 
the higher spin currents and the background gauge fields. 
The structure of these couplings is uniquely determined from the action of the underlying (super)gravity 
theory. We will perform a general analysis of this structure at the leading order of the 
long-wavelength expansion.

The precise identification
of this structure is not only interesting as an academic exercise in fluid dynamics, it is also 
one of the first steps towards a systematic long-wavelength analysis of black hole solutions in general backgrounds.
For example, it can be useful in problems that involve black holes in the background of curved geometries, e.g. 
black holes in the vicinity of other black holes, problems that involve extremal, and non-extremal, brane solutions
in backgrounds with fluxes in the context of string theory (describing the gravitational backreaction of 
massive configurations of D/M-branes), and problems with real time dynamics where the background
is forcing a black hole solution to evolve dynamically in time. 

In the context of the blackfold formalism, most of the developments have focused so far on black hole solutions
in flat spacetimes \cite{Emparan:2009vd, Armas:2015kra, Armas:2015nea}. Preliminary aspects of black brane solutions in AdS spacetimes have been studied in 
\cite{Caldarelli:2008pz,Armas:2010hz,Armas:2012bk,Armas:2013ota}  (see also \cite{Camps:2008hb}). 
For specific AdS flux backgrounds, thermal probe brane techniques based on the blackfold
approach have been applied in \cite{Armas:2012bk,Armas:2013ota} to construct thermal giant gravitons. 
However, a general treatment of blackfolds in general backgrounds with fluxes in (super)gravity
theories has not been performed and part of our motivation is to initiate such study.

Another motivation for this work is the recent proposal \cite{Niarchos:2015moa} 
(see also \cite{Grignani:2016bpq} for related independent work building on \cite{Grignani:2010xm,Grignani:2011mr,Grignani:2013ewa})
that the effective hydrodynamic description of black brane solutions in the blackfold formalism is connected to 
the underlying microscopic description of D/M-branes in string/M-theory via a general open/closed string duality
that works in many cases in gravity as a tomographic principle.
In the proposal of Ref.\ \cite{Niarchos:2015moa}  the abelian hydrodynamic blackfold equations are conjectured 
to be effective equations of 
singleton dynamics. They provide a strong-coupling description of the effective long-wavelength dynamics of the 
abelian, center-of-mass degrees of freedom of D/M-branes. In accordance with this expectation
it was demonstrated in specific examples in \cite{Niarchos:2015moa} that the 
leading order blackfold equations of extremal $p$-brane configurations in string theory are equivalent to the 
DBI equations that describe long-wavelength dynamics of D-branes. 

The formulation of the blackfold equations
in backgrounds with arbitrary fluxes, that we venture here, will allow us to probe this conjecture further, in more detail and in more generic
situations. As an immediate forthcoming task one can test the expected equivalence of the extremal forced blackfold 
equations derived in this paper with the full set of well-known open-closed string couplings in the DBI action. The
proof of this equivalence would provide a pure supergravity derivation of the complete DBI action 
(including all open-closed string couplings). A similar exercise could be performed in 
M-theory to re-derive from supergravity the PST action of M5-branes \cite{Pasti:1997gx}, which is a theory of
a self-dual 3-form field. Earlier related work in this direction
has appeared in \cite{Niarchos:2012pn,Niarchos:2014maa}.

\subsubsection*{Brief summary of technical results and outline of the paper}

The main task of this paper is to provide a general treatment of the leading order collective mode equations of $p$-brane
solutions in arbitrary backgrounds for generic (super)gravity theories. We follow the general strategy of the
blackfold approach and work in a long-wavelength derivative expansion around an exact or an approximate solution
of the supergravity equations of motion (see Sec.~\ref{conclusions} for an explicit discussion of this distinction).

In the (super)gravity analysis of the blackfold approach one attempts to construct a perturbatively deformed black hole solution in a scheme of matched asymptotic expansions (MAEs). A separate perturbative analysis of the gravity equations is performed in the {\it far-zone} (near the fixed asymptotic background), and in the {\it near-zone} (in the vicinity of the black hole horizon). At the end the solutions in the two zones are matched order-by-order in the perturbative expansion. Explicit applications of MAEs in the context of blackfolds can be found in \cite{Emparan:2007wm,Camps:2012hw,Gath:2013qya,DiDato:2015dia}, to which we refer the reader for further details.

In Sec.~\ref{far} we discuss how an arbitrary asymptotic background affects the constraint equations (and associated
conservation equations) in the far-zone analysis. Working in the linearized approximation, in this region
we describe the leading order modifications of the asymptotic geometry in the presence of a bulk source. As such, this 
treatment is very general and conceptually straightforward. On its own, it applies to very general black hole solutions and 
requires no assumptions of a long-wavelength expansion. The obtained conservation equations, which arise as Bianchi identities in gravity and abelian gauge theory, are exact and valid for generic configurations with arbitrary derivatives, in space and/or in time. We formulate them in terms of 
generic currents without reference to specific black holes and specific constituent relations.
The key elements of the analysis are exhibited first in a simplified model of an Einstein-dilaton theory with a
$(q+1)$-form gauge field. Then, the logic is applied in more general settings that include the ten-dimensional type-IIA/B, and eleven-dimensional supergravity theories to obtain the general forced blackfold equations in string/M-theory. Besides the standard Lorentz coupling of the schematic form $F \cdot J$ between background gauge field strengths $F$ and currents $J$, as well as their electromagnetic duals, one can also see, in the forced equations of the stress-energy tensor, couplings of the schematic form $F \cdot C \cdot J$ arising from Chern-Simons interactions in the supergravity theory. One can also see explicitly the corresponding 
background-dependent modifications of the conservation equations of the abelian currents.

In Sec.~\ref{sec:near} we initiate the study of the same equations in the near-zone region. At this point we need
to be very specific about the type of solution that we consider. We can only proceed in a systematic way
by setting up a perturbative expansion scheme. This is the point where the separation of scales and the small
derivatives (inherent in the blackfold approach) are most needed to set up the matched asymptotic expansion
that will allow us to extend the solution in the near-horizon region and to incorporate the full non-linearities of
gravity beyond the linearized approximation. Generalizing earlier results for neutral black
$p$-brane solutions, we consider the cases of charged black branes in Einstein-dilaton theories and D$p$-F1
bound states in type II supergravity. We focus on the constraint equations of gravity that lead to
the corresponding effective blackfold equations. The D$p$-F1 case is particularly interesting because it
demonstrates very explicitly how couplings between currents of different spin and background fields arise 
simultaneously in the force terms. Of course, in all cases we reproduce the results of the asymptotic linearized
treatment of Sec.~\ref{far}. The added benefit of the near-zone analysis is that it gives us a first taste of the ans\"atze that need to be implemented to find the near-zone deformations of the supergravity solutions. Eventually these solutions should be matched to the solutions of the far-zone analysis. 
 
In Sec.~\ref{partition} we present yet another derivation of the forced blackfold equations by studying how external 
couplings arise in hydrostatic partition functions. After a general discussion we focus on the cases of solutions with Maxwell charge and top-form
charged black $p$-brane solutions. This approach is based on the analysis of 
the on-shell gravitational action. It applies only to stationary configurations and leads to a natural connection with 
standard relations in thermodynamics. The forced blackfold equations arise in this case, in a standard fashion, 
as Ward identities of the on-shell gravitational action.

Finally, in Sec.~\ref{conclusions} we conclude with a general overview of the approach and a summarizing discussion 
of some of the key conceptual issues that can arise in generic long-wavelength treatments of black holes in 
non-trivial background geometries. We point out that in ideal situations (that we dub the `exact brane' application of
blackfolds) one has exact information about a specific (leading order) 
homogeneous black brane configuration that is subsequently perturbed in small derivatives along the 
homogeneous directions. Since such exact information is mostly absent in situations with arbitrary 
asymptotic geometries we highlight the role that is usually played by parallel expansions of the leading order order solution.
These are expansions in small numbers that are typically ratios of quantities of the leading order order solution over 
quantities characterizing the background. The disadvantage of this approach is the implementation of further
approximations; the advantage is the possibility of a wider, more flexible application of the method. We briefly 
comment on the open-closed string interpretation of these additional approximations in the context of the 
proposal \cite{Niarchos:2015moa}. We conclude Sec.~\ref{conclusions} with a list of interesting open problems
that could be treated in a natural continuation of this work.

We provide four appendices. In App.~\ref{app:notation}, we give the form notation that we employ throughout the paper. In App.~\ref{app:AB}, we provide the equations of motion and probe brane equations derived in Sec.~\ref{far} for the special case of type II A/B supergravity and in App.~\ref{eqn:DpF1charges}, we provide the effective charges and currents for the brane solutions considered in Secs.~\ref{sec:near} and \ref{partition}.
Finally, in App.~\ref{app:entropy}  an entropy current analysis of the forced fluids considered in Sec.~\ref{sec:hydroqp} is given. 

\paragraph{Notation:}
Throughout this paper we will use Greek letters, $\mu,\nu,...$, to denote the $D$-dimensional spacetime directions $x^\mu$, and small Latin letters, $a,b,...$, to express the $p+1$ (worldvolume) directions $\sigma^a$, along which a $p$-brane solution is infinitely extended. The Minkowski metric is denoted by $\eta_{\mu\nu}$, $G$ is Newton's gravitational constant and $\star$ denotes the $D$-dimensional Hodge star operator. Further details on our notation can be found in App.~\ref{app:notation}.

\section{Far-zone analysis}  
\label{far}

In this section we concentrate on the asymptotic region far from the black hole horizon at radial distances $r\gg r_H$, where
$r_H$ is the location of the black hole horizon. In this region the gravitational fields are small deformations, 
weighted by positive powers of the small ratio $\nicefrac{r_H}{r}$, of the asymptotic solution. In the asymptotic solution the 
gravitational field (and any other matter field) are assumed arbitrary.
In particular, at this point, no assumption of weak field or small derivatives is made. The leading deformations 
induced by the black hole in the bulk can be studied in this region by analyzing the linearized Einstein equations
\beq \label{eq:Einstein1}
\left(G_{\mu\nu}-8\pi G\textbf{T}_{\textbf{M}\mu\nu}\right)|_{\textbf{linear}}=0 ~~,
\eeq
where $\textbf{T}_{\textbf{M}\mu\nu}$ denotes the stress-energy contribution of any sources of energy/matter 
that are present in the gravitational theory. The subscript $\textbf{linear}$ in \eqref{eq:Einstein1} is there to remind 
that the Einstein equations are linearized around the asymptotic background. 

The method of equivalent sources%
\footnote{See for example \cite{Myers:1986un} and \cite{Harmark:2002tr,Kol:2003if}.}
 allows us to replace the complicated details of the bulk with an effective stress-energy 
tensor $\btd$ localized in the bulk
\beq \label{eq:Einstein}
\left(G_{\mu\nu}-8\pi G\textbf{T}_{\textbf{M}\mu\nu}\right)|_{\textbf{linear}}=8\pi G \btd ~~,
\eeq
where $\btd$, which is supported on a $(p+1)$-dimensional hypersurface for a $p$-brane solution, 
is sourcing the gravitational field of interest at large distances. Solving \eqref{eq:Einstein} for all $r$ and 
extracting the result for $r\gg r_H$ is equivalent to solving \eqref{eq:Einstein1}. Using the Bianchi identity 
$\nabla_\mu G^{\mu\nu}=0$ in \eqref{eq:Einstein} we find that the total stress-energy tensor is conserved, namely 
\beq \label{eq:probe}
\nabla_{\mu}\bt=-\nabla_{\mu}\textbf{T}_{\textbf{M}}^{\mu\nu}~~.
\eeq
In what follows we will use the full set of gravitational equations to re-express the r.h.s.\ of this equation in terms of 
the asymptotic profile of the gravitational fields and other currents characterizing the source. At the end, the r.h.s.\ will
be recast as a force term driven by the non-trivial asymptotic profiles of the gravitational fields. 

The equations derived in this way from \eqref{eq:probe} are equations describing a probe brane in the 
asymptotic background. For example, when $\bt$ is the stress-energy tensor of a charged point particle, 
and $\textbf{T}_{\textbf{M}}^{\mu\nu}$ is the stress-energy tensor of a $U(1)$ gauge field, 
Eq.\ \eqref{eq:probe} provides a derivation of the Lorentz force acting on the particle \cite{rindler2001relativity}. Similar modified conservation equations of other currents will be derived from the Bianchi identities
of bulk gauge fields.

In the rest of this section, we will use this strategy to derive the 
equations of motion of generic probe branes in Einstein-dilaton theory with a $(q+1)$-form gauge field, type IIA/B 
and eleven dimensional supergravity. The results cover the most general type of brane bound states that 
can be encountered in string theory.

The above derivation based on the linearized approximation \eqref{eq:Einstein} is (in a sense) straightforward, 
yet it turns out to provide very general expressions with very few assumptions. This will be most appreciated in the 
next section when we try to track $\btd$ in the bulk of the solution away from the asymptotic region. The same
$\btd$ (and other currents) with the same modified conservation equations will arise there as constraint equations
of the non-linear Einstein equations. Nevertheless, the treatment of the solution at finite radius $r$ will be much 
harder and a systematic analysis will require more assumptions and more case-specific data. 

It is also worth stressing that from the point of view of the full solution across the whole spacetime, the effective
currents and their conservation equations are not merely probe data and probe equations, in fact, they are describing 
a full-fledged backreacted solution of the gravitational equations (see Sec.~\ref{sec:near}).


\subsection{Einstein-dilaton theory with a \texorpdfstring{$(q+1)$}{(q+1)}-form gauge field \label{sec:2.1}}

We start with the simplest system of Einstein-dilaton theory with a $(q+1)$-form gauge field with action
\beq \label{eq:emd}
I=\frac{1}{16\pi G}\int_{\mathcal{M}_D}\left[\star R-\frac{1}{2}d\phi\wedge\star d\phi-\frac{1}{2}e^{a_q\phi}F_{q+2}\wedge \star F_{q+2} \right]~~,
\eeq
where the dilaton coupling $a_q$ is arbitrary. We wish to couple a probe brane to field configurations which are solutions to the equations of motion that arise from \eqref{eq:emd}. We consider a probe brane carrying an electric current $\jt_{q+1}$, a magnetic current $\jmt_{D-q-3}$ and a dilaton current $\textbf{j}_{\phi}$ coupled to the fields of the theory \eqref{eq:emd}. The presence of the magnetic current modifies the Bianchi identity for the $(q+2)$-form field strength $F_{q+2}$ so that
\beq \label{eq:emdbianchi}
dF_{q+2}=16\pi G \star \jmt_{D-q-3}~~,~~F_{q+2}=dC_{q+1}+16\pi G \star D_{D-q-2}~~,
\eeq
where $D_{D-q-2}$ is the Dirac brane defined by $\star \jmt_{D-q-3}=d\star D_{D-q-2}$. The equations of motion for the gauge field $C_{q+1}$ and the dilaton $\phi$ for the theory \eqref{eq:emd} coupled to sources are
\begin{align} \label{eq:emdeom}
\begin{split}
&d\star\left(e^{a_q \phi}F_{q+2}\right)=(-1)^{D+q^2-1}16\pi G\star \jt_{q+1}~~,\\
& \square \phi-\frac{a_q}{2}e^{a_q\phi}\star(F_{q+2}\wedge \star F_{q+2})=-16\pi G \textbf{j}_{\phi}~~.
\end{split}
\end{align}

In order to evaluate the r.h.s. of Eq.\ \eqref{eq:probe} we require the explicit form of the energy/matter contributions to $\textbf{T}_{\textbf{M}}^{\mu\nu}$. Splitting $\textbf{T}_{\textbf{M}}^{\mu\nu}$ as $\textbf{T}_{\textbf{M}}^{\mu\nu}=\textbf{T}_{(F)}^{\mu\nu}+\textbf{T}_{(\phi)}^{\mu\nu}$ we obtain the bulk stress-energy tensor for the $(q+1)$-form gauge field and the dilaton
\begin{align} \label{eq:emdstress}
\begin{split}
&16\pi G \textbf{T}_{(F)}^{\mu\nu}=\frac{e^{a_q\phi}}{(q+1)!}\left(F_{q+2}^{\mu \mu_1...\mu_{q+1}}{{F_{q+2}}^{\nu}}_{\mu_1...\mu_{q+1}}-\frac{1}{2(q+2)}g^{\mu\nu} F_{q+2}^2 \right)~~, \\
&16\pi G\textbf{T}_{(\phi)}^{\mu\nu}=\partial^{\mu}\phi\partial^{\nu}\phi-\frac{1}{2}g^{\mu\nu}\partial_\lambda\phi\partial^{\lambda}\phi ~~.
\end{split}
\end{align}

Inserting these expressions into the r.h.s. of Eq.~\eqref{eq:probe} we find 
\begin{align} \label{eq:aux1}
  16\pi G\nabla_{\mu}\textbf{T}_{\textbf{M}}^{\mu\nu}=&\frac{1}{(q+1)!}F_{q+2}^{\nu \mu_1...\mu_{q+1}}\nabla_\mu\left(e^{a_q\phi}{F^{\mu}}_{q+2\mu_1...\mu_{q+1}}\right)-\frac{e^{a_q\phi}}{(q+2)!}F^{\mu_1...\mu_{q+2}}_{q+2}{dF^{\nu}}_{q+2\mu_1...\mu_{q+2}}\nonumber\\
&+\left(\square \phi-\frac{a_q}{2}e^{a_q\phi}\star(F_{q+2}\wedge \star F_{q+2})\right)\partial^{\nu}\phi~~.
\end{align}
This result simplifies further by using the equations of motion \eqref{eq:emdeom}.  Inserting the final expression into \eqref{eq:probe} leads to the modified conservation equation 
\begin{align} \label{eq:forceemd}
\begin{split}
\nabla_{\mu}\bt&=\frac{1}{(q+1)!}{F_{q+2}^{\nu\mu_1...\mu_{q+1}}}\jt_{q+1\mu_1...\mu_{q+1}}\\
&+\frac{(-1)^{qD+1}e^{a_q\phi}}{(D-q-3)!}{F}_{D-q-2}^{\nu\mu_1...\mu_{D-q-3}}\jmt_{D-q-3\mu_1...\mu_{D-q-3}}+\textbf{j}_{\phi}\partial^{\nu}\phi~~,
\end{split}
\end{align}
where we have defined the dual field strength via the relation ${F}_{D-q-2}=\star F_{q+2}$. In the case of a non-dilatonic electrically charged point-particle ($q=0$), Eq.~\eqref{eq:forceemd} yields the Lorentz force. In general, Eq.~\eqref{eq:forceemd} describes the electric coupling to the field strength $F_{q+2}$ and to its dual ${F}_{D-q-2}$. In addition, we obtain from general principles the force due to the presence of background dilaton fields, which is proportional to the gradient of $\phi$. This type of force has been encountered previously, for example, in the context of forced fluid dynamics \cite{Bhattacharyya:2008ji} in the fluid/gravity correspondence. 

The conservation equations for the currents $\jt_{q+1}$ and $\jmt_{D-q-3}$ can be obtained similarly as Bianchi identities from the first equation in \eqref{eq:emdbianchi} and \eqref{eq:emdeom}. They yield
\begin{align} \label{eq:currentemd}
\begin{split}
&d\star \jt_{q+1}=0~~,\\
&d\star\jmt_{D-q-3}=0~~.
\end{split}
\end{align}
The dilaton current $\textbf{j}_\phi$ does not obey any conservation equation, as can be seen from its equation of motion \eqref{eq:emdeom}.

The dilaton in the case above (and all others considered in this paper) describes primary hair. 
We remark that in some setups there can be extra equations, that do not originate from conservation equations, but as a consequence of boundary conditions. This is for example the case in forced superfluid dynamics considered in \cite{Bhattacharya:2011eea} where the conservation equations are supplemented by an extra equation coming from requiring AdS asymptotics,  which in the dual fluid description becomes the zero curl condition
on the superfluid velocity.  

%
\subsubsection*{Equations for localized stress-energy tensor and currents}

The equations of motion \eqref{eq:forceemd}-\eqref{eq:currentemd} were obtained for arbitrary stress-energy tensor and currents. However, in most of this work we are interested in localized stress-energy tensor and currents describing a $(p+1)$-dimensional probe. In this particular case, 
\begin{equation} \label{eq:localt}
\begin{split}
&\bt=T^{\mu\nu}\tilde\delta^{(n+2)}(x^{\mu}-X^{\mu})~~,~~\jt_{q+1}=J_{q+1}\tilde\delta^{(n+2)}(x^{\mu}-X^{\mu})~~,\\
&\jmt_{D-q-3}=\mathcal{J}_{D-q-3}\tilde\delta^{(n+2)}(x^{\mu}-X^{\mu})~~,~~\textbf{j}_\phi=j_{\phi}\tilde\delta^{(n+2)}(x^{\mu}-X^{\mu})~~,
\end{split}
\end{equation}
where $\tilde\delta^{(n+2)}(x^{\mu}-X^{\mu})$ is the reparametrization invariant delta function localized in the $(n+2)=(D-p-1)$-transverse directions, $x^{\mu}$ are spacetime coordinates and $X^{\mu}$ the set of mapping functions describing the position of the object in the ambient spacetime with metric $g_{\mu\nu}$. All indices in \eqref{eq:localt} are tangential to the probe's worldvolume $\mathcal{W}_{p+1}$, e.g., $T^{\mu\nu}=T^{ab}\partial_a X^{\mu}\partial_b X^{\nu}$ where $\partial_a X^{\mu}$ acts as a projector onto the worldvolume. Projecting Eq.~\eqref{eq:forceemd} along the worldvolume one obtains
\begin{align} \label{eq:intemd}
\begin{split}
\nabla_{a}T^{ab}&=\frac{1}{(q+1)!}{\FF_{q+2}^{b a_1...a_{q+1}}}J_{q+1a_1...a_{q+1}}\\
&+\frac{(-1)^{qD+1}e^{a_q\phi}}{(D-q-3)!}{\FF}_{D-q-2}^{b a_1...a_{D-q-3}}\mathcal{J}_{D-q-3a_1...a_{D-q-3}}+j_{\phi}\partial^{b}\varphi~~,
\end{split}
\end{align}
expressing the conservation of the stress-energy tensor along the worldvolume. 
Here, and in the following, we have introduced $\FF_{q+2}$, $\varphi$ to denote the pull-back of the background fields
$F_{q+2}$, $\phi$ onto the worldvolume of the brane. 
Defining the transverse projector ${n^{i}}_\mu$ onto the transverse $(n+2)$-dimensional space such that ${n^{i}}_\mu\partial_aX^{\mu}=0$ we can project Eq.~\eqref{eq:forceemd} along the transverse direction to obtain
\begin{align} \label{eq:extemd}
\begin{split}
T^{ab}{K_{ab}}^{i}&=\frac{1}{(q+1)!}{\FF_{q+2}^{i a_1...a_{q+1}}}J_{q+1a_1...a_{q+1}}\\
&+\frac{(-1)^{qD+1}e^{a_q\phi}}{(D-q-3)!}{\FF}_{D-q-2}^{i a_1...a_{D-q-3}}\mathcal{J}_{D-q-3a_1...a_{D-q-3}}+j_{\phi}\partial^{i}\varphi~~,
\end{split}
\end{align}
where ${K_{ab}}^{i}={n_{\mu}}^{i}\nabla_a\partial_bX^{\mu}$ is the extrinsic curvature of the worldvolume. Eq.~\eqref{eq:extemd} expresses the mechanical balance of forces on the worldvolume. 

Finally, the current conservation equations \eqref{eq:currentemd} lead to
\beq \label{eq:cemd}
\nabla_{a_1} J^{a_1...a_{q+1}}_{q+1}=0~~,~~\nabla_{a_1} \mathcal{J}^{a_1...a_{D-q-3}}_{D-q-3}=0~~,
\eeq
which express the conservation of electric and magnetic currents along the worldvolume.\footnote{We have not assumed the existence of any boundaries $\partial\mathcal{W}_{p+1}$. In case they are present these equations must be supplemented by the boundary conditions $T^{ab} \eta_a=J^{a_1...a_{q+1}}_{q+1}\eta_{a_1}= \mathcal{J}^{a_1...a_{D-q-3}}_{D-q-3}\eta_{a_1}|_{\partial\mathcal{W}_{p+1}}=0$.}

So far we do not assume any specific form for the currents. 
In Sec.~\ref{sec:near} we will present explicit cases of these equations that arise as constraint equations for perturbations of particular types of black brane solutions.

\subsection{Type IIA/B supergravity}

Next we consider the slightly more non-trivial cases of type IIA/B supergravity. Both of these cases can be treated in the same framework using the democratic formulation introduced in \cite{Bergshoeff:2001pv}, in which the number of degrees of freedom of the RR sector are doubled.\footnote{One may consider the more general case in which the degrees of freedom of the NSNS sector are also doubled \cite{Dall'Agata:1998va, Bandos:2003et}, however, these formulations introduce several new structures which are not required for the purposes of this paper.} The action for the bosonic sector of ten-dimensional supergravity in the Einstein frame can be written as \cite{Bergshoeff:2001pv},
\beq \label{eq:type}
I=\frac{1}{16\pi G}\int_{\mathcal{M}_{10}}\left[\star R-\frac{1}{2}d\phi\wedge\star d\phi-\frac{1}{2}e^{-\phi}H_{3}\wedge \star H_{3}-\frac{1}{4}\sum_{q}e^{a_{q}\phi}\tilde F_{q+2}\wedge \star \tilde F_{q+2}\right]~,
\eeq
where the dilaton coupling $a_q$ is given by $a_q=(3-q)/2$ and $D=10$. Here the index $q$ runs over the values $q=0,2,4,6$ for type IIA\footnote{We are not including here the case $q=8$, 
 however, the results presented in this section could be easily generalized to include this case.} and the values $q=-1,3,5,7$ for type IIB. 

In this theory, we consider a brane probe with electric current $\textbf{j}_{2}$, which sources the NSNS field strength $H_{3}$, and a set of electric currents $\jt_{q+1}$, which source the RR field strengths $\tilde F_{q+2}$. We also assume that the brane is characterized by a dilatonic current $\textbf{j}_\phi$ and several magnetic currents, namely, a non-zero magnetic current ${\boldsymbol{\mathfrak{j}}}_{6}$, which modifies the Bianchi identity for $H_{3}$, and a set of magnetic currents $\jmt_{D-q-3}$, which modify the Bianchi identities for the RR fields, i.e.,
\begin{equation} \label{eq:Bianchidemocratic}
\begin{split}
&dH_{3}=16\pi G\star \boldsymbol{{\mathfrak{j}}}_{6}~~, \\
&d\tilde F_{q+2}-H_{3}\wedge \tilde F_{q}=16\pi G\left( \star \jmt_{D-q-3}-(\star\boldsymbol{{\mathfrak{j}}}_{6}\wedge C_{q-1})\right) ~~.
\end{split}
\end{equation}
This implies that the field strengths $H_{3}$ and $\tilde F_{q+2}$ are given by the expressions
\beq
H_{3}=dB_{2}+16\pi G\star \mathcal{D}_{7}~~,~~\tilde F_{q+2}=F_{q+2}-H_{3}\wedge C_{q-1}~~,
\eeq
where we have defined $F_{q+2}=dC_{q+1}+16\pi G \star D_{D-q-2}$. We have introduced the Dirac brane $\mathcal{D}_{7}$ satisfying $\star\boldsymbol{{\mathfrak{j}}}_{6}=d\star \mathcal{D}_{7}$, as well as the Dirac branes $D_{D-q-2}$ satisfying $\star\jmt_{D-q-3}=d\star D_{D-q-2}$ . One may now couple the currents of the 
charged probe brane to the fields of the theory \eqref{eq:type} via the sourced equations of motion 
\begin{align} \label{eq:typeeom}
\begin{split}
&d\left(\star (e^{-\phi}H_{3})-\sum_q\frac{1}{2!}e^{a_q\phi}\left[\star \tilde F_{q+2}\wedge C_{q-1}\right]\right)=-16\pi G\star  \textbf{j}_{2}~~,\\
&d\star(e^{a_q}\tilde F_{q+2})+(-1)^{q}e^{a_{q+2}\phi}\left[\star \tilde F_{q+4}\wedge H_{3}\right]=(-1)^{q}16\pi G\star \jt_{q+1}~~,\\
& \square \phi+\frac{1}{2}e^{-\phi}\star(H_{3}\wedge \star H_{3})-\sum_q \frac{a_q}{4}e^{a_q\phi}\star(\tilde F_{q+2}\wedge \star \tilde F_{q+2})=-16\pi G \textbf{j}_{\phi}~~.
\end{split}
\end{align}
We note that the second equation in \eqref{eq:typeeom} is in fact a set of equations, one for each $q$ present in \eqref{eq:type}. As noted in \cite{Bergshoeff:2001pv}, the action \eqref{eq:type} is a pseudo-action and the equations of motion \eqref{eq:typeeom} must be supplemented by the duality relations
\beq \label{eq:duality}
\tilde F_{q+2}=(-1)^{[q/2]+1}e^{-a_q\phi}\star \tilde F_{D-q-2}~~,
\eeq
in order to account for the correct number of degrees of freedom. Once imposing \eqref{eq:duality}, the equations of motion for the higher-form fields $\tilde F_{q+2}$ with $q=4,...,7$ yield the Bianchi identities \eqref{eq:Bianchidemocratic} for the lower form fields $q=-1,...,3$, while the Bianchi identities for the higher-form fields yield the equations of motion for the lower form fields. This means that the 8 currents (4 electric and 4 magnetic) associated with the higher-form fields are given in terms of the currents associated with the lower form fields, i.e.,
\begin{align}
\begin{split}
&\star \jt_{D-q-3}=(-1)^{q+1+[q/2]}\left(\star \jmt_{D-q-3}-(\star\boldsymbol{\mathfrak{j}}_{6}\wedge C_{q-1})\right)~~,\\
&\star  \jt_{q+1}=(-1)^{[q/2]}\left(\star \jmt_{q+1}-(\star\boldsymbol{\mathfrak{j}}_{6}\wedge C_{D-q-5})\right)~~,~~q=4...,7~~,
\end{split}
\end{align}
and furthermore, due to the self-duality relation $\tilde F_{5}=\star \tilde F_{5}$ we also have that
\beq \label{eq:selfcurrrent}
\star \jt_{4}=-\left(\star \jmt_{4}-(\star\boldsymbol{\mathfrak{j}}_{6}\wedge C_{2})\right)~~.
\eeq

Using the explicit form of the stress-energy tensor contributions \eqref{eq:emdstress}, with appropriate factors\footnote{The stress-energy tensor for the RR fields in the action \eqref{eq:type} is given by $1/2$ of \eqref{eq:emdstress} due to the $1/4$ factor in the action \eqref{eq:type} for the RR fields.}, one may readily derive 
as before the equations of motion of a probe brane in type IIA/B supergravity. The final expression is
\begin{align} \label{eq:forcetype}
\nabla_{\mu}\bt=&~\frac{1}{2!}{H_{3}^{\nu\mu_1\mu_2}}\textbf{j}_{2\mu_1\mu_{2}}+\frac{e^{-\phi}}{6!}{{H}_{7}^{\nu\mu_1...\mu_6}}\boldsymbol{\mathfrak{j}}_{6\mu_1...\mu_{6}}+\textbf{j}_\phi\partial^{\nu}\phi \nonumber\\
&+\sum_q\frac{1}{(q+1)!}\left({\tilde F_{q+2}^{\nu\mu_1...\mu_{q+1}}}+{(-1)}^{q+1}\frac{q(q+1)}{2!}H_{3}^{\nu \mu_1\mu_2}C_{q-1}^{\mu_3...\mu_{q+1}}\right)\jt_{q+1\mu_1...\mu_{q+1}} \nonumber\\
&+\sum_q\frac{e^{a_q\phi}}{(\tilde q+1)!}\left(\tilde{{F}}_{\tilde q+2}^{\nu\mu_1...\mu_{\tilde q+1}}+(-1)^{\tilde q+1}\frac{\tilde q (\tilde q-1)}{2!}H_{3}^{\nu \mu_1\mu_2}C_{\tilde q-1}^{\mu_3...\mu_{q+1}}\right) {{{\jmt}_{\tilde q+1}}}_{\mu_1...\mu_{\tilde q+1}} \nonumber \\
&+\frac{1}{4!}\left({\tilde F_{5}^{\nu\mu_1...\mu_{4}}}+3H_{3}^{\nu \mu_1\mu_2}C_{2}^{\mu_3...\mu_{4}}\right)\jt_{4\mu_1...\mu_{4}}
 \nonumber \\
&-\sum_q\frac{e^{a_q\phi}}{(q+2)!}\tilde F_{q+2}^{\mu_1...\mu_{q+2}}{\left[\star\boldsymbol{\mathfrak{j}}_{6}\wedge C_{q-1}\right]^{\nu}}_{\mu_1...\mu_{q+2}}~~,
\end{align}
where we have introduced the dual of $H_{3}$ via the relation ${H}_{7}=\star H_{3}$ and the duals $\tilde{{F}}_{D-q-2}=\star \tilde{{F}}_{q+2}$. The sums over $q$ take values only over $q=-1,...,2$ and we have introduced $\tilde q=D-q-4$. Note that since we have imposed the duality conditions \eqref{eq:duality} we assume in \eqref{eq:duality} that $C_{q}=0$ for $q\ge3$. The type of force terms involved here include Lorentz forces due to the presence of $H_{3},{H}_{7}$, $F_{q+2}$ and $\tilde{{F}}_{\tilde q+2}$, the force term due to the non-trivial dilaton as well as several force term 
originating from Chern-Simons terms. 

One can also obtain the conservation equations for the remaining currents using Eqs.~\eqref{eq:Bianchidemocratic} and \eqref{eq:typeeom}. They are
\begin{align} \label{eq:current}
\begin{split}
&d\star \jt_{q+1}+(-1)^{q+1}\star \jt_{q+3}\wedge H_{3}+(-1)^{q+1}e^{a_{q+2}\phi}\star \tilde F_{q+4}\wedge\star\boldsymbol{\mathfrak{j}}_{6}=0~~,~~q=0,...,3~~\\
&d\star \jmt_{D-q-3}=0~~,~~q=-1,0~~,~~d\star {\jmt}_{D-q-3}=H_{3}\wedge\star\jmt_{D-q-1}~~,~~q=1,2~~,\\
&d\star \textbf{j}_{2}=0~~,~~d\star \boldsymbol{\mathfrak{j}}_{6}=0~~.
\end{split}
\end{align}
The dilaton current $\textbf{j}_{\phi}$ does not obey any conservation equation neither does the current $\jt_{(0)}$. 

The reader can find the more specific form of the above expressions for each of the type IIA/B cases explicitly in 
App.~\ref{app:AB}. 

\subsection{Eleven-dimensional supergravity}

Finally we consider probe branes coupled to field configurations which are solutions of the eleven-dimensional supergravity action \cite{becker2006string}\footnote{It is possible, instead, to use the democratic formulation of eleven-dimensional supergravity \cite{Bandos:1997gd} but this necessitates introducing auxiliary variables. While such endeavor might be of interest, it goes beyond the purposes of this paper.}
\beq \label{eq:mtheo}
I=\frac{1}{16\pi G}\int_{\mathcal{M}_{11}}\left[\star R-\frac{1}{2}F_{4}\wedge \star F_{4}-\frac{1}{6} \int C_3 \wedge F_4 \wedge F_4\right] ~~.
\eeq
The probe branes are assumed to
carry an electric current $\jt_{3}$ and a magnetic current ${\jmt}_{6}$. The presence of the magnetic current modifies the Bianchi identity for the field strength $F_{4}$ so that
\beq
dF_{4}=16\pi G\star {\jmt}_{6}~~,~~F_{4}=dC_{3}+16\pi G \star D_{7}~~,
\eeq
where we have introduced the Dirac brane $D_{7}$ satisfying $\star {\jmt}_{6}=d\star D_{7}$. The electric coupling modifies the supergravity equations of motion 
\beq
d\star F_{4}+\frac{1}{2}\left(F_{4}\wedge F_{4}\right)=-16\pi G\star \jt_{3}~~.
\eeq

Repeating the procedure of the previous sections, we deduce the following modified conservation equations
\begin{align} \label{eq:forcemtheo}
\begin{split}
&\nabla_{\mu}\bt=\frac{1}{3!}{F_{3}^{\nu\mu_1\mu_2\mu_3}}\jt_{3\mu_1\mu_{2}\mu_3}+\frac{1}{6!}{{F}_{7}^{\nu\mu_1...\mu_6}}\jmt_{6\mu_1...\mu_{6}}~~,\\
&d\star \jt_{3}+\star \jmt_{6}\wedge F_{4}=0~~,\\
&d\star \jmt_{6}=0~~,
\end{split}
\end{align}
where we have defined the dual field strength ${F}_{7}$ such that ${F}_{7}=\star F_{4}$. In contrast with the 
type IIA/B cases, here the force equations \eqref{eq:forcemtheo} include only the Lorentz force terms.


\section{Near-zone analysis}
\label{sec:near}

We are now in position to move to the more involved part of our analysis. Our main task is to study the gravitational equations in the bulk near the horizon of the putative solution, and to identify there the role of the modified conservation equations of the previous section. This entails concrete information about the
form of the bulk solution and the physics near the horizon. In order to proceed systematically
it is convenient to restrict the scope of the exercise in two ways. 

First, it is useful to consider solutions that are long-wavelength deformations of a leading order 
homogeneous solution (with the proper asymptotics at infinity). If $\RR$ is the characteristic length scale of the 
deformation, then, by assumption, we consider the long-wavelength regime $r_{H} \ll \RR$ where $r_{H}$ is the smallest scale associated to the brane (such as the horizon size).\footnote{More precisely, in order to determine the regime of validity of these long-wavelength deformations, it is required that the magnitude of all scalar invariants (including scalars associated with the background) appearing in an effective action at a given perturbative order is much smaller than those of the preceding orders. Specific details for stationary perturbations and uncharged branes can be found in \cite{Armas:2015kra}.} The near-zone 
analysis focuses on the region $r\ll \RR$. In the discussion of Sec.~\ref{conclusions} we refer to this long-wavelength
expansion as the `exact brane' application of blackfolds. This expansion assumes an exact (partially) homogeneous solution 
of the gravitational equations with the required large-$r$ asymptotics.

Since an exact leading order solution is not always known it is frequently useful to resort to a further parallel expansion 
of the leading order solution in powers of the small ratio $\nicefrac{r_H}{L}$, where $L$ is the characteristic length scale of the 
background. In this regime, $r_H \ll \RR, L$, the perturbative expansion assumes small derivatives not only of the
black brane solution we are searching for, but also small derivatives of the asymptotic background solution. 
At leading order in $\nicefrac{r_H}{L}$ we can approximate the leading order solution in the near-zone region by a black $p$-brane 
solution in flat space. That is the basis of a concrete ansatz for the gravitational fields in the near-zone region.
Further qualitative features of the above two expansions are reviewed and re-discussed in the Sec.~\ref{conclusions}. 

In this section we proceed with the assumption of the double-perturbative regime 
$\nicefrac{r_H}{\RR}\ll 1$, $\nicefrac{r_H}{L}\ll 1$, and determine the role of the modified conservation equations
at leading order in the perturbation. 
We show in specific examples that equations like \eqref{eq:forceemd} arise as constraint equations in gravity 
in exact analogy to the derivations in the fluid/gravity correspondence. Conservation equations of the type \eqref{eq:intemd} and \eqref{eq:cemd} arise from worldvolume (intrinsic) perturbations of the brane solution, while the equilibrium equations \eqref{eq:extemd}, that determine $X^{\mu}$, arise from elastic-type (extrinsic) perturbations which break the symmetries of the transverse space to the worldvolume. We consider separately both types of perturbations.

Extending  the analysis of  \cite{Camps:2012hw} for neutral branes we treat in this section the case of perturbations of
dilatonic black $p$-branes charged under a $(p+1)$-form gauge potential and the D$p$-F1 bound state in type II string theory, both in the presence of external background fields. 
These two representative cases serve as illustrative examples, which can be further extended to more complicated brane configurations and/or to theories with additional matter fields.

\subsection{Constraint equations for charged black branes}
\label{sec:charged}

We start by considering the example of charged $p$-brane solutions in theories of gravity with 
spacetime action \eqref{eq:emd} with $q=p$. Besides the metric, the theory includes a dilaton and a $(p+1)$-form gauge field. We focus on the class of charged dilatonic $p$-brane solutions to the equations of motion that arise from \eqref{eq:emd} obtained in \citep{Horowitz:1991cd}. 
Incorporating a boost velocity $u^a$ on the worldvolume, the metric, dilaton $\phi$ and the $(p+1)$-form gauge field $C_{p+1}$ of the corresponding 
charged $p$-brane take the form
\begin{align}
\begin{split}
\label{eq:pbranesol}
\td s^2 &= H^{-\frac{Nn}{D-2}}\left(P_{ab}-f u_a u_b\right)\td \sigma^a\td \sigma^b+H^{\frac{N(p+1)}{D-2}}\left(f^{-1}\td r^2
+r^2\td \Omega_{(n+1)}^2 \right) \ , \\
\phi &=  \frac{a_{p} N}{2} \log H \ , \quad 
C_{p+1} = \sqrt{N}\coth \alpha \left(H^{-1}-1 \right) \star_{(p+1)} 1 \ .
\end{split}
\end{align}
Here $P_{ab}=\eta_{ab} + u_a u_b$ is the projector on the worldvolume in directions orthogonal to the constant unit timelike vector $u^a$, while $\star_{(p+1)} 1 = \td t \wedge \td x^1 \wedge ... \wedge \td x^p$ is the induced volume form on the worldvolume. The functions $f \equiv f(r)$ and $H \equiv H(r)$ are given by
\begin{equation}
\label{eq4:281112}
f(r)=1-\left(\frac{r_0}{r}\right)^n \ , \quad
H(r)=1+\left(\frac{r_0}{r}\right)^n\sinh^2\alpha ~.
\end{equation}
The dilaton coupling constant is arbitrary and related to $N, p, D$ through
\beq
\label{chargedaa}
n= D-p-3 \ , \quad a_{p}^2 = \frac{4}{N} - \frac{2(p+1)n}{D-2} \ .
\eeq
The effective currents and charges for this particular solution are given in App.~\ref{app:charged}.

We have also introduced the induced metric on the worldvolume
$\eta_{ab} = \eta_{\mu\nu}\, \d_a X^\mu \d_b X^\nu$ with trivial embedding scalars $X^\mu$, such that $\d_a X^\mu=\delta^{\mu}_{a}$. If the embedding scalars $X^{\mu}$, which determine the position of the worldvolume geometry, are non-trivial, they will induce a spontaneous breaking of the isometry group of the asymptotic background solution. In the case of flat space we have the breaking 
\begin{equation*}
\label{neutralad}
\text{SO}(1,D-1) \to \text{SO}(1,p) \times \text{SO}(n+2) ~~.
\end{equation*}

\subsubsection*{Leading order ansatz in the presence of background fields}

Starting from the exact solution above, we wish to construct classes of solutions to the action \eqref{eq:emd} that asymptote at large $r$ to a given background solution
\begin{equation} \label{eq:bgfields}
\td s^2 = g_{\mu\nu}(x)  \td x^{\mu}\td x^{\nu} \ , \quad \phi  (x) \ , \quad C_{p+1} (x) ~.
\end{equation}
Here, the asymptotic background fields $g_{\mu\nu}$, $\phi$ and $C_{p+1}$ are a priori arbitrary profiles.
We now assume that we put a brane in this background with embedding coordinates $X^\mu ( \sigma)$ and that
the (shortest) length scale $L$ characterizing the background 
is much larger than the (largest) length scale $r_H$ that enters the black brane solution \eqref{eq:pbranesol}, i.e. $r_H \ll L$. 
Thus to leading order we can ignore the variations of the background fields close to the brane. 
Given the brane with embedding coordinates $X^\mu (\sigma)$  and a foliation in the transverse coordinates, which
we  denote collectively by $y$, we can parametrize the spacetime with coordinates $X^\mu (\sigma,y)$ 
where we choose $X^\mu (\sigma,0) = X^\mu (\sigma)$.
This  means that in the overlap region, i.e. the asymptotic region of the brane, the background can  be approximated
as
\begin{equation} \label{eq:bgfields2}
\td s^2 \simeq  \gamma_{a b}(\sigma ,y ) \, \td \sigma^{a} \td \sigma^{b} + \td r^2 + r^2\td \Omega^2_{(n+1)}  
 \ , \quad \phi \simeq \varphi(\sigma, y) \ , \quad C_{p+1} \simeq \mc{C}_{p+1}(\sigma, y) ~.
\end{equation}
Here
\begin{equation}
\label{assfields}
\gamma_{ab} (\sigma ,y) =  \d_a X^\mu \d_b X^\nu g_{\mu\nu} (X (\sigma ,y) ) \ , \quad 
\varphi(\sigma, y) = \phi(X (\sigma ,y))  \ , \quad \mathcal{C}_{p+1}( \sigma, y )  = {C}_{p+1}  (X  (\sigma , y))
 \end{equation}
are the pull-backs of the background fields \eqref{eq:bgfields} onto the foliated hypersurfaces spanning the
spacetime. In particular for suitable chosen transverse coordinates (see below) at $y=0$ these expressions are the  pull-back onto the  worldvolume $\WW_{p+1}$ of the brane.   

To leading order we can make an ansatz for the full solution that at short distances is described by \eqref{eq:pbranesol} and at large distances by \eqref{eq:bgfields}. This can be achieved by  considering perturbations to the exact solution \eqref{eq:pbranesol} in which we promote the 
free constant parameters $r_0, \alpha, u^a, \d_a X^{\mu} $  to slowly varying functions of the worldvolume coordinates $\sigma^a$,
providing the leading order ansatz 
\begin{align}
\begin{split}
\label{eq3:281112}
\td s^2&=H^{-\frac{Nn}{D-2}}\left(P_{ab}(\sigma,y)-f u_a(\sigma) u_b(\sigma)\right)\td \sigma^a\td \sigma^b+H^{\frac{N(p+1)}{D-2}}\left(f^{-1}\td r^2
+r^2\td \Omega_{(n+1)}^2 \right)
\ , \\
C_{p+1} &= \mc C_{p+1}( \sigma,y) + e^{-a_{p}\varphi(\sigma,y)/2} \sqrt{N}\coth \alpha(\sigma) \left(H^{-1}-1 \right) \star_{(p+1)} 1 \ , \\
\phi &= \varphi(\sigma,y) + \frac{a_{p} N}{2} \log H \ ,
\end{split}
\end{align}
Here  $P_{ab}(\sigma,y ) = \gamma_{ab} (\sigma,y) + u_a (\sigma) u_b (\sigma)$
and $ \gamma_{ab} (\sigma ,y)$,  $\mc{C}_{p+1}(\sigma,y)$,  $ \varphi(\sigma,y)$ are the pull-backs of the background
fields given
in \eqref{assfields}. We also note that the functions $f$ and $H$ depend on $\sigma$ through their dependence on $r_0, \alpha$.  Furthermore, we have only kept the $\sigma$-dependence in $u^a (\sigma)$, absorbing possibly $y$-dependent
terms into higher order corrections to the metric. 

This ansatz has the property that it asymptotes at large $r$ to the background \eqref{eq:bgfields2} in the overlap region and, moreover it solves the equations of motion with constant $\gamma_{ab} = \eta_{ab},~ \mc{C}_{p+1},~ u^a,~ \varphi,~ r_0$, and $\alpha$ as it essentially reduces to \eqref{eq:pbranesol}. To see this, note first that we can always add a constant $\mc{C}_{p+1}$ to the gauge field $C_{p+1}$ as it corresponds to a gauge transformation and second that we can always shift the dilaton with a constant $\varphi$ if we also rescale the field strength with a constant factor $e^{-a_{p}\varphi/2}$. Indeed, both of these transformations correspond to symmetries of the action \eqref{eq:emd}.

Once the brane parameters are promoted to functions of $\sigma^{a}$ as is done in the ansatz \eqref{eq3:281112}, it is necessary to add corrections to all of the fields, which are determined by solving the equations of motion of the action. In a derivative expansion, these corrections can be of two types:  intrinsic and extrinsic. To first order, intrinsic corrections are derivative corrections in the brane parameters $r_0, \alpha, u^{a}, \varphi$. Denoting $\lambda$ as the minimum length scale associated with fluctuations of these fields, we introduce the small expansion parameter $\varepsilon=r_H/\lambda\ll1$ which controls intrinsic perturbations. 

Extrinsic perturbations, in turn, are fluctuations in the extrinsic geometry of the brane worldvolume. To first order they appear due to the non-zero extrinsic curvature ${K_{ab}}^{i}$ of the induced metric. Denoting $\RR$ as the length scale associated to such fluctuations we introduce the expansion parameter $\tilde\varepsilon=r_H/\RR\ll 1$ which controls extrinsic perturbations. In general, the corrections appear as a multipole expansion in the transverse sphere $S^{(n+1)}$.
For example, for the metric perturbation, we can write to first order in derivatives 
\begin{equation}
h_{\mu\nu}(r,\theta)= \varepsilon f_{\mu\nu}(r) + \tilde\varepsilon\cos \theta\, d_{\mu\nu}(r) + \dots ~~,
\end{equation}
where $f_{\mu\nu}(r)$ is the monopole part and purely intrinsic and $d_{\mu\nu}(r)$ is the dipole part and purely extrinsic. 
Thus at first order intrinsic and extrinsic deformations decouple, which will be used below. 
To higher orders the two types of perturbations couple generically. It is important to stress that our focus will not be to determine the corrections $h_{\mu\nu}$ (and the corresponding
ones for the gauge field and dilaton), but instead to extract the subset of field equations that are exactly independent of these corrections.
\subsubsection*{Intrinsic equations}
\label{intCharged}

We first consider the intrinsic perturbations $f_{\mu\nu}$, which do not break the symmetries of the transverse space, 
i.e. $\d_a X^\mu=\delta^{\mu}_{a}$ so that the induced metric becomes trivial. 
 It follows that the leading order background metric  \eqref{eq:bgfields2} in the overlap region is given by 
 \begin{equation}
\td s^2 = \eta_{a b} \, \td \sigma^{a} \td \sigma^{b} + \td r^2 + r^2\td \Omega^2_{(n+1)} \, , \quad	
\phi = \varphi(\sigma) \ , \quad C_{p+1} = \mc{C}_{p+1}(\sigma) ~,
\end{equation}
and we proceed by considering the derivative expansion of the leading order solution \eqref{eq3:281112}. 
Because the background fields only depend on the worldvolume coordinates $\sigma^{a}$, the intrinsic perturbations do not couple to extrinsic perturbations.
The analysis of these perturbations is very similar to that of the well-known fluid/gravity correspondence \cite{Bhattacharyya:2008ji} (see also  \cite{Erdmenger:2008rm,Banerjee:2008th} which treats the case for charged
branes).

The expansion is controlled by the parameter $\varepsilon$ which is assumed to be of the order of the inverse length scale of the variation of the fields $u^a, \varphi, r_0$, and $\alpha$ over the length scale $\lambda$ of the fluctuations. We can then write, up to first order, the expansions
\begin{equation} \label{eq:intexpansions}
	\td s^2 = \td s_{(0)}^2 + \varepsilon f_{\mu\nu} \td x^{\mu} \td x^{\nu} \ , \quad
	\phi = \phi^{(0)}+\varepsilon \phi^{(1)}  \ , \quad
	C_{p+1} = C_{p+1}^{(0)}+\varepsilon C_{p+1}^{(1)} \ ,
\end{equation}
where we regard \eqref{eq3:281112} as the first term in the expansion, i.e., as $\td s_{(0)}^2, \phi^{(0)}$, and $C^{(0)}_{p+1}$ and the (intrinsic) fields $u^a, \varphi, r_0$, and $\alpha$ are expanded at a given point $\mathcal{P}$ such that
\begin{align} \label{eq:intexp}
\begin{split}
u^a(\sigma) &= u^a|_{\mc{P}} + \varepsilon \sigma^b \partial_b u^a|_{\mc{P}} + \mc{O}(\varepsilon^2) ~\, , \quad
\varphi(\sigma) = \varphi|_{\mc{P}} + \varepsilon \sigma^a \partial_a \varphi|_{\mc{P}} + \mc{O}(\varepsilon^2) \\
r_0(\sigma) &= r_0|_{\mc{P}} + \varepsilon \sigma^a \partial_a r_0|_{\mc{P}} + \mc{O}(\varepsilon^2) ~\, , \quad
\alpha(\sigma) = \alpha|_{\mc{P}} + \varepsilon \sigma^a \partial_a \alpha|_{\mc{P}} + \mc{O}(\varepsilon^2) ~.
\end{split}
\end{align}
We note that since $\mc C_{p+1}$ does not depend on the transverse space coordinates it simply corresponds to a gauge transformation. It therefore plays an irrelevant role for flat extrinsic geometry. The expansions \eqref{eq:intexpansions} are then inserted into the field equations that arise from the action \eqref{eq:emd} with the stress-energy tensor for the energy/matter fields given in \eqref{eq:emdstress}. The resulting equations are regarded as \quoting{ultralocal} in the worldvolume coordinates and are as such simply linear ordinary differential equations for radial fluctuations of the metric, dilaton, and gauge field in the background \eqref{eq3:281112}.

It is interesting to note that the first order source terms appearing in the ODE's for the metric do not involve derivatives of the background scalar $\varphi$. The first order metric corrections $f_{\mu\nu}$ are therefore not altered by the non-trivial background to first order. In fact, they were determined in \cite{DiDato:2015dia}. The constraint equations constitute a subset of the field equations, coming from the $(rb)$ component of the metric EOM
and the $(a_1 \ldots a_p)$ component of the gauge field EOM.  To first order they take the form
\begin{equation} \label{eq:chargedint}
	\nabla_a T^{a b} = j_{\phi} \partial^b \varphi \ , \quad \nabla_a J_{p+1}^{a a_1 ... a_p} = 0 ~, 
\end{equation}
where we have introduced the modified current and dilaton current
\begin{equation} \label{eq:modcurrent}
J_{p+1} = \star_{(p+1)} \left( e^{a_{p}\varphi/2} Q_{p} \right) ~\, , \quad
j_{\phi} = \frac{a_{p}}{2} \Phi_p Q_{p} ~.
\end{equation}
The stress-energy tensor $T^{ab}$, the total electric charge $Q_{p}$ and electric potential $\Phi_p$ are given in App.~\ref{app:charged}. The equations \eqref{eq:chargedint} are valid for all values of $r$ up to first order in derivatives. One could now proceed order-by-order in the expansion, but since we are interested in simultaneously considering extrinsic perturbations of the worldvolume geometry we truncate at first order.

\subsubsection*{Extrinsic equations}
\label{extCharged}

We now repeat the same steps as in the analysis for the extrinsic perturbations performed for neutral branes%
\footnote{As noted in \cite{Camps:2012hw} the analysis of that paper still holds for charged branes and/or backgrounds with a non-trivial
metric. The generalization below thus pertains mainly the fact that we allow for further non-trivial background fields.}
 in \cite{Camps:2012hw},  but with a non-trivial background. 
 To consider the extrinsic perturbations  it is useful to introduce Fermi normal coordinates adapted to the worldvolume $\mc W_{p+1}$. 
 In these coordinates, the metric  is parametrized by the coordinates $(\sigma^{a},y_{i})$ where $y_{i}=0$ denotes the position of the worldvolume $\mc W_{p+1}$ in the transverse $(n+2)$-dimensional space. 
 Thus we have that  $ \delta X^{i}(\sigma,y) = y^{i}$ for the transverse scalars around the flat embedding $\eta_{ab}$ and
 under the variation $X^{i} \rightarrow X^{i}+\delta X^{i}$ one finds that the  induced metric transforms as
$\delta \gamma_{ab}=-2K_{ab}{}^{i}\delta X_{i} = -2K_{ab}{}^{i}y_{i}$. 
 
  The perturbations along any of the coordinates $y_i$ decouple 
 \cite{Gorbonos:2004uc} and therefore we restrict the analysis to a given $i=\hat{i}$. Introducing the direction cosine $y_{\hat i}=r\cos\theta$ and noting that $r^2=y_i y^{i}$, a curved worldvolume metric $\gamma_{ab}$ to first order in derivatives in these coordinates can then be expanded as
\begin{equation}
 \gamma_{ab}=\eta_{ab}-2K_{ab}^{\phantom{ab}\hat{i}} \; r \cos\theta + \mathcal{O}(\tilde \varepsilon^2) ~,
\end{equation}
where ${K_{ab}}^{\hat i}$ is a one-derivative term of $\mathcal{O}(\tilde \varepsilon)$. 
 This means that 
the leading order background metric  \eqref{eq:bgfields2} in the overlap region is given by 
\begin{align}
\begin{split}
\label{eq:bgr}
& \td s^2 = \left(\eta_{ab} - 2{K_{ab}}^{\hat i} \, r \cos\theta \right)\td \sigma^a\td \sigma^b +
\td r^2 + r^2\left(\td \theta^2+\sin^2\theta \thinspace \td  \Omega_{(n)}^2\right) \, , \\
& C_{p+1} = \mc{C}_{p+1}(y) \ , \quad \phi = \varphi(y) ~.
\end{split}
\end{align}
Note that in this case we only allow the background fields to have dependence on the transverse coordinates $y$, so that the extrinsic
perturbations are decoupled from the intrinsic ones.  
We also remark that we can assume that the background field strength is non-zero only in the singled-out direction $y^{\hat{i}}$. 
In principle, $\mc{F}_{p+2} = \td \mc{C}_{p+1}$ could have non-zero components in two or more transverse directions, however, it is not difficult to realize that such components do not play a role to leading order and only become important at higher orders. 

We now consider a dipole-type perturbation to first order in $\tilde\varepsilon$ and write the expansion of the fields as
\begin{align} \label{eq:asympExp}
\begin{split}
\td s^2 &= \td s^2_{(0)} + \tilde\varepsilon\cos\theta d_{\mu\nu}(r) \td x^{\mu} \td x^{\nu} \ , \quad \phi = \phi^{(0)}+\tilde \varepsilon \phi^{(1)} \, , \quad
F_{p+2} =  \td C^{(0)}_{p+1} + \tilde\varepsilon \td C^{(1)}_{p+1} \ , \\
\end{split}
\end{align}
where we regard \eqref{eq3:281112} as the first term in the expansion (denoted by subscript $(0)$), 
and we note that $\td y^{\hat i} = \left(\cos \theta \td r-r\sin \theta \td \theta \right)$. 
Analogous to  \cite{Camps:2012hw} we focus on the large $r$ region of the near-zone solution. 
In this region, the metric takes the form 
\begin{equation}\label{overlapmetric}
\begin{split}
\td s^2=& \left(\eta_{ab}-2  {K_{ab}}^{\hat i} \, r \cos\theta +\frac{16\pi G}{n\Omega_{(n+1)}}\left(T_{ab}({\sigma})-\frac{T({\sigma})}{D-2} \eta_{ab}\right)\frac{1}{r^n} \right)\td \sigma^a\td \sigma^b \\ & +
\left(1 - \frac{16\pi G}{\Omega_{(n+1)}}\frac{1}{D-2}\frac{T({\sigma})}{r^n}\right)\td r^2 + r^2\left(\td \theta^2+\sin^2\theta \thinspace \td  \Omega_{(n)}^2\right) \\
& + \tilde\varepsilon \cos \theta\, d_{\mu\nu}(r) \td x^\mu \td x^\nu + \mathcal{O}(\tilde\varepsilon^2)+ \mathcal{O}(T_{ab}^2/r^{2n}) ~.
\end{split}
\end{equation}
while the field strength and dilaton are given by
\begin{equation}\label{eq:overlapzonefields}
\begin{split}
F_{p+2}&=F^{\text{(M)}}_{p+2} + \tilde\varepsilon \mc{F}_{p+2} + \mc{O}(\tilde\varepsilon^2) +\mathcal{O}\left(T_{ab}^2/r^{2n}\right) \ , \\
\td \phi&=\td \phi^{\text{(M)}} + \tilde\varepsilon\td \varphi + \mc{O}(\tilde\varepsilon^2) + \mathcal{O}\left(T_{ab}^2/r^{2n}\right) \ .
\end{split}
\end{equation}
The first order corrections $\td C^{(1)}_{p+1}$ and $\td \phi^{(1)}$ drop out as they appear at higher order in the large $r$ expansion. This also means that only the monopole part - by definition of order $\mathcal{O}(T_{ab}/r^n)$ - of $\td C^{(0)}_{p+1}$ and $\td\phi^{(0)}$ plays a role, which we denote by $F^{\text{(M)}}_{p+2}$ and $\td \phi^{\text{(M)}}$ and are given by
\begin{equation}
\begin{split}
\label{Monparts}
F_{p+2}^{(M)}= -e^{-a_{p}\varphi/2}\frac{16\pi G}{\Omega_{(n+1)}}  \frac{Q_{p}}{r^{n+1}} \ts \td r\wedge \star_{(p+1)}1 \ , \quad
\td \phi^{(M)}=-\frac{16\pi G}{\Omega_{(n+1)}} \frac{a_{p} \Phi_p Q_{p}}{2r^{n+1}} \ts \td r \ .
\end{split}
\end{equation}
and are thus determined by the current source \eqref{eq:modcurrent} of the brane. 

In the presence of non-trivial background fields, we therefore see that the bulk stress tensor $\textbf{T}_{\textbf{M}\mu\nu}$ does not vanish at large $r$ distances. Indeed, from \eqref{eq:overlapzonefields} the stress-energy tensor consists of a simple pole-dipole term and in particular does not involve the metric and background field perturbations. The combination
\begin{equation}\label{eq:exteq1}
(n+1) \thinspace \text{csc} \thinspace \theta \thinspace \left(G_{r\theta}-8\pi G \thinspace \textbf{T}_{\textbf{M}r\theta}\right)- r \thinspace \text{sec} \thinspace\theta \thinspace \left( G_{rr}-8\pi G \thinspace \textbf{T}_{\textbf{M}r r}\right)=0 \ ,
\end{equation}
is therefore still a constraint equation\footnote{When no background fields are present ($\textbf{T}_{\textbf{M}\mu\nu}=0$), the constraint equation \eqref{eq:exteq1} was obtained in \cite{Camps:2012hw}.} with $\mathbf{T}_{\textbf{M}\mu\nu}=\mathbf{T}_{\textbf{M}\mu\nu}^{(F)}+\mathbf{T}_{\textbf{M}\mu\nu}^{(\phi)}$ and takes the modified form
\begin{equation}
\frac{ n+2}{r^{n}}\frac{8\pi G}{\Omega_{(n+1)}}\left(T^{ab}K_{ab}^{\ \ \hat i}-\mathcal{F}^{\hat i}\right)=0 \ ,
\end{equation}
where $\mathcal{F}^{\hat i}$ is the induced force term given by the pole-dipole interaction term
\begin{equation}\label{eq:forcetermwithT}
\mathcal{F}^{\hat i}=\frac{\Omega_{(n+1)}}{n+2}r^n \left((n+1) \text{csc} \thinspace \theta \thinspace \textbf{T}_{\textbf{M}r \theta} - r \thinspace \text{sec} \thinspace \theta \thinspace  \textbf{T}_{\textbf{M}r r} \right) \ .
\end{equation}
To evaluate the force term \eqref{eq:forcetermwithT} we use the expressions \eqref{eq:emdstress} for the stress-energy tensor, insert
the expansion \eqref{eq:overlapzonefields} keeping only the pole-dipole contribution and then use  \eqref{Monparts}
 along with \eqref{eq:modcurrent}.  Covariantizing the result for any $i=\hat i$, we arrive at the following constraint equation
\begin{equation}\label{eq:fullconsteq}
T^{ab}K_{ab}^{\phantom{ab}i} = n^{i}_{\phantom{i} \mu}  \left(\frac{1}{(p+1)!} \mc{F}^{\mu\mu_1\dotsc\mu_{p+1}}_{p+2}J_{p+1\mu_1\dotsc\mu_{p+1}}+ j_{\phi}\partial^{\mu} \varphi \right) \ ,
\end{equation}
where the currents $J_{p+1}$ and $j_{\phi}$ are given by Eq.~\eqref{eq:modcurrent}. 
We emphasize that the quantities appearing in the constraint equation \eqref{eq:fullconsteq} are evaluated in the region $r_H\ll r \ll {\rm min}(\mathcal{R},L)$. 
Together \eqref{eq:chargedint} and \eqref{eq:fullconsteq} are in agreement with Eqs.~\eqref{eq:intemd}-\eqref{eq:cemd} for a localized stress-energy tensor and current.

\subsection{Constraint equations for the D\texorpdfstring{$p$}{p}-F1 bound state} 
\label{sec:DpF1}

We now apply the same procedure to the non-extremal D$p$-F1 bound state, which will exhibit some new features. 
The non-extremal D$p$-F1 bound state in the Einstein frame has the metric \cite{Harmark:2000wv}
\begin{align} \label{eqn:DpF1sol}
\begin{split}
  \td s^2 =& \; D^{\frac{1-p}{8}} H^{\frac{p-7}{8}} \left( -f u_a u_b + v_a v_b \right) \td \sigma^a \td \sigma^b +
	D^{\frac{9-p}{8}} H^{\frac{p-7}{8}} \perp_{ab}\td \sigma^a \td \sigma^b + \\
	&\; D^{\frac{1-p}{8}} H^{\frac{p+1}{8}} \left( f^{-1} \td r^2 + r^2\td \Omega^2_{8-p} \right) ~,
\end{split}
\end{align}
where $u_a$ is a normalized timelike vector (the boost velocity), $v_a$ is a normalized spatial vector characterizing the direction of the F1-string satisfying the orthogonality condition $u^{a} v_a = 0$ and the projector onto the worldvolume directions orthogonal to the string is $\perp^a_{\phantom{a}b} \equiv \delta^a_{\phantom{a}b} + u^a u_b - v^a v_b$. The dilaton is given by
\begin{equation}
	e^{2(\phi - \varphi)} = D^{\frac{p-5}{2}} H^{\frac{3-p}{2}} ~,
\end{equation}
and the gauge fields are\footnote{Beware of the solution in Ref.~\cite{Harmark:2000wv} which has a typo in the $C_{p+1}$ field that has been correct in Ref.~\cite{Grignani:2013ewa} for the D3-F1 solution.}
\begin{align} 
	B_2 &= e^{ - a_{\text{F1}} \varphi / 2} \sin\xi \left( H^{-1} - 1 \right) \coth \alpha \; u \wedge v ~, \label{eqn:DpF1potentials1} \\
	C_{p-1} &= (-1)^p \; e^{ -a_{p-2} \varphi / 2} \tan \xi \left( D H^{-1} - 1 \right) \; \star_{(p+1)} \left( u \wedge v \right) ~, \label{eqn:DpF1potentials2} \\	
	C_{p+1} &= (-1)^p \; e^{ -a_{p} \varphi / 2} \cos\xi \left( H^{-1} - 1 \right) \coth \alpha \; \star_{(p+1)} \mathbf{1} ~, \label{eqn:DpF1potentials3}	
\end{align}
where $B_2$ is the NSNS two-form, and $C_{p-1}$ and $C_{p+1}$ are the $(p-1)$-form and $(p+1)$-form RR fields, respectively. The dilaton coupling constants are $a_q = (3-q)/2$ and $a_{\text{F1}}= -1$. The Hodge star operator $\star_{(p+1)}$ is defined with respect to the $(p+1)$-dimensional worldvolume metric. The structure functions $f\equiv f(r)$, $D\equiv D(r)$ and $H\equiv H(r)$ are
\begin{equation}
	f(r) = 1 - \frac{r_0^n}{r^n} \, , \quad H(r) = 1 + \frac{r_0^n}{r^n} \sinh^2\alpha \, , \quad D^{-1}(r) = \cos^2\xi + H^{-1} \sin^2\xi ~,
\end{equation}
with $n = 7-p$. The solution depends on three real parameters; $r_0 > 0$, $\alpha$, and the angle $\xi \in [0, 2\pi[$. For $\xi = 0$, the solution reduces to the $p$-brane solution \eqref{eq:pbranesol} in ten dimensions. 
Following the previous subsection, we have already included in the expressions above the redundant shift $\varphi$ of the dilaton to make this particular symmetry of the action manifest. The field strengths are\footnote{For $p=3$, the composite five-form field strength $\tilde{F}_5$ should be made self-dual.}
\begin{equation} \label{eqn:DpF1inv}
  H_3 = \td B_2 \, , \quad F_{p} = \td C_{p-1} \, , \quad \tilde{F}_{p+2} = \td C_{p+1} - H_3 \wedge C_{p-1} ~ .
\end{equation}
For the $\tilde{F}_{p+2}$ to be invariant under the gauge transformation $\delta C_{p-1} = \td \Lambda_{p-2}$, the $(p+1)$-form potential should transform as $\delta C_{p+1} = \td \Lambda_{p-2} \wedge B_2$. The $(p+2)$-form is invariant under the gauge transformation $\delta C_{p+1} = \td \Lambda_{p}$.
The effective currents and charges are given in App.~\ref{eqn:DpF1charges}.

\subsubsection*{Ansatz and background fields}

The perturbation ansatz is constructed from the bound-state solution \eqref{eqn:DpF1sol}-\eqref{eqn:DpF1inv} by promoting the parameters $u^a, v^a, r_0, \alpha, \xi$ to functions of the worldvolume coordinates $\sigma^{a}$, as well as promoting the worldvolume metric $\eta_{ab} \rightarrow \gamma_{ab}({\sigma})$. In addition, we now include the general background gauge fields $\mathcal{B}_2, \mathcal{C}_{p-1}$,  $\mathcal{C}_{p+1}$
and dilaton $\varphi$ in the overlap region (cf. \eqref{assfields}).  Explicitly, they enter through the associated field strengths
\begin{equation}
	\mathcal{H}_3 = \td \mathcal{B}_2 \, , \quad 
	\mathcal{F}_p = \td \mathcal{C}_{p-1} \, , \quad 
	\tilde{\mathcal{F}}_{p+2} = \td \mathcal{C}_{p+1} - \mathcal{H}_3 \wedge \mathcal{C}_{p-1} ~.
\end{equation}
In analogy with \eqref{eq3:281112}, the ans\"atze for the field strengths are thus composed by taking
\begin{equation}
	H_3 \rightarrow H_3 + \mathcal{H}_3 \, , \quad 
	F_p \rightarrow F_p + \mathcal{F}_p \quad \text{and} \quad 
	\tilde{F}_{p+2} \rightarrow \tilde{F}_{p+2} + \tilde{\mathcal{F}}_{p+2} ~.
\end{equation}
In this way, the background fields do not affect the solution at leading order in the derivative expansion and the ansatz asymptotes at large $r$ to the pull-back of the background fields
\begin{align}
\begin{split}	
	H_3 &= \mathcal{H}_3(\sigma,y) \, , \quad
	F_{p} = \mathcal{F}_{p}(\sigma, y) \, , \quad
	\tilde{F}_{p+2} = \tilde{\mathcal{F}}_{p+2}(\sigma, y) \, , \quad 
	\phi = \varphi(\sigma, y) ~.
\end{split}
\end{align}
The asymptotic charges are altered accordingly
\begin{equation}
  \mathbf{Q}_{\text{F1}} = e^{a_{\text{F1}} \varphi / 2} \mathcal{Q}_{\text{F1}} + (-1)^{p} e^{a_{p} \varphi / 2} Q_p \mathcal{C}_{p-1} \, , \quad
	\mathbf{Q}_{p-2} = e^{a_{p-2} \varphi / 2} \mathcal{Q}_{p-2} \, , \quad
	\mathbf{Q}_p = e^{a_{p} \varphi / 2} {Q}_p ~,
\end{equation}
where the charge densities $\mathcal{Q}_{\text{F1}},~\mathcal{Q}_{p-2}$  and total charge $Q_p$ are given by Eq.~\eqref{eqn:DpF1charges0} and the currents take the expected form
\begin{equation} \label{eq:DpF1currents}
	j_2 = \mathbf{Q}_{\text{F1}} \; u \wedge v \, , \quad J_{p-1} = \mathbf{Q}_{p-2} \star_{(p+1)} (u \wedge v) \, , \quad J_{p+1} = \mathbf{Q}_p \star_{(p+1)} \mathbf{1} ~.
\end{equation}
In the following we consider the intrinsic and extrinsic perturbations separately. 

\subsubsection*{Intrinsic equations}

We restrict the dependence of the background fields to the worldvolume coordinates $\sigma^a$ and consider each field in a derivative expansion analogous to Eq.~\eqref{eq:intexpansions} allowing in principle for the construction of a multi-charged bound state solution order-by-order. We note that since the fields are restricted to worldvolume coordinate dependence only, the background field $\mathcal{C}_{p+1}$ simply corresponds to a pure gauge transformation analogous to Sec.~\ref{intCharged}.

We are interested in the subset of field equations Eqs.~\eqref{eq:IIAeom} (and \eqref{eq:IIBeom}) that constitute the set of constraint equations. Taking particular combinations of the constraint equations, which we list in App.~\ref{sec:intrcomb}, these can be expressed as stress-energy and current conservation equations on the worldvolume 
\begin{align} \label{eqn:DpF1bgrcons}
\begin{split}
	\nabla_a T^{ab} =& \; 
	 \frac{1}{(p-1)!} \left[ \mathcal{F}_{p}^{b a_1 ... a_{p-1}} J_{p-1 a_1...a_{p-1}}
	  + (-1)^{p+1} \frac{1}{2} \mathcal{H}^{b a_1 a_2}_3 \mathcal{C}_{p-1}^{a_3 ... a_{p+1}}  J_{p+1 a_1 ... a_{p+1}}  \right] \\
	&\;  +	\frac{1}{2} \mathcal{H}^{b a_1 a_2}_3 j_{2a_1 a_2}
	+	j_{\phi} \partial^b \varphi \, , \\
	\nabla_a j_{2}^{ab} =&\; 0 \, , \quad
	\nabla_{a} J^{a b_1 ... b_{p-2}}_{p-1} = \frac{1}{3!} \mathcal{H}_{3 abc}  J_{p+1}^{abc b_1 ... b_{p-2}}  \, , \quad
	\nabla_a J_{p+1}^{a b_1 ... b_{p}} = 0 ~,
\end{split}
\end{align}
where the dilaton current is related to the charges such that
\begin{equation} \label{eq:dilatoncurrentDpF1}
	j_{\phi} = \frac{1}{2} \left( a_{\text{F1}} \mathcal{Q}_{\text{F1}}\Phi_{\text{F1}} + a_{p} Q_p \Phi_p \right) ~.
\end{equation}	
These are worldvolume equations and are satisfied for all values of $r$ up to first order in derivatives. The stress-energy tensor is given in Eq.~\eqref{eq:DpF1stress} and the currents in Eq.~\eqref{eq:DpF1currents}.

\subsubsection*{Extrinsic equations}

For the extrinsic perturbations we restrict the dependence of the background fields to the transverse space spanned by the coordinates $y^{i}$ and consider the perturbation to first order in a single direction $y^{\hat{i}} = r \cos\theta$. The $r$-asymptotics of the near-zone field strengths therefore follow a similar form as \eqref{eq:overlapzonefields}, explicitly
\begin{align}
H_3 &= H^{(M)}_{3} + \tilde\varepsilon \mathcal{H}_{3} + \mathcal{O}(\tilde\varepsilon^2) + \mathcal{O}(T_{ab}^2/r^{2n}) \, , \quad
F_{p} = F^{(M)}_{p} + \tilde\varepsilon \mathcal{F}_{p} + \mathcal{O}(\tilde\varepsilon^2) + \mathcal{O}(T_{ab}^2/r^{2n}) \, , \\
\tilde{F}_{p+2} &= \tilde{F}^{(M)}_{p+2} + \tilde\varepsilon \tilde{\mathcal{F}}_{p+2} + \mathcal{O}(\tilde\varepsilon^2) + \mathcal{O}(T_{ab}^2/r^{2n}) \, , \quad
\td \phi = \td \phi^{\text{(M)}} + \tilde\varepsilon\td \varphi + \mathcal{O}(\tilde\varepsilon^2) + \mathcal{O}\left(T_{ab}^2/r^{2n}\right) \ , \nonumber
\end{align}
while the asymptotic metric is again of the form given by Eq.~\eqref{overlapmetric}. The monopole parts of the leading order fields are
\begin{align}
\begin{split}
H^{(M)}_{3} &= - e^{-a_{\text{F1}} \varphi/2} \frac{16 \pi G}{\Omega_{(n+1)}} \frac{\mathcal{Q}_{\text{F1}}}{r^{n+1}}  \; \td r \wedge u \wedge v \, , \\
F^{(M)}_{p} &= (-1)^{p+1} e^{-a_{p-2} \varphi/2} \frac{16 \pi G}{\Omega_{(n+1)}} \frac{\mathcal{Q}_{p-2}}{r^{n+1}} \, \td r \wedge \star_{(p+1)} (u\wedge v) \, , \\
\tilde{F}^{(M)}_{p+2} &= (-1)^{p+1} e^{-a_{p} \varphi/2} \frac{16 \pi G}{\Omega_{(n+1)}} \frac{Q_p}{r^{n+1}} \, \td r \wedge \star_{(p+1)} \textbf{1} \, , \quad 
\td \phi^{(M)}=-\frac{16\pi G}{\Omega_{(n+1)}} \frac{j_{\phi}}{r^{n+1}} \, \td r ~,
\end{split}
\end{align}
where $j_{\phi}$ is given by Eq.~\eqref{eq:dilatoncurrentDpF1}.

The constraint equations can again be extracted through the combination of the bulk stress-energy tensor given by Eq.~\eqref{eq:exteq1}, where the force terms arising from the presence of the background fields are given by Eq.~\eqref{eq:forcetermwithT}. In particular, the individual covariantized contributions are
\begin{align}
\begin{split}
\mathcal{F}^{\mu}_{H} &= \frac{1}{2} \mathcal{H}_3^{\mu \mu_1 \mu_2} \left( j_{2\mu_1\mu_2} + (-1)^{p+1} \frac{1}{(p-1)!} \mathcal{C}^{\mu_3 ... \mu_{p+1}}_{p-1} J_{p+1\mu_1 ... \mu_{p+1}} \right) \, , \\
\mathcal{F}^{\mu}_{F_p} &= \frac{1}{(p-1)!} \mathcal{F}_p^{\mu \mu_1 ... \mu_{p-1}} J_{p-1\mu_1 ... \mu_{p-1}} \, , \\ 
\mathcal{F}^{\mu}_{\tilde{F}_{p+2}} &= \frac{1}{(p+1)!} \tilde{\mathcal{F}}_{p+2}^{\mu \mu_1 ... \mu_{p+1}} J_{p+1\mu_1 ... \mu_{p+1}} \, , \quad 
\mathcal{F}^{\mu}_{\phi} = j_{\phi} \partial^{\mu} \varphi ~.
\end{split}
\end{align}
We note that the currents $j_2, J_{p-1}$ and $J_{p+1}$ are given by Eq.~\eqref{eq:DpF1currents} in terms of the modified charges. The extrinsic equations thus take the form
\begin{align} \label{eq:DpF1extrinsic}
\begin{split}
& T^{ab} K_{ab}^{\phantom{ab}i} = n^{i}_{\phantom{i} \mu} 
\left(
\mathcal{F}^{\mu}_{{H}} +
\mathcal{F}^{\mu}_{{F}_p} + 
\mathcal{F}^{\mu}_{{F}_{p+2}} + 
\mathcal{F}^{\mu}_{\phi}
\right) ~.
\end{split}
\end{align}
This is exactly the stress-energy tensor conservation equation in the perpendicular directions to the worldvolume. 

Finally, we note that the intrinsic equations \eqref{eqn:DpF1bgrcons} together with the extrinsic equations \eqref{eq:DpF1extrinsic} are in agreement with the far-region analysis \eqref{eq:forcetype} for localized stress-energy tensor and currents (see App.~\ref{app:AB}) 
once we replace in those expressions the fields with their pull-back onto the worldvolume (analogous to the discussion at the end of Sec.~\ref{sec:2.1}), i.e. $\tilde F_q \to \tilde{\mathcal{F}}_q $ for all $q$ and also $\phi \to \varphi$. Notice that in this procedure it is understood that both the fields and their derivatives are pulled-back onto the worldvolume, so e.g.  $\partial_i \varphi = [\partial_i \phi (X^\mu (\sigma,y) ) ]\vert_{y= 0}$.

It was shown in \cite{Niarchos:2015moa} that the combined set of equations \eqref{eqn:DpF1bgrcons},
\eqref{eq:DpF1extrinsic} in a trivial flat space background at extremality are equivalent to the equations of
motion of the $(p+1)$-dimensional DBI action with an electric field constraint (see eq.\ (7.34) in 
\cite{Niarchos:2015moa}). It would be interesting to extend the analysis of \cite{Niarchos:2015moa} to 
non-trivial backgrounds using the general form of \eqref{eqn:DpF1bgrcons}, \eqref{eq:DpF1extrinsic}
derived here.


\section{External couplings from hydrostatic partition functions}
\label{partition}
In this section we consider another method by which one can derive the couplings to backgrounds fields and their consequences in the form of the equations of motion. This method relies on obtaining the hydrostatic partition function from the Euclidean on-shell action for black holes and requires local measurements of temperature and chemical potentials as well as the measurement of stress-energy, electric and dilaton currents. As is well-known, the existence of a hydrostatic partition function requires the existence of a timelike Killing vector field along which the fluid velocity is aligned \cite{Emparan:2009at}.\footnote{It may be possible to relax the requirement of stationarity or, more generally, of non-dissipative flows, as advertised in \cite{Crossley:2015evo, Haehl:2015uoc}, by doubling the number of degrees of freedom.} Therefore, throughout this section we focus on fluids in stationary equilibrium. Nevertheless, the final form of the equations of motion that arise from the partition function, including force terms, is completely general. We begin with general considerations on gravitational partition functions and then derive the external couplings for black $p$-branes carrying Maxwell charge as well as for black $p$-branes charged under a $p$-form gauge field. The method can be applied to any bound state such as the D$p$-F1 analyzed in the previous section and we leave this longer endeavor for future work.

\subsection{General considerations on partition functions}
In a semi-classical approximation, the partition function $\mathcal{Z}$ for a given black hole solution can be obtained by evaluating the Euclidean on-shell action $I_{\text E}$ for that particular solution such that \cite{Gibbons:1976ue}
\beq
\ln \mathcal{Z}=i I_{\text E}~~,
\eeq
where we have Wick rotated the time coordinate $t\to i t$. In this approximation, the Euclidean action yields the Gibbs free energy of the black hole, which takes the generic form
\beq \label{eq:iEu}
i I_{\text E}=\beta\left(M-TS-\Omega J-\Phi_{\text{H}}Q_{p}\right)~~,
\eeq
where $\beta=T^{-1}$ is the radius of the time circle, $T$ is the global temperature, $M$ the total energy, $\Omega$ the angular velocity, $J$ the angular momentum, $\Phi_{\text{H}}$ the global chemical potential and $Q_{p}$ the total electric charge of the black hole. The generic form of \eqref{eq:iEu} is a consequence of stationarity. Therefore, the partition function obeys the relation
\beq
d\left(T\ln \mathcal{Z}\right)=SdT+Jd\Omega+Q_{p}d\Phi_{\text{H}}~~,
\eeq
consistent with the first law of thermodynamics, which leads to the following thermodynamic identities
\beq \label{eq:thermoid}
S=\frac{\partial\left(T\ln \mathcal{Z}\right)}{\partial T}~~,~~J=\frac{\partial\left(T\ln \mathcal{Z}\right)}{\partial \Omega}~~,~~Q_{p}=\frac{\partial\left(T\ln \mathcal{Z}\right)}{\partial \Phi_{\text{H}}}~~.
\eeq
In the following, we will consider the partition function of a generic black $p$-brane with a definite temperature, chemical potential (or equivalently horizon radius $r_0$ and charge parameter $\alpha$), boost velocity $u^{a}$ and boundary values of the external gauge fields $\mathcal{C}_{q+1}$ and dilaton field $\varphi$. Furthermore, due to stationarity, we assume that the boost velocity is aligned with a worldvolume Killing vector field $\textbf{k}^{a}$ such that $u^{a}=\textbf{k}^{a}/\textbf{k}$ with $\textbf{k}$ being its norm. Due to the translational invariance of the partition function along the worldvolume $\mathcal{W}_{p+1}$, the partition function must factorize,
\beq
\mathcal{Z}=\prod_{\mathcal{W}_{p+1}} \mathcal{Z}_{V_p}~~,
\eeq
where $\mathcal{Z}_{V_p}$ denotes the partition function of the arbitrarily small $p$-dimensional volume. We now consider perturbing the black brane to a new stationary solution by allowing the black brane parameters to depend on the worldvolume coordinates $\sigma^{a}$.  In this case, the partition function takes the form
\beq
\mathcal{Z}=\prod_{\mathcal{W}_{p+1}} \mathcal{Z}_{V_p}(\sigma)+\mathcal{O}\left(\varepsilon, \tilde\varepsilon\right)~~,
\eeq
where the derivative corrections of order $\mathcal{O}\left(\varepsilon, \tilde\varepsilon\right)$ are evaluated from solving the equations of motion for a given theory and determining the corresponding perturbations. However, in order to derive the constraint equations obtained in the previous sections, it is not necessary to consider higher-order corrections. Therefore, in what follows we will restrict the analysis to the leading order case for which the partition function of the leading order solution takes the form
\beq \label{eq:partfunct}
\ln \mathcal{Z}=\int_{\mathcal{W}_{p+1}}\star_{(p+1)} \ln \mathcal{Z}_{0}[\gamma_{ab},\textbf{k}^{b},X^{i},\mathcal{C}_{q+1},\varphi]~~,
\eeq
where $\ln \mathcal{Z}_0$ denotes the partition function of the uncorrected $p$-brane as a function of the external background sources. Here, the worldvolume Killing vector field $\textbf{k}^{a}$ should be understood as the pull-back of a background Killing vector field $\textbf{k}^{\mu}$ and $X^{i}$ the transverse embedding scalars denoting the position of the worldvolume in the ambient space. The external gauge field and dilaton in \eqref{eq:partfunct} should be understood either as the pull-back onto the worldvolume of the external fields or as the components of the external fields projected and evaluated on the worldvolume.

Except for possible gauge or gravitational anomalies, the partition function \eqref{eq:partfunct} must be invariant under diffeomorphisms and gauge transformations. Under worldvolume diffeomorphisms the induced metric transforms as $\delta_{||} \gamma_{ab}=2\nabla_{(a}\xi_{b)}$, the gauge field transform as $\delta_{||} \mathcal{C}_{q+1}=\xi^{a}\nabla_a \mathcal{C}_{q+1\mu_1...\mu_{q+1}}+\nabla_{\mu_1}\xi^{\nu}\mathcal{C}_{q+1\nu\mu_2...\mu_{q+1}}+...+\nabla_{\mu_{q+1}}\xi^{\nu}\mathcal{C}_{q+1\mu_1\mu_2...\nu}$ and the dilaton transforms as $\delta_{||} \varphi=\xi^{a}\partial_a \varphi$. The Killing vector is held fixed, due to stationarity, as well as the transverse scalars since we are performing a Lagrangian type variation.\footnote{Lagrangian type variations are variations in which the worldvolume position, characterized by the transverse scalars $X^{i}$, is held fixed and the background metric is displaced. Alternatively, one may consider a variational scheme in which the background metric is held fixed and the transverse scalars are displaced. These two types of variations are equivalent, even to higher order in derivatives \cite{Armas:2013hsa}.} Assuming that no boundaries are present, this leads to the variation of the partition function
\beq \label{eq:var1}
\delta_{||} \ln \mathcal{Z}=\int_{\mathcal{W}_{p+1}}\star_{(p+1)} \left(\nabla_{a}T^{ab}-\frac{1}{(q+1)!}{\mathcal F_{q+2}^{b}}_{a_1...a_{q+1}}J^{a_1...a_{q+1}}_{q+1}-j_\phi\partial ^{b}\varphi \right)\xi_{b}~~,
\eeq
where we have used the fact that invariance of \eqref{eq:partfunct} under gauge transformations $\delta \mathcal{C}_{q+1}=d\Lambda_{q}$ leads to the current conservation equation
\beq \label{eq:c11}
\nabla_{a_1}J^{a_1...a_{q+1}}_{q+1}=0~~.
\eeq
In \eqref{eq:var1} we have introduced the worldvolume stress-energy tensor, electric current and dilaton current via the expressions
\beq \label{eq:id}
T^{ab}=-\frac{2}{\beta\sqrt{-\gamma}}\frac{\delta\ln\mathcal{Z} }{\delta \gamma_{ab}}~~,~~ J^{a_1...a_{q+1}}_{q+1}=-\frac{1}{\beta\sqrt{-\gamma}}\frac{\delta\ln\mathcal{Z} }{\delta \mathcal{C}_{q+1a_1...a_{q+1}}}~~,~~j_\phi=-\frac{1}{\beta\sqrt{-\gamma}}\frac{\delta\ln\mathcal{Z} }{\delta \varphi}~~.
\eeq
Since \eqref{eq:var1} must hold for arbitrary $\xi_b$ we find that we must have
\beq \label{eq:intpart}
\nabla_{a}T^{ab}=\frac{1}{(q+1)!} \mathcal{F}_{q+2 \, a_1...a_{q+1}}^{b} J^{a_1...a_{q+1}}_{q+1}+j_\phi\partial ^{b}\varphi~~.
\eeq
Analogously, considering a diffeomorphism in the directions orthogonal to the worldvolume the induced metric transforms as $\delta_\perp \gamma_{ab}=-2{K_{ab}}^{i}\xi_i$ and one finds the equation of motion
\beq \label{eq:extpart}
T^{ab}{K_{ab}}^{i}=\frac{1}{(q+1)!}\mathcal F^{i}_{q+2 \, a_1...a_{q+1}}J^{a_1...a_{q+1}}_{q+1}+j_\phi\partial ^{i}\varphi~~.
\eeq
Once the partition function \eqref{eq:partfunct} is given in terms of the background sources, one may just use \eqref{eq:id} to obtain the stress-energy tensor and currents while under direct variation with respect to $X^{i}$ one obtains the equation of motion \eqref{eq:extpart} and hence the non-trivial form of the force terms. 

It is interesting to note that the partition function \eqref{eq:partfunct} can be written as a localized integral over the spacetime, i.e.
\beq 
\ln \mathcal{Z}=\int_{\mathcal{M}_{D}}\star \ln \mathcal{Z}_{0}[\gamma_{ab},\textbf{k}^{b},X^{i},\mathcal{C}_{q+1},\varphi]\tilde \delta^{(n+2)}(x^{i}-X^{i})~~,
\eeq
where $x^{i}$ are spacetime coordinates and $\tilde\delta^{(n+2)}$ the reparametrization invariant delta function in the transverse $(n+2)$-dimensional space. Written in this form, one can extract the spacetime stress tensor, which takes the general form
\beq \label{eq:localstress}
\bt=-\frac{2}{\sqrt{-g}}\frac{\delta\left(T\ln\mathcal{Z}\right)}{\delta g_{\mu\nu}}=T^{ab}\partial_{a}X^{\mu}\partial_{b}X^{\nu}\tilde \delta^{(D-p-1)}(x^{i}-X^{i})~~,
\eeq
where $T^{ab}$ is the worldvolume stress-energy tensor obtained using \eqref{eq:id} and $\partial_{a}X^{\mu}$ is a projector along the worldvolume directions. Here, $X^{\mu}$ is the set of mapping functions which includes the $(p+1)$-worldvolume directions besides the transverse scalars $X^{i}$. Therefore, to leading order, spacetime stress-energy tensors that arise from \eqref{eq:partfunct} represent localized objects in the ambient background with metric $g_{\mu\nu}$.

We note that even though stationarity will impose some restrictions on the form of the currents or configurations that solve \eqref{eq:intpart}-\eqref{eq:extpart}, the form of the equations of motion \eqref{eq:intpart}-\eqref{eq:extpart}, obtained by demanding diffeomorphism and gauge invariance of the partition function, is completely general and matches exactly those obtained in \eqref{eq:intemd} and \eqref{eq:extemd} derived from \eqref{eq:forceemd} for a stress-energy tensor of the form \eqref{eq:localstress} as well as with localized electric and dilaton currents and without a magnetic current.

\subsection{External couplings for black branes carrying Maxwell charge}
In this section we consider the case of black $p$-branes carrying Maxwell charge, which were not analyzed in the previous sections. These are also solutions to the action \eqref{eq:emd} but with a two form field strength $F_{2}$ ($q=0$) and for simplicity we consider the case where no dilaton field is present. The metric and gauge field can be found in \cite{Gath:2013qya} and read
\begin{align} \label{eq:dsmax}
\begin{split}
ds^2&=H^{N-2}\left( \left(P_{ab}-H^{-N}fu_{a}u_{b}\right)\td\sigma^{a}\td\sigma^{b}+f^{-1}dr^2+r^2\td\Omega_{(n+1)}^2\right)~~,\\
C_{1}&=\frac{\sqrt{N}}{H}\left(\frac{r_0}{r}\right)^{n}\sinh\alpha\cosh\alpha \; u_a \td\sigma^{a} \, , \quad N = \frac{2(n+p+1)}{n+p} ~~,
\end{split}
\end{align}
where the functions $f$ and $H$ were given in \eqref{eq4:281112}. Here $r_0$ and $\alpha$ are the horizon radius and charge parameter, respectively. In order to evaluate the Euclidean on-shell action we need to add appropriate boundary counter-terms to the action \eqref{eq:emd}. Using the fact that on-shell (for general $q$) one has
\beq \label{eq:onshell}
\star R= \frac{1}{2}d\phi\wedge\star d\phi+e^{a_q\phi}\left(\frac{1}{2}-\frac{q+1}{D-2}\right)F_{q+2}\wedge\star F_{q+2}~~,
\eeq
the Euclidean action over a $D$-dimensional region $Y$ with boundary $\partial Y$ for $q=0$ and $a_q=0$ becomes
\beq \label{eq:Eaction}
I_{\text E}=\frac{1}{8\pi G}\int_{\partial Y}\star [K] -\frac{1}{16\pi G}\frac{1}{(D-2)}\int_{\partial Y}C_{1}\wedge\star F_{2}~~,
\eeq
where the boundary $\partial Y$ is chosen to be a constant radial slice at infinity in the geometry of \eqref{eq:dsmax} and where $[K]$ denotes the difference between the extrinsic curvature of a constant radial slice in \eqref{eq:dsmax} and the analogous radial slice in flat spacetime written in the same coordinates as in \eqref{eq:dsmax} but with an appropriate temperature redshift at infinity.

Analogous to the case considered in Sec.~\ref{sec:charged} of $p$-branes with a top-form, we construct the leading order ansatz by adding a constant gauge field $\mathcal{C}_{1}$ to $C_{1}$ via a local gauge transformation such that, without loss of generality, the boosted gauge field \eqref{eq:dsmax} is given by
\beq
C_{1}\to C_{1}+\mathcal{C}_{1b}u^{b}u_ad\sigma^{a}~~.
\eeq
However, such gauge transformation does not affect the evaluation of \eqref{eq:Eaction}. This is because in order to evaluate the Euclidean action one must require the gauge field to be regular at the horizon by subtracting its value at the horizon as in \cite{Gibbons:1976ue}. In practice, this means that we must perform the shift $C_{1}\to C_{1}-(\Phi_p+ \mathcal{C}_{1b}u^{b})u_ad\sigma^{a}$ via a local gauge transformation and therefore removing potential contributions due to $\mathcal C_{1}$. Given this, explicit evaluation of \eqref{eq:Eaction} leads to the Euclidean on-shell action
\beq \label{eq:E1}
I_{\text E}=-i\beta \frac{\Omega_{(n+1)}}{16\pi G}\int_{\mathcal{B}_{p}}\star_{(p)}r_0^{n}~~,
\eeq
where we have taken the worldvolume geometry to be $\mathcal{W}_{p+1}=\mathbb{R}\times\mathcal{B}_{p}$ with $\mathcal{B}_{p}$ being the spatial part and $\star_{(p)}1=\sqrt{-\gamma} d\sigma^{1}\wedge...\wedge d\sigma^{p}$ (see App.~\ref{app:notation}). We note that \eqref{eq:E1} is simply proportional to the pressure $P$ as written explicitly in App.~\ref{app:charged0} and that it must be extremized at fixed global temperature and chemical potential.

The integrand in \eqref{eq:E1} is a local version of the Gibbs free energy of the brane. In order to obtain the partition function one must re-express it in terms of the background sources as in \eqref{eq:partfunct} for stationary configurations. In order to do so, we must demand gauge invariance, which from \eqref{eq:c11} implies that the electric current must be conserved $\nabla_a J^{a}_1=0$. Furthermore, we must also demand worldvolume diffeomorphism invariance which, in this case, using \eqref{eq:intpart}, requires that
\beq \label{eq:cc0}
\nabla_{a}T^{ab}={\mathcal{F}_{2}^{b}}_{a}J^{a}_{1}~~,
\eeq
where both the stress-energy tensor and electric current are given in App.~\ref{app:charged0}. The projection of this equation along the fluid flows $u_{a}$ is automatically satisfied due to the fact that the r.h.s. vanishes by symmetry while the l.h.s. vanishes due to the fact that the electric current is conserved and assuming the existence of a conserved entropy current $J^{a}_s=su^{a}$. The projection of Eq.~\eqref{eq:cc0} perpendicular to the fluid flows leads to
\beq \label{eq:dyn1}
\mathcal{T}s{P^{c}}_b\left(\partial^{b}\ln\mathcal{T}+a^{b}\right)+\mathcal{Q}(\Phi_p-\mathcal{C}_{1a}u^{a}){P^{c}}_b\left(\partial^{b}\ln(\Phi_p-\mathcal{C}_{1a}u^{a})+a^{b}\right)=0~~,
\eeq
where $a^{b}=u^{a}\nabla_a u^{b}$ is the fluid acceleration. Here we have assumed that the fluid velocity $u^{a}$ is aligned with a worldvolume Killing vector field $\textbf{k}^{a}$ with modulus $\textbf{k}$, that is, $u^{a}=\textbf{k}^{a}/\textbf{k}$ and furthermore that for any stationary configuration one must have that $\mathcal{L}_{\textbf{k}} \mathbb{T}=0$ for an arbitrary tensor $\mathbb{T}$. Eq.~\eqref{eq:dyn1} is solved by requiring the local temperature $\mathcal{T}$ and chemical potential $\Phi_p$ to satisfy
\beq \label{eq:sol1}
T=\textbf{k}\mathcal{T}~~,~~\Phi_{\text{H}}=\textbf{k}\left(\Phi_p-\mathcal{C}_{1a}u^{a}\right)~~,
\eeq
where $T$ is the constant global temperature and $\Phi_{\text{H}}$ the constant global chemical potential. When the background gauge field $\mathcal{C}_1$ vanishes, this reduces to the solution found in \cite{Caldarelli:2010xz}. 

Using the relation between local temperature, horizon radius and chemical potential given in Eqs.~\eqref{eq:chargedcharges} and \eqref{eq:ts} in App.~\ref{app:charged} together with \eqref{eq:sol1} and \eqref{eq:E1}, the partition function \eqref{eq:partfunct} takes the form
\beq \label{eq:part4}
\ln \mathcal{Z}=\beta \frac{\Omega_{(n+1)}}{16\pi G}\left(\frac{n}{4\pi T}\right)^{n}\int_{\mathcal{B}_{p}}\star_{(p)} \textbf{k}^{n}\left(1-\frac{\Phi_\text{H}^2}{N\textbf{k}^2}-\frac{\left(\mathcal{C}_{1a}u^{a}\right)^{2}}{N}\right)^{\frac{Nn}{2}}~~,
\eeq
where we have used \eqref{eq:sol1} to replace terms containing $\Phi_p$. For consistency we note that, using \eqref{eq:part4}, we can easily extract the electric current, i.e.,
\beq
J_{1}^{a}=-\frac{\partial \ln\mathcal{Z}}{\partial \Phi_{p}}\frac{\partial \Phi_{p}}{\partial \mathcal{C}_{1a}}=\mathcal{Q}u^{a}~~,
\eeq
and also the correct perfect fluid stress-energy tensor
\beq
T^{ab}=P\gamma^{ab}+(nP+\mathcal{Q}\Phi_p)u^{a}u^{b}=P\gamma^{ab}+(\epsilon+P)u^{a}u^{b}~~,
\eeq
in agreement with App.~\ref{app:charged0}. The equations of motion that follow from varying \eqref{eq:part4} by an arbitrary diffeomorphism are exactly those of \eqref{eq:intpart} and \eqref{eq:extpart} for $q=0$. 

Finally, we consider changing to another ensemble that resembles the usual coupling of charged point particles moving in an external electric field where the total electric charge is held constant. This can be done by performing a global Legendre transformation by adding a term of the form $Q_{1}\Phi_{\text{H}}$ to \eqref{eq:part4} with $Q_{1}$ being the total electric charge given by
\beq
Q_{1}=\int_{\mathcal{B}_{p}}\mathcal{Q}u^{a}n_{a}~~,~~n_{a}=\frac{\xi_a}{R_0}~~,
\eeq
where $\xi^{a}\partial_a$ is the worldvolume Killing vector associated with time translations and $R_0$ its norm. The partition function \eqref{eq:part4} becomes
\beq \label{eq:part5}
\ln \mathcal{Z}=\beta\left(-\int_{\mathcal{B}_{p}}\star_{(p)}P+Q_1\Phi_{\text{H}}\right)=\beta\left(\int_{\mathcal{B}_{p}}\star_{(p)}(\epsilon-\mathcal{T}s)+\int_{\mathcal{B}_{p}}\star_{(p)}\mathcal{Q}u^{a}\mathcal{C}_{1a}\right)~~,
\eeq
where the energy density $\epsilon$, temperature $\mathcal{T}$ and entropy density $s$ are given in Eqs.~\eqref{eq:e0} and \eqref{eq:chargedcharges}. The partition function \eqref{eq:part5} yields the same equations of motion as \eqref{eq:part4} as long as variations are taken at constant global electric charge $Q_{1}$.
\subsection{External couplings for black branes carrying \texorpdfstring{$q=p$}{q=p}-brane charge} \label{sec:hydroqp}
In this section we focus on the more complicated case of black brane solutions \eqref{eq:pbranesol} to the action \eqref{eq:emd} with external sources of gauge $\mathcal{C}_{q+1}$ and dilaton field $\varphi$. In order to evaluate the Euclidean on-shell action we consider the general result \eqref{eq:onshell} for $q=p$. In such case the Euclidean action over a $D$-dimensional region $Y$ with boundary $\partial Y$ becomes
\beq \label{eq:Eaction1}
I_{\text E}=\frac{1}{8\pi G}\int_{\partial Y}\star [K] -\frac{1}{16\pi G}\frac{(p+1)}{(D-2)}\int_{\partial Y}e^{a_p\phi}C_{p+1}\wedge\star F_{p+2}~~,
\eeq
where the boundary $\partial Y$ is chosen to be a constant radial slice at infinity in the geometry of \eqref{eq:pbranesol}. Since this case is qualitatively different than the previous example, we first consider the situation for which no background gauge or dilaton fields are present, i.e., $\mathcal{C}_{q+1}=\varphi=0$. In order to evaluate \eqref{eq:Eaction} we consider the gauge field and its field strength near infinity using \eqref{eq:pbranesol}, 
\begin{equation} \label{eq:expansion}
\begin{split}
C_{p+1}&=\left(-\frac{16\pi G Q_{p}}{n r^{n}}-\Phi_p\right)\star_{(p+1)}1+\mathcal{O}\left(\frac{r_0}{r}\right)^{2n}~~,\\
F_{p+2}&=16\pi G\frac{Q_{p}}{r^{(n+1)}}dr\wedge\star_{(p+1)}1+\mathcal{O}\left(\frac{r_0}{r}\right)^{2n+1}~~,
\end{split}
\end{equation}
where we have shifted the gauge field, via a gauge transformation, by subtracting the horizon chemical potential $\Phi_p=\sqrt{N}\tanh\alpha$ so that the gauge field is regular at the horizon \cite{Horowitz:1991cd} as in the example of the previous section. 

Direct evaluation of \eqref{eq:Eaction} and using \eqref{eq:expansion} leads to the same form of the Euclidean on-shell action \eqref{eq:E1} as in the previous example. In this case, the integrand of \eqref{eq:E1} is identified with the local Gibbs free energy $\mathcal{G}$ of the brane and not with the pressure (see Eq.~\eqref{eq:pcharged}). In order to re-express it in terms of the background sources as in \eqref{eq:partfunct}, we demand gauge invariance and worldvolume diffeomorphism invariance. The former implies that the electric current $J_{p+1}$ as given in \eqref{eq:tcp} is conserved which in turn implies that the total dipole charge $Q_{p}$ is constant on the worldvolume, i.e.,
\beq \label{eq:chargec}
\partial_a Q_{p}=0~~.
\eeq
This condition suggests that the thermodynamic ensemble of the Euclidean action \eqref{eq:E1} at fixed global temperature and global chemical potential is not the appropriate one in order to directly deal with the worldvolume conservation equations. For this reason, one changes to a new ensemble where the total charge $Q_{p}$ is held constant instead of the global chemical potential. As in \cite{Emparan:2011hg}, we perform a local Legendre transformation by adding $\Phi_p Q_{p}$ to the Gibbs free energy $\mathcal{G}$. By doing so, the Euclidean action \eqref{eq:E1} at fixed $Q_p$, which we refer to as $\tilde I_{\text{E}}$, is given by
\beq \label{eq:E2}
\tilde I_{\text E}=-i\beta\int_{\mathcal{B}_{p}}\star_{(p)}\left( \frac{\Omega_{(n+1)}}{16\pi G}r_0^{n}+\Phi_p Q_p\right)=-i\beta\int_{\mathcal{B}_{p}}\star_{(p)}P ~~,
\eeq
where the pressure $P$ is given in Eq.~\eqref{eq:pcharged} of App.~\ref{app:charged}. However, since \eqref{eq:E2} must hold globally, we readily identify the global chemical potential $\Phi_{\text{H}}$ as%
\footnote{Since in order to obtain the global chemical potential one must integrate over the local one, $\Phi_p$ is better understood as a density of chemical potential on the brane, analogously to the energy density $\epsilon$ or the entropy density $s$.}
\beq \label{eq:global1}
\Phi_{\text H}=\int_{\mathcal{B}_{p}}\star_{(p)}~\Phi_p~~.
\eeq
We now turn to the requirement of worldvolume diffeomorphism invariance, which in the absence of external backgrounds fields is given by \eqref{eq:intpart}, i.e., $\nabla_a T^{ab}=0$. Assuming the conservation of the entropy current $J^{a}_s=su^{a}$, this set of equations is solved by
\beq \label{eq:sol2}
T=\textbf{k}\mathcal{T}~~,~~u^{a}=\frac{\textbf{k}^{a}}{\textbf{k}}~~.
\eeq
We note that $\Phi_p$ is not a local degree of freedom of the fluid because it is completely determined by the condition \eqref{eq:chargec} and therefore does not contribute to the solution \eqref{eq:sol2}. Given the solution \eqref{eq:sol2} and using the relation between the local temperature $\mathcal{T}$ of the black brane \eqref{eq:pbranesol} in terms of the horizon radius $r_0$ and chemical potential $\Phi_p$ (Eqs.\eqref{eq:chargedcharges} and \eqref{eq:ts} in App.~\ref{app:charged}) the partition function \eqref{eq:partfunct} takes the form
\beq \label{eq:part1}
\ln \mathcal{Z}=\beta \frac{\Omega_{(n+1)}}{16\pi G}\left(\frac{n}{4\pi T}\right)^{n}\int_{\mathcal{B}_{p}}\star_{(p)} \textbf{k}^{n}\left(1-\frac{\Phi_p^2}{N}\right)^{\frac{Nn}{2}}\left(1+\frac{n\Phi_p^2}{1-\frac{\Phi_p^2}{N}}\right)~~,
\eeq
and should be extremized at fixed $T$ and $Q_p$.\footnote{It may not be evident from \eqref{eq:part1} but the correct perfect fluid stress-energy tensor \eqref{eq:tcp} follows from \eqref{eq:part1} using \eqref{eq:id} and noting that $(\delta \ln \mathcal{Z}/\delta\Phi_p)|_{T,Q_p}=0$.} Alternatively, since we have identified the global chemical potential \eqref{eq:global1} one may perform the inverse Legendre transformation of \eqref{eq:E2} in order to obtain a variational principle at fixed $T$ and $\Phi_{\text{H}}$ as in \cite{Emparan:2011hg}.

\subsubsection*{Adding external background fields}
We now consider introducing background values for the dilaton and gauge fields. As explained in Sec.~\ref{sec:charged}, a constant shift in the dilaton field is a symmetry of the action \eqref{eq:emd} and leads to a rescaling of the gauge field, i.e.,
\beq
\phi\to \phi+\varphi~~,~~C_{p+1}\to e^{-\frac{a_p}{2}\varphi}C_{p+1}~~. 
\eeq
In turn this implies that the chemical potential and the electric charge are rescaled according to
\beq \label{eq:newgauge}
\boldsymbol{\Phi}_p= e^{-\frac{a_p}{2}\varphi}\Phi_p~~,~~\boldsymbol{Q}_{p}= e^{\frac{a_p}{2}\varphi}Q_{p}
\eeq
such that the product $\boldsymbol{\Phi}_p \boldsymbol{Q}_{p}=\Phi_pQ_{p}$ remains invariant. We then add a constant gauge field $\mathcal{C}_{q+1}$ such that $C_{p+1}\to C_{p+1}+\mathcal{C}_{q+1}$. Analogously to the previous case, constant shifts of the gauge field do not affect the evaluation of the Euclidean on-shell action while constant shifts of the dilaton only modify the result via \eqref{eq:newgauge}. This is expected since both these shifts are symmetries of the action \eqref{eq:emd}. Therefore, once again, the Euclidean on-shell action is given by \eqref{eq:E1}.

The presence of background fields, however, changes significantly the analysis. Demanding gauge invariance now implies that the modified charge is conserved along the worldvolume, i.e., $\partial_a\boldsymbol{Q}_{p}=0$ since the electric current is now given by \eqref{eq:modcurrent}. As in the case where no external background fields were present, one should now change to a new ensemble where $\boldsymbol{Q}_{p}$ is held constant instead of the global chemical potential but prior to do so we will consider the requirement of diffeomorphism invariance \eqref{eq:intpart} which in this case reads
\beq \label{eq:c12}
\nabla_{a}T^{ab}=j_\phi\partial ^{b}\varphi~~,
\eeq 
where $j_\phi$ is given in \eqref{eq:modcurrent}. We note that there is no Lorentz force because $\mathcal C_{p+1}$ is a top-form from the worldvolume point of view. As we will see below, this implies that the background field $\mathcal{C}_{p+1}$ will not affect the requirements due to worldvolume diffeomorphism invariance but it will contribute to changes in the global chemical potential. We now proceed and try to find a stationary solution to \eqref{eq:c12} for arbitrary background sources. Projecting \eqref{eq:c12} along $u_{b}$ leads to a vanishing l.h.s. assuming the conservation of the entropy current and therefore we obtain the condition $u^{a}\partial_a\varphi=0$. This condition suggests that one must, as in the previous cases, choose $u^{a}=\textbf{k}^{a}/\textbf{k}$ and in fact we demonstrate in App.~\ref{app:entropy} that this must indeed be the case in order to have a fluid configuration that does not dissipate. On the other hand, the projection of \eqref{eq:c12} perpendicular to the fluid flows leads to
\beq \label{eq:c23}
\mathcal{T}s{P^{c}}_b\left(\partial^{b}\ln\mathcal{T}+a^{b}\right)=j_\phi{P^{c}}_b\partial^{b}\varphi~~.
\eeq
We see that the driving force due to a spatially varying dilaton must be compensated by a modification of the local temperature $\mathcal{T}$ compared to \eqref{eq:sol2} since the dilaton current $j_\phi$ is non-vanishing at leading order. Note that this case is qualitatively different than the case considered in \cite{Bhattacharyya:2008ji} since there $j_\phi$ appears at first order in derivatives and hence plays no role at leading order. In order to solve \eqref{eq:c23} we denote the $m$ worldvolume directions perpendicular to $u_a$ collectively by $\tilde \sigma=\{\tilde\sigma_1,...,\tilde\sigma_m\}$ and make the ansatz
\beq \label{eq:ansatz}
\mathcal{T}=\frac{T}{\textbf{k}}f(\tilde\sigma)~~,
\eeq
for constant global temperature $T$ and for some function $f(\tilde\sigma)$ to be determined. Introducing this ansatz into \eqref{eq:c23} leads to\footnote{Note that $\varphi$ can also depend on the transverse scalars $X^{i}$ without affecting the analysis that we are carrying out.}
\beq \label{eq:solve1}
\mathcal{T}s\partial_{c}\ln f(\tilde\sigma)=j_\phi\partial_c \varphi(\tilde\sigma)~~.
\eeq
It is now imperative to note that the product $\mathcal{T}s$, as well as the dilaton current $j_\phi$, is a function of the temperature $\mathcal{T}$ and the global charge $\boldsymbol{Q}_{p}$ for the specific case that we are considering. Obtaining the dependence of these quantities in terms of $\mathcal{T}$ is not straightforward as it demands obtaining $\boldsymbol{\Phi}_p$ in terms of $\boldsymbol{Q}_{p}$. This can be done analytically in appropriate small or large charge limits as in \cite{Armas:2012bk, Armas:2013ota} and for specific values of $n$. Nevertheless, the final result is always dependent on $\mathcal{T}$ and therefore from \eqref{eq:ansatz} and \eqref{eq:solve1} the function $f(\tilde\sigma)$ will always depend non-trivially on the global temperature $T$. When such dependence is introduced in the partition function \eqref{eq:partfunct} then direct evaluation of the entropy using \eqref{eq:id} would lead to modifications to the entropy current $J^{a}_s=su^{a}$ which are not present at leading order.\footnote{Besides this general argument, we have not been able to find solutions to \eqref{eq:solve1} even in simple cases of constant driving forces.} This forces us to conclude that there are no stationary solutions to \eqref{eq:c12} with current $j_\phi$ as given in \eqref{eq:modcurrent} and with a spatially varying dilaton $\varphi(\tilde\sigma)$ along the worldvolume directions. In turn, we conclude that there are no regular stationary black holes constructed from fluid-type deformations of \eqref{eq:pbranesol} with a spatial varying dilaton background field along worldvolume directions, though it can depend non-trivially on the transverse scalars $X^{i}$. Therefore, the solution \eqref{eq:sol2} still holds with the further requirement that $\partial_a\varphi=0$ when the temperature $\mathcal{T}$ is non-zero. At extremality ($\mathcal{T}=0$) the result is different. In this case, the stress-energy tensor becomes $T^{ab}=-e^{-a_p\varphi/2}\boldsymbol{Q}_{p}\gamma^{ab}$ and therefore \eqref{eq:c12} is automatically satisfied.

Given the stationary solution at finite temperature just obtained, we have all the necessary ingredients to write the partition function in terms of the background sources. However, as mentioned earlier, the presence of a background top-form gauge field does not affect \eqref{eq:sol2} but can contribute to changes in the global chemical potential \eqref{eq:global1}. We parametrize this ignorance by considering an additional contribution $\tilde{\boldsymbol{\Phi}}_p$ to the global chemical potential \eqref{eq:global1} such that
\beq \label{eq:global11}
\Phi_{\text H}=\int_{\mathcal{B}_{p}}\star_{(p)}~{\boldsymbol{\Phi}}_p+\int_{\mathcal{B}_{p}}\tilde{\boldsymbol{\Phi}}_p~~.
\eeq
We now consider performing a global Legendre transformation in the Euclidean action \eqref{eq:E1} by adding a term of the form $\boldsymbol{Q}_{p}\Phi_{\text{H}}$. The partition function becomes
\beq \label{eq:part3}
\ln \mathcal{Z}=\beta\left(\int_{\mathcal{B}_{p}}\star_{(p)}\mathcal{G}+{\Phi}_{\text{H}} \boldsymbol{Q}_{p}\right)=-\beta\int_{\mathcal{B}_{p}}\left(\star_{(p)}P-\boldsymbol{Q}_{p}\tilde{\boldsymbol{\Phi}}_p\right)~~.
\eeq
Since the background gauge field $\mathcal{C}_{p+1}$ did not affect \eqref{eq:sol2}, the pressure $P$ has no dependence on $\mathcal{C}_{p+1}$. Therefore, demanding consistency with \eqref{eq:id} we must require that
\beq \label{eq:cdil}
J_{p+1}^{a_1...a_{p+1}}=-\frac{1}{\beta}\frac{\delta\ln\mathcal{Z}}{\delta \mathcal{C}_{p+1a_1...a_{p+1}}}=-\boldsymbol{Q}_{p}\frac{\delta\tilde{\boldsymbol{\Phi}}_p}{\delta \mathcal{C}_{p+1a_1...a_{p+1}}}=\boldsymbol{Q}_{p}\epsilon^{a_1...a_{p+1}}~~,
\eeq
where $\epsilon^{a_1...a_{p+1}}$ is the Levi-Civita tensor on the $(p+1)$-dimensional worldvolume. Eq.~\eqref{eq:cdil} has a unique straightforward solution, namely, $\tilde{\boldsymbol{\Phi}}_p=-\mathbb{P}[\mathcal{C}_{p+1}]$ where $\mathbb{P}[\mathcal{C}_{p+1}]$ is the pull-back of the gauge field onto the worldvolume. Finally, using the relations between local temperature and chemical potential in App.~\ref{app:charged}, the solution \eqref{eq:sol2}, $\partial_a\varphi=0$ and $\tilde{\boldsymbol{\Phi}}_p=-\mathbb{P}[\mathcal{C}_{p+1}]$ we obtain the partition function in terms of the background sources
\beq \label{eq:part10}
\ln \mathcal{Z}=-\beta\int_{\mathcal{B}_{p}}\left(\star_{(p)}P+\boldsymbol{Q}_{p}\mathbb{P}[\mathcal{C}_{q+1}]\right)~~,
\eeq
where the pressure $P$ is given by
\beq
P=-\frac{\Omega_{(n+1)}}{16\pi G}\left(\frac{n}{4\pi T}\right)^{n}\textbf{k}^{n}\left(1-\frac{e^{a_p\varphi}\boldsymbol{\Phi}_p^2}{N}\right)^{\frac{Nn}{2}}\left(1+\frac{ne^{a_p\varphi}\boldsymbol{\Phi}_p^2}{1-\frac{e^{a_p\varphi}\boldsymbol{\Phi}_p^2}{N}}\right)~~.
\eeq
When no background dilaton fields are present, the partition function \eqref{eq:part10} was the one used in \cite{Armas:2012bk, Armas:2013ota} to study giant graviton configurations at finite temperature. As a consistency check, consider obtaining the dilaton current from \eqref{eq:part10} using \eqref{eq:id}, one finds\footnote{Note that by consistency we also find $\Phi_{\text H}=-(\partial \ln\mathcal{Z}/\partial \boldsymbol{Q}_p)|_{T}$ where $\Phi_{\text{H}}$ is given in \eqref{eq:global11} with $\tilde{\boldsymbol{\Phi}}_p=-\mathbb{P}[\mathcal{C}_{p+1}]$.}
\beq
j_\phi=-\frac{\delta \ln \mathcal{Z}}{\delta \varphi}|_{T,\boldsymbol{Q}_{p}}=\frac{a_p}{2}Q_{p}{\Phi}_p~~,
\eeq
in agreement with \eqref{eq:modcurrent}.

\subsubsection*{The extremal limit}
The partition function \eqref{eq:part10} must reduce to the DBI action in the presence of a background dilaton and gauge field once the extremal limit $\mathcal{T}\to0$ is taken. This limit requires that $r_0\to0$, $\alpha\to\infty$ and $T\to0$ while keeping the total charge $\boldsymbol{Q}_{p}$ fixed. Therefore, at extremality, we find the following limiting behavior for the fluid pressure $P\to-e^{-a_p\varphi/2}\boldsymbol{Q}_{p}$.
Identifying $\boldsymbol{Q}_{p}$ with the brane tension $T_{p}$ such that $\boldsymbol{Q}_{p}=T_p$ we obtain the DBI action in the form 
\beq
S=-\frac{1}{\beta}\int dt \ln \mathcal{Z}=-T_p\int_{\mathcal{W}_{p+1}}\!\!\!d^{p+1}\sigma~ e^{-a_p\varphi/2}\sqrt{-\gamma}+T_p\int_{\mathcal{W}_{p+1}}\!\!\!\mathbb{P}[\mathcal{C}_{q+1}]~~,
\eeq
as expected. We note that this action is valid for arbitrary background dilaton field $\varphi(\sigma,X^{i})$ contrary to the finite temperature case where $\varphi$ cannot have any dependence on the coordinates $\sigma$ in stationary equilibrium.


\section{Conclusions}
\label{conclusions}

The ultimate objective of this work is a systematic construction of black hole solutions in 
appropriate long-wavelength expansion schemes in arbitrary (super)gravity theories. In the 
present paper we continued work in this direction in the context of the blackfold formalism to include  
generic effects of the asymptotic background that arise from curvature, and/or fluxes of general matter fields. 
Focusing on the constraint equations of the gravitational system at first order in the long-wavelength expansion 
we derived an effective hydrodynamic description that involves fluids propagating under the influence of external 
forces. The resulting equations are the dynamical equations of \emph{forced blackfolds}. These equations describe
a part of the full dynamics of the putative complete (super)gravity solution. We conclude with a few
remarks on the (super)gravity problem, the regimes of the sought-after perturbative expansions, and some of the key issues 
that arise in the presence of generic asymptotic backgrounds (that are less elaborated upon in the existing literature).

\subsection{Comments on the long-wavelength expansions of the blackfold approach}
\label{properetc}

\subsubsection*{Elements and formulation of the (super)gravity problem}
The specific problem that we want to consider in (super)gravity starts with the following ingredients:
\begin{itemize}
\item[$(a)$] We are given an arbitrary gravitational action in $D$ spacetime dimensions $(D>4)$. 
Besides the metric, this action may involve a variety of other fields, e.g.\ matter fields and abelian gauge fields 
that are common in supergravities. 

\item[$(b)$] An exact (black) $p$-brane solution of the equations of motion of this action with specified, but arbitrary asymptotics, is known. We assume that the $p$-brane solution has Killing isometries along $m\leq p+1$ 
non-compact worldvolume directions, and is characterized by $\ell$ independent free parameters, 
$e.g.$ thermodynamic parameters like the mass, or other charges. 
The $m$ symmetries of the solution are a subset of the symmetries of the asymptotic background. 
To avoid potential backreaction issues to the asymptotic background we 
also assume that the codimension of the $p$-brane solution is appropriately high, typically greater than two, $D-p-1 >2$.  
\end{itemize}
Our goal is to construct a larger class of inhomogeneous $p$-brane solutions with the same asymptotics,
where the $m$ symmetries of the solution in $(b)$ are broken. The new solutions are continuous deformations of 
the (partially) homogeneous solution in point $(b)$.

\subsubsection*{Long-wavelength deformations}
The $m$ non-compact symmetric directions imply the potential existence of symmetry-breaking deformations
in a long-wavelength regime.
A natural subclass of such deformations can be attacked with semi-analytic methods that promote
the $\ell$ free parameters of the leading order exact solution $(b)$ to arbitrary slowly-varying functions of the $m$ 
spacetime coordinates along which we seek to break the symmetry. With an appropriate ansatz for the
(super)gravity fields based on an order-by-order deformation of the leading order solution one aims to construct
less symmetric solutions perturbatively in a scheme of small derivative expansions. The best studied
and most successful application of this logic in gravity has focused on large AdS black holes, where it leads
to the fluid/gravity correspondence. Applications in a wider setting constitute the basis of the blackfold formalism.

Clearly, the extent of the gravitational dynamics that can be captured in the above analysis depends on the
form of the leading order solution and the ansatz that is employed to study deformations around it. 

The ans\"atze that were described in the main text and are usually employed in the context of the blackfold formalism
may not cover in general the full set of available long-wavelength dynamics. 
In \cite{Niarchos:2015moa} it was emphasized that the blackfold approach 
captures the effective long-wavelength dynamics
of abelian singleton degrees of freedom, and may contain only partial information about long-wavelength dynamics
of the microscopic non-abelian degrees of freedom. From a gravitational point of view, a better understanding of the
degrees of freedom that dominate the long-wavelength dynamics can be obtained by studying the spectrum
of quasinormal modes of the black brane solution. A complete analysis of quasinormal modes of black branes
in general spacetimes is currently missing.

\subsubsection*{`Exact brane' applications of the blackfold approach} 
The (super)gravity problem that was formulated above provides from the start two exact gravitational solutions: 
a solution that fixes the asymptotic spacetime ({\it background solution}), and a $p$-brane solution 
with a suitable codimension ({\it leading order solution} of the subsequent expansion scheme). 
The leading order solution asymptotes to the background solution at large distances along a radial direction. 
In general, the two solutions have different symmetries and are characterized by different characteristic scales. 
Let us call $L$ the characteristic scale of the background solution, and $r_H$ the characteristic scale of the 
leading order solution. $r_H$ is a characteristic horizon scale, e.g.\ it can be the charge radius of a charged 
$p$-brane (the AdS radius for a solution with an AdS near-horizon region), or the Schwarzschild radius for 
a black $p$-brane at finite temperature.\footnote{The generic situation may involve further characteristic 
scales with a more complicated pattern of regimes. We will shortly address such an example below. For the 
moment we keep a minimum number of scales to exhibit clearly the basics of the construction.}
Carrying out the long-wavelength deformation analysis exactly in the ratio $\nicefrac{r_H}{L}$ will be 
dubbed {\it the `exact brane' application of blackfolds}.

In this exercise
the deformation of the leading order solution that we seek introduces a third scale into the problem: the scale $\RR$ of
deformations. Since we are interested in long-wavelength deformations, by definition we require the hierarchy 
of scales
\beq
\label{introaa}
r_H \ll \RR
~.
\eeq
Notice that the background scale $L$ does not enter in this inequality, because the leading order solution is exactly known
for all values of $r_H$ and $L$.

The perturbative construction of long-wavelength deformations of the leading order solution can be pursued with 
the use of a suitable scheme of matched asymptotic expansions (MAEs) (see Refs.\ \cite{Emparan:2007wm,Camps:2012hw,Gath:2013qya,DiDato:2015dia} for concrete applications of this method in the context
of blackfold constructions).
In a MAE the (super)gravity equations are solved separately in a {\it near-zone} region $(r\ll \RR)$, and a 
{\it far-zone} region $(r\gg r_H)$. The large hierarchy of scales \eqref{introaa} guarantees the existence of 
a large intermediate {\it overlap} region $(r_H \ll r \ll \RR)$ where the integration constants of the near- and far-zone
solutions are matched.

In this context, part of the gravitational equations (constraint equations) result naturally to  
a $(p+1)$-dimensional effective hydrodynamic description of the collective mode dynamics of the resulting 
$p$-brane solution. This description, which was the main theme in this paper, is formulated as a set of conservation 
equations for appropriate currents. 

These currents are evaluated in the 
overlap zone, deep in the asymptotic region where one can position the screen of the effective description. 
In general, they depend non-trivially on the details of the background solution and its scale $L$.
As one proceeds order-by-order in the derivative expansion these currents receive higher-derivative corrections, 
but the conservation equations are always formulated as equations of a $(p+1)$-dimensional fluid on a dynamical 
(elastic) hypersurface propagating in the {\it fixed} asymptotic supergravity background that does not get any 
corrections in the expansion.

\subsubsection*{`Exact brane' applications and open/closed string duality}
Ref.\ \cite{Niarchos:2015moa} recently argued that this effective description of collective mode dynamics in gravity 
is related holographically to the effective description of a dual non-gravitational higher-spin theory
(open string field theory in the case of D-branes) via a general open/closed string duality.
For extremal $p$-brane solutions in flat space, it was further anticipated that the derivative corrections of the 
effective hydrodynamic equations in gravity are dual at all orders to the higher-derivative corrections of the 
abelian Dirac-Born-Infeld (DBI) action, which can be computed independently in classical open string theory.

Three interesting practical aspects of this connection are the following. First, it is interesting that gravitational solutions
in asymptotically flat space are carrying information about the higher-spin degrees of freedom of the dual open 
string field theory. Second, exact solutions in gravity or the dual open string theory are 
providing non-perturbative completions of the hydrodynamic blackfold derivative expansion.
Third, when the exercise is performed in curved/fluxed asymptotic backgrounds in the `exact brane' application
(i.e.\ exactly in $r_H/L$), the effective long-wavelength description of the collective mode dynamics is expected to provide an interesting deformation of DBI-like actions that incorporates information about open-closed string couplings beyond the standard ones in weak coupled open string theory. Finite temperature corrections are producing further deformations of DBI-like actions.

\subsubsection*{Multiple expansions and further approximations in the blackfold approach}
Frequently, in practical applications one has to deal with complicated gravitational configurations where some of the 
exact solution prerequisites of the above (super)gravity problem are not known. For example, the exact 
leading order solution is not a priori known, and has to be constructed from scratch. 
In that case we cannot proceed with the `exact brane' application of the blackfold formalism that was outlined above. 
Instead, one can attempt to employ a secondary parallel expansion scheme that reconstructs the leading order 
solution perturbatively around a solution in a different asymptotic background \cite{Emparan:2009at}.
As an illustration, let us consider two examples emphasizing the interplay of different scales and the multiple
expansions associated with them.

As a first concrete example consider the construction of a black string solution in AdS. In this case the 
background characteristic scale is the radius $L$ of the asymptotic AdS solution. Since an exact leading order black string solution in AdS is not known it was pointed out in \cite{Caldarelli:2008pz} 
that one could proceed perturbatively in the limit
\beq
\label{introba}
r_H \ll \min(\RR, L)
~.
\eeq
Besides the derivative expansion in the limit $\nicefrac{r_H}{\RR} \ll 1$, \eqref{introaa}, which is characteristic of the
`exact brane' application of the blackfold formalism, the inequality \eqref{introba} allows us to 
implement a second parallel expansion in the small ratio $\nicefrac{r_H}{L}$. At first order in this second expansion
the leading order solution of the blackfold expansion can be approximated locally (in the transverse space) 
by the well-known uniform black string solution in flat space. From the point of view of the leading order 
blackfold equations, in this
regime one describes how a black string probe propagates on the AdS background. For details of this
approximation we refer the reader to Refs.\ \cite{Caldarelli:2008pz,Armas:2010hz}.
More generally, the cases that we considered in Sec.~\ref{sec:near} are of this type, since the charged black brane
solutions we use as input are asymptotically flat. 

As another example consider the case of a double-centered D3-brane solution in ten-dimensional type IIB supergravity.
Viewing one of the centers as the background spacetime we can ask whether it is possible to add perturbatively 
the second center to obtain general solutions describing how two stacks of D3-branes interact in supergravity. 
We note immediately that this is a case where the single-centered D3-brane solution cannot be viewed as a proper 
asymptotic background according to the definition of the problem posed in the beginning.
It is clear that the second center backreacts to deform the first center. 

In the `exact brane' application of the blackfold formalism in this example the true background solution is the asymptotic 
flat space, and one would have to begin with an exact double-centered leading order solution.
For example, extremal solutions could be constructed perturbatively around the planar double-centered supersymmetric
solution with harmonic function
\beq
\label{introbb}
H = 1 +  \frac{L^4}{| \vec x - \vec \Delta |^4} + \frac{r_H^4}{|\vec x|^4}
~.
\eeq
In this expression, we call $L$ the near-horizon AdS radius of the first center at $\vec x = \vec \Delta$, and $r_H$ the 
near-horizon AdS radius of the second center at $|\vec x|=0$. $\vec x$ is a 6-vector parameterizing the 
six-dimensional space transverse to the planar D3-brane worldvolumes of the leading order solution.
The blackfold derivative expansion would proceed by promoting the scales $L$, $r_H$, and
the vector of relative positions $\vec\Delta$ to slowly varying functions of the 3-brane worldvolume coordinates.
For this expansion we would simply require the hierarchy of scales
\beq
\label{introbc}
L,\, r_H,\, |\vec\Delta| \ll \RR
~.
\eeq

It is clear from \eqref{introbb} that in regions where $|\vec x|\ll |\vec \Delta|$ (the vicinity of the second center) 
we can approximate
the leading order solution in terms of a single-centered uniform 3-brane in flat-space. In the region $r_H \ll |\vec x| \ll |\vec \Delta|$
the space asymptotes to the background of the first center. Then, in the spirit of \eqref{introba} we could 
employ a multiple expansion in the limit
\begin{subequations}
\label{introbd}
\beq
\label{introbda}
r_H \ll |\vec \Delta|~,
\eeq
\beq
\label{introbdb}
r_H \ll \min(\RR, L)
~.
\eeq
\end{subequations}
At first order in the power-series expansion in terms of the small ratios $\nicefrac{r_H}{|\vec \Delta|}$ and 
$\nicefrac{r_H}{L}$ we can 
phrase the blackfold expansion (namely the expansion in powers of $\nicefrac{r_H}{\RR}$) in terms of a deformed flat-space D3-brane 
(representing the second center) that propagates in the background of the first center. However, unlike the 
case of the inequality \eqref{introba}, at higher orders of $\nicefrac{r_H}{|\vec \Delta|}$ and $\nicefrac{r_H}{L}$ the backreaction of the 
second-center D3-blackfold to the first center background has to be included.

\subsubsection*{Multiple expansions versus open $+$ closed string theory}
As we mentioned previously,
in the case of an exactly known leading order solution and the single expansion in \eqref{introaa}, the
collective mode (blackfold) equations are phrased as a lower-dimensional theory on a screen propagating
in a fixed background. This lower-dimensional theory has a conjectured open string dual \cite{Niarchos:2015moa}. 
The open and closed string pictures are complementary equivalent descriptions of the same dynamics.

In contrast, the above-mentioned multiple expansions, e.g.\ when \eqref{introba} holds, have a closer
resemblance to an interacting system of both open and closed strings. In the associated derivative expansion
schemes in gravity, the effective theory on the lower-dimensional screen interacts order-by-order
with a {\it dynamical} gravitational theory in the bulk. 
The dynamics of the `open/closed string' couplings in the gravity-induced effective description are expressing the 
parallel expansion in $\nicefrac{\RR}{L}$ and the associated backreaction effects of the probe to the bulk.
It would be very interesting to understand better how such multiple expansions proceed in gravity, and how 
the bulk-boundary interactions are encoded in the effective long-wavelength description. From a purely 
effective field theory point of view, when the backreaction effects are included
one has to deal with related self-gravity effects. Typically, such effects lead to divergences. 
Since in the context of the expansions of the full gravitational equations one has a concrete 
underlying system of equations that characterize well-defined gravity solutions, it is natural
to expect that a proper understanding of the gravity-induced effective description knows how to deal 
properly with such divergences. It is interesting to examine this aspect in detail.
We emphasize again, that no backreaction effects are expected in 
the `exact brane' application of the blackfold formalism where the asymptotic gravity solution is fixed and non-dynamical
at all orders in the long-wavelength derivative expansion.

\subsection{Open problems}
\label{openproblems}

Let us conclude with the summary of a few interesting open problems that constitute a natural continuation of 
the work presented in this paper.

\subsubsection*{Effective actions for arbitrary background field configurations}
In Sec.~\ref{partition} we have derived, using diffeomorphism and gauge invariance, the general form of the constraint equations of systems coupled to a background gauge field $\mathcal C_{q+1}$ and a dilaton $\varphi$. This, however, does not exhaust in any way the possible types of couplings and force terms that have been derived in Sec.~\ref{far}. In certain cases, it is straightforward to extend the analysis of Sec.~\ref{partition} to include further couplings. For example, in the case of type IIA/B supergravity, one may consider background solutions with $H_3\ne0$ and probes without magnetic currents. In these cases, the hydrostatic partition function is easily generalized by adding further gauge fields of different ranks. However, once magnetic currents are turned on, it is necessary to consider couplings to the several Dirac branes involved as in \cite{Bandos:1997gd}. Furthermore, if $H_3$ is non-vanishing, further work will be necessary in order to obtain force terms of the schematic form $H.C.J$. 
It would be of great interest to understand such examples in detail since they would allow us to study, for example, the DBI action and the PST action \cite{Pasti:1997gx} at finite temperature in an arbitrary background in the spirit of 
\cite{Grignani:2010xm, Grignani:2011mr,Grignani:2012iw,Niarchos:2012cy, Niarchos:2012pn, Armas:2012bk, Niarchos:2013ia, Armas:2013ota, 
Grignani:2013ewa,Armas:2014nea, Giataganas:2014mla}.

\subsubsection*{Effective actions at extremality and new supergravity solutions}
The construction of new extremal supergravity solutions is particularly opportune technically, especially if some 
amount of supersymmetry is present. It was pointed out in \cite{Niarchos:2014maa} that a perturbative
construction of supersymmetric solutions in a long-wavelength regime may lead to an interesting framework
of $G$-structure deformations where the Killing spinor version of the constraint equations is related to $\kappa$-symmetry conditions. 

It was further pointed out in \cite{Niarchos:2015moa}, as we emphasized above, that the 
long-wavelength treatment of extremal (but not necessarily supersymmetric) $p$-brane solutions leads naturally to the 
DBI equations, well-known from open string theory. It would be interesting to extend this connection 
beyond the examples of \cite{Niarchos:2015moa} to include general open-closed string couplings of the DBI.
It would also be interesting to explore what kind of deformations of the DBI action are induced in the 
supergravity-derived blackfold effective action in 'exact brane' applications of blackfolds where one goes beyond
the usual probe equations derived from the use of approximations based on flat space $p$-brane solutions.

Further development of this formalism in supergravity could be useful in many applications that require
the construction of complicated extremal supergravity solutions. A particular problem of interest, is the 
construction of new solutions describing the gravitational backreaction of massive configurations of D/M-branes. 
Recent applications to brane intersections in string/M-theory in flat space include \cite{Grignani:2010xm,Niarchos:2012pn,Grignani:2012iw,Niarchos:2012cy, Niarchos:2012pn, Armas:2012bk, Niarchos:2013ia, Armas:2013ota, 
Grignani:2013ewa}. Solutions in supergravity backgrounds with fluxes have not been studied in this manner
and it would be interesting to do so. For example, it would be interesting to examine the backreaction 
of anti-brane configurations in backgrounds with fluxes along the lines of recent work in this direction (see, for 
instance \cite{Bena:2009xk,Blaback:2011pn,Bena:2016fqp}). For example, one can ask about extremal D3-D5 blackfold constructions in the Klebanov-Strassler
background \cite{Klebanov:2000hb}  (in analogy to \cite{Kachru:2002gs,Bena:2009xk,Bena:2011wh}), or anti-M2 blackfolds at the tip of Stenzel space in M-theory (in analogy to 
\cite{Klebanov:2010qs,Bena:2010gs,Massai:2011vi}). 
In all these cases, the real problem, which is also the central issue in the recent literature, has to do with 
the construction of the leading order solution (in the language of the previous subsection). The combination of recent
results in the literature of anti-brane backreaction with MAE techniques frequently 
used in the blackfold formalism might be fruitful.

\subsubsection*{More examples of `exact brane' applications; similarities with the tachyon-DBI derivation}
In the formulation of the general supergravity problem in the beginning of subsection \ref{properetc}
we purposely included the case of leading order $p$-brane solutions with $m<p+1$ symmetric worldvolume directions.
In such cases the leading order solution is already inhomogeneous in $p+1-m$ directions. For example, it could
be time-dependent in a time-independent background.
It would be interesting to find and examine an explicit example of this type. At this point one cannot help but notice
the analogy of such a case with $S$-brane and rolling tachyon solutions in string theory 
\cite{Gutperle:2002ai,Sen:2002nu} 
and the corresponding derivation of the tachyon-DBI action as an open string long-wavelength effective 
field theory around the rapidly changing rolling tachyon solution \cite{Kutasov:2003er}.

\subsubsection*{Forcing and time-dependence}
In the presence of a curved/fluxed background the typical solutions will be non-stationary solutions with a 
small number of symmetries, with the non-stationarity being driven by the external forcing. These effects will 
be manifested in the exact leading order solution and/or in the forced blackfold equations in multiple expansions.
It is interesting to understand further the physics of such effects, and their implications in the construction of
perturbative time-evolving black brane solutions in explicit cases; for example, in cases of black brane solutions
moving in the vicinity of other black hole solutions.

\subsubsection*{Higher order corrections to effective actions}
The type of effective actions considered in this paper were derived at leading order in a long-wavelength expansion. At higher orders, one must take into account derivative corrections due to fluid and elastic deformations. These corrections can be taken into account in a systematic way following \cite{Armas:2013hsa, Armas:2013goa, Armas:2014rva} and it would be extremely interesting to consider these in the case of multi-charged bound states in the presence of external backgrounds fields as well as in the presence of boundaries as in \cite{Armas:2015ssd}. As advertised in \cite{Armas:2013aka}, these corrections would account for the polarization properties of relativistic fluids encoded in the form of the electric and magnetic susceptibilities in the stress-energy tensor and electric/magnetic currents besides the Young modulus \cite{Armas:2011uf} and the piezoeletric moduli \cite{Armas:2012ac}. Of considerable interest would be to consider corrections due to possible quantum anomalies for theories with higher-form gauge fields.

We also note that to compute the  response coefficients corresponding to such higher order corrections, one needs to
have access to the full first-order corrected solution, i.e. solve all the field equations in the near-zone to first order (and not the subset of constraint equations considered in Sec.~\ref{sec:near} for particular examples). Performing this analysis (which was
done for various types of blackfold constructions in e.g.  \cite{Emparan:2007wm,Caldarelli:2008pz,Camps:2008hb,
Camps:2012hw,Gath:2013qya,Emparan:2013ila,DiDato:2015dia}), even though challenging, would be interesting in its own right in order to 
fully show that  to this order a solution exists that is regular on the horizon.

\subsubsection*{First law of thermodynamics in arbitrary background fields}
The effective actions studied in this paper can be used to construct new stationary black hole solutions by solving the constraint equations in the presence of background fields for specific configurations. As exemplified in \cite{Armas:2015qsv}, if the resulting solutions are characterized by length scales associated with the background spacetime, these can be allowed to vary leading to new terms in the first law of thermodynamics involving integrated brane tensions - the dual thermodynamic quantities to the background length scales. However, here we have generalized these actions to also include background gauge and dilaton fields, which can in principle be characterized by several non-trivial length scales, e.g. as in the case of a non-trivial black hole solution playing the role of the background field configuration. These length scales can now be allowed to vary leading to new terms in the first law for which their dual thermodynamic quantities may be of interest to study.

 \section*{Acknowledgments}
We would like to thank Jyotirmoy Bhattacharya, Matthias Blau, Ori Ganor, Gianluca Grignani, Adolfo Guarino, Troels Harmark, Jelle Hartong, Gianluca Inverso, Cindy Keeler, Ricardo Monteiro and Marta Orselli for useful discussions.
The work of VN was supported in part by European Unions Seventh Framework Programme under
grant agreements FP7-REGPOT-2012-2013-1 no 316165, and by European Unions Horizon
2020 Programme under grant agreement 669288-SM-GRAV-ERC-2014-ADG. The work of JA was supported by the ERC Starting Grant 335146 HoloBHC. JA acknowledges two short term mission grants to NBI, one by the Holograv network and another by the COST network. JG is supported by VILLUM FONDEN, research grant VKR023371. The work of NO is supported in part by the Danish National Research Foundation project ``New horizons in particle and condensed matter physics from black holes". The work of AVP has been supported by The Danish Council for Independent Research - Natural Sciences (FNU), DFF-4002-00307. AVP gratefully acknowledge support from UC Berkeley where some of the research for this paper was carried out.

 \appendix
 
 \section{Notation and conventions} \label{app:notation}

In this section we collect the notation and convention used throughout this paper. We define a generic $p$-form with components $A_{(p)\mu_1...\mu_p}$ and its Hodge dual as
 \begin{align}
 \begin{split}
& A_{(p)}=\frac{1}{p!}A_{(p)\mu_1...\mu_p}dx^{\mu_1}\wedge...\wedge dx^{\mu_p}~~,\\
 &\star A_{(p)}=\frac{1}{p!(D-p)!}{\epsilon_{\mu_1...\mu_{D-p}}}^{\nu_1...\nu_p}A_{(p)\nu_1...\nu_p}dx^{\mu_1}\wedge...\wedge dx^{\mu_{D-p}}~~.
 \end{split}
 \end{align}
 Furthermore, the wedge product of a $p$- and $q$-form is defined as
 \beq
 A_{(p)}\wedge B_{(q)}=\frac{1}{p!q!}A_{(p)[\mu_1...\mu_p}B_{(q)\nu_1...\nu_q]}dx^{\mu_1}\wedge...\wedge dx^{\mu_p}\wedge dx^{\nu_1}\wedge...\wedge dx^{\nu_q}~~,
 \eeq  
 while the exterior derivative of a $p$-form is given by
 \beq
 dA_{(p)}=\frac{(p+1)}{p!}\nabla_{[\mu_1}A_{(p)\mu_2...\mu_{p+1}]}dx^{\mu_1}\wedge...\wedge dx^{\mu_{p+1}}~~.
 \eeq 
We define the square of the D-dimensional $\star$ operator acting on a $p$-form in space-times with Lorentzian signature as
 \beq
 \star^2=(-1)^{p(D-p)+1}~~.
 \eeq
 We also introduce the star operator on the worldvolume $\star_{(p+1)}$ such that $\star_{(p+1)}1=\sqrt{-\gamma}d\sigma^{0}\wedge...\wedge d\sigma^{p}$ as well as the star operator $\star_{(p)}$ on the spatial part of the worldvolume $\mathcal{B}_{p}$ such that $\star_{(p)}1=\sqrt{-\gamma}d\sigma^{1}\wedge...\wedge d\sigma^{p}$. We also assume that the worldvolume topology is $\mathcal{W}_{p+1}=\mathbb{R}\times\mathcal{B}_{p}$ and hence that the determinant of the induced metric on the worldvolume can be decomposed as $\sqrt{-\gamma}d\sigma^{1}\wedge...\wedge d\sigma^{p}=R_0dV_{(p)}$ where $R_0$ is the modulus of the Killing vector field $\xi^{a}\partial_a$ associated with worldvolume time translations.

 \section{Explicit form of type IIA/B probe brane equations} \label{app:AB}
 In this appendix we restrict the equations of motion \eqref{eq:typeeom}-\eqref{eq:duality} and the probe brane equations \eqref{eq:forcetype}-\eqref{eq:current} to the type IIA/B cases individually. 
 
 \subsection*{Type IIA supergravity}
We consider type IIA supergravity by restricting the equations of motion \eqref{eq:typeeom}-\eqref{eq:duality} to $q=0,2,4,6$. When all currents vanish, these equations can be obtained from the action\footnote{We are using the conventions of \cite{becker2006string} but we used the equivalence of Chern-Simons terms $\int B_{2}\wedge F_{4}\wedge F_{4}=\int C_{3}\wedge H_{3}\wedge F_{4}$ up to boundary terms. }
\begin{align} \label{eq:typeA}
\begin{split}
I=&~\frac{1}{16\pi G}\int_{\mathcal{M}_{10}}\left[\star R-\frac{1}{2}d\phi\wedge\star d\phi-\frac{1}{2}e^{-\phi}H_{3}\wedge \star H_{3}-\frac{1}{2}\sum_{q=0,2}e^{a_q\phi}\tilde F_{q+2}\wedge \star \tilde F_{q+2}\right] \\
& -\frac{1}{32\pi G} \int_{\mathcal{M}_{10}} C_{3} \wedge H_{3} \wedge F_{4}~~,
\end{split}
\end{align}
while in the presence of sources the equations of motion become\footnote{We note that before coupling sources to the equations of motion, which can be obtained from \eqref{eq:typeA}, we use the Bianchi identities $dH_{3}=d\tilde F_{q+2}=0~,~q=0,2$ in vacuum in order to simplify them. This simplification is such that the equations of motion for type IIA are those which can be obtained from \eqref{eq:typeeom} by taking all current sources to be zero and restricting to $q=0,2,4,6$. We proceed similarly for type IIB.}
\begin{align} \label{eq:IIAeom}
\begin{split}
&d\left(e^{-\phi}\star H_{3}-e^{\phi/2}\star \tilde F_{4}\wedge C_{1}-\frac{1}{2}C_{3}\wedge F_{4}\right)=-16\pi G \star \textbf{j}_{2}~~, \\
&d\left(e^{3\phi/2}\star F_{2}\right)+e^{\phi/2}H_{3}\wedge \star \tilde F_{4}=16\pi G \star \jt_{1}~~,\\
&d\left(e^{\phi/2}\star \tilde F_{4}\right)+H_{3}\wedge \tilde F_{4}=16\pi G \star \jt_{3}~~.
\end{split}
\end{align}
In turn, the equations of motion for the probe brane can be obtained by restricting \eqref{eq:forcetype} and hence we obtain
\begin{align} \label{eq:forcetypeA}
\begin{split}
\nabla_{\mu}\bt=&~\frac{1}{2!}{H_{3}^{\nu\mu_1\mu_2}}{\textbf{j}}_{2\mu_1\mu_{2}}+\frac{e^{-\phi}}{6!}{{H}_{7}^{\nu\mu_1...\mu_6}}{\boldsymbol{\mathfrak{j}}}_{6\mu_1...\mu_{6}}+\textbf{j}_\phi\partial^{\nu}\phi\\
&+F_{2}^{\nu\mu_1}\jt_{1\mu_1}+\left(\frac{1}{3!}{\tilde F_{4}^{\nu\mu_1...\mu_{3}}}-\frac{1}{2!}H_{3}^{\nu \mu_1\mu_2}C_{1}^{\mu_3}\right)\jt_{3\mu_1...\mu_{3}}\\
&+e^{\phi/2}\left(\frac{1}{5!}\tilde{{F}}_{6}^{\nu\mu_1...\mu_{5}}-\frac{1}{2\cdot 3!}H^{\nu\mu_1\mu_2}_3 C_{3}^{\mu_3...\mu_5}\right){{{\jmt}_{5}}}_{\mu_1...\mu_{5}}\\
&+\frac{e^{3\phi/2}}{7!}\tilde{{F}}_{8}^{\nu\mu_1...\mu_{7}}{{{\jmt}_{7}}}_{\mu_1...\mu_{7}}-\frac{e^{\phi/2}}{4!}\tilde F_{4}^{\mu_1...\mu_4}{\left[\star{\boldsymbol{\mathfrak{j}}}_{6}\wedge C_{1}\right]^{\nu}}_{\mu_1...\mu_4}~~,
\end{split}
\end{align}
where we have defined $\tilde F_{6}=\star \tilde F_{4}$ and $\tilde F_{8}=\star \tilde F_{2}$, while the current conservation equations \eqref{eq:current} lead to
\begin{align}
\begin{split}
&d\star \jt_{1}-\star \jt_{3}\wedge H_{3}-e^{\phi/2}\star \tilde F_{4}\wedge\star{\boldsymbol{\mathfrak{j}}}_{6}=0~~,\\
&d\star \jt_{3}-\star{\boldsymbol{\mathfrak{j}}}_{6}\wedge F_{4}-H_{3}\wedge \star \jmt_{5}=0~~,\\
&d\star \textbf{j}_{2}=0~~,~~d\star {\boldsymbol{\mathfrak{j}}}_{6}=0~~,~~d\star {\jmt}_{5}=H_{3}\wedge \star \jmt_{7}~~,~~d\star {\jmt}_{7}=0~~.\\
\end{split}
\end{align}

\subsection*{Type IIB supergravity}
We now consider restricting the equations of motion \eqref{eq:typeeom}-\eqref{eq:duality} to $q=-1,1,3$. The resulting equations of motion can be obtained from the action\footnote{We are following the conventions of \cite{becker2006string} but have redefined $C_{4}\to C_{4}-(1/2)B_{2}\wedge C_{2}$ for convenience. }
\begin{align} \label{eq:typeB}
\begin{split}
I=&~\frac{1}{16\pi G}\int_{\mathcal{M}_{10}}\left[\star R-\frac{1}{2}d\phi\wedge\star d\phi-\frac{1}{2}e^{-\phi}H_{3}\wedge \star H_{3}-\frac{1}{2}\sum_{q=-1,1}e^{a_q\phi}\tilde F_{q+2}\wedge \star \tilde F_{q+2}\right] \\
&-\frac{1}{16\pi G}\int_{\mathcal{M}_{10}}\left[\frac{1}{4}\tilde F_{5}\wedge \star \tilde F_{5} +\frac{1}{2} C_{3} \wedge H_{3} \wedge F_{4}\right]~~,
\end{split}
\end{align}
supplemented with the self-duality relation $\tilde F_{5}=\star \tilde F_{5}$. In the presence of sources, the equations of motion read
\begin{align} \label{eq:IIBeom}
\begin{split}
&d\left(e^{-\phi}\star H_{3}-e^{\phi}\star \tilde F_{3}\wedge C_{0}-\frac{1}{2}\star \tilde F_{5}\wedge C_{2}+\frac{1}{2}C_{4}\wedge F_{3}\right)=-16\pi G \star \textbf{j}_{2}~~, \\
&d\left(e^{2\phi}\star F_{1}\right)+e^{\phi}H_{3}\wedge \star  \tilde F_{3}=-16\pi G \star \jt_{0}~~,\\
&d\left(e^{\phi}\star \tilde F_{3}\right)+H_{3}\wedge \star \tilde F_{5}=-16\pi G \star \jt_{2}~~,\\
&d\left(\star \tilde F_{5}\right)-H_{3}\wedge F_{3}=-16\pi G \star \jt_{4}~~.\\
\end{split}
\end{align}
The equations of motion for the probe brane can be obtained by restricting \eqref{eq:forcetype} take the form
\begin{align} \label{eq:forcetypeB}
\nabla_{\mu}\bt=&~\frac{1}{2!}{H_{3}^{\nu\mu_1\mu_2}}{\textbf{j}}_{2\mu_1\mu_{2}}+\frac{e^{-\phi}}{6!}{{H}_{7}^{\nu\mu_1...\mu_6}}{\boldsymbol{\mathfrak{j}}}_{6\mu_1...\mu_{6}}+\textbf{j}_\phi\partial^{\nu}\phi \nonumber\\
&+F_{1}^{\nu}\jt_{0}+\frac{1}{2!}\left({\tilde F_{3}^{\nu\mu_1...\mu_{2}}}+H_{3}^{\nu \mu_1\mu_2}C_{0}\right)\jt_{2\mu_1\mu_{2}}\\
&+\frac{1}{4!}\left({\tilde F_{5}^{\nu\mu_1...\mu_{4}}}+3H_{3}^{\nu \mu_1\mu_2}C_{2}^{\mu_3...\mu_{4}}\right)\jt_{4\mu_1...\mu_{4}}\\
&+e^{\phi}\left(\frac{1}{6!}\tilde{{F}}_{7}^{\nu\mu_1...\mu_{6}}+\frac{1}{2\cdot 4!}H^{\nu\mu_1\mu_2}_3 C_{4}^{\mu_3...\mu_6}\right){{{\jmt}_{6}}}_{\mu_1...\mu_{6}} \\
&+\frac{e^{2\phi}}{8!}\tilde{{F}}_{9}^{\nu\mu_1...\mu_{8}}{{{\jmt}_{8}}}_{\mu_1...\mu_{8}}-\frac{e^{\phi}}{3!}\tilde F_{3}^{\mu_1...\mu_3}{\left[\star{\boldsymbol{\mathfrak{j}}}_{6}\wedge C_{0}\right]^{\nu}}_{\mu_1...\mu_3} \nonumber ~~,
\end{align}
where we have defined $\tilde F_{7}=\star \tilde F_{3}$ and $\tilde F_{9}=\star \tilde F_{1}$, while the current conservation equations \eqref{eq:current} lead to
\begin{align}
\begin{split}
&d\star \jt_{2}+H_{3}\wedge\star \jt_{4}+\star{\boldsymbol{\mathfrak{j}}}_{6}\wedge \tilde F_{5}=0~~,\\
&d\star J_{4}-\star{\boldsymbol{\mathfrak{j}}}_{6}\wedge F_{3}+\star{\jmt}_{6}\wedge \tilde F_{5}=0~~,\\
&d\star j_{2}=0~~,~~d\star {\boldsymbol{\mathfrak{j}}}_{6}=0~~,~~d\star {\jmt}_{6}=H_{3}\wedge\star\jmt_{8}~~,~~d\star {\jmt}_{8}=0~~.\\
\end{split}
\end{align}
Note that the magnetic force associated to $F_{5}$ in \eqref{eq:forcetypeB} was exchanged by a Lorentz type force as a consequence of the self-duality relation \eqref{eq:selfcurrrent}. Also note that there is no conservation equation associated with $\jt_0$.

\section{Effective currents, charges and constraint equations} \label{eqn:DpF1charges}

In this appendix we provide the effective currents and charges for the several black brane solutions used in the main text.

\subsection{Black branes carrying \texorpdfstring{$q=p$}{q=p}-brane charge}
\label{app:charged}

The asymptotic stress-energy tensor and current of the charged black $p$-brane solution \eqref{eq:pbranesol} take the form
\begin{equation}  \label{eq:tcp}
	T_{ab} = \epsilon \, u_a u_b + P P_{ab} ~, \quad J_{p+1} = Q_p \star_{(p+1)} \mathbf{1} ~.
\end{equation}
The stress-energy tensor is of the form of a perfect fluid with the energy density and (negative) pressure given by
\begin{align} \label{eq:pcharged}
\begin{split}
	\epsilon &= \frac{\Omega_{(n+1)}}{16 \pi G} r_0^n ( n+1+nN\sinh^2\alpha) ~, \quad
	P = -\frac{\Omega_{(n+1)}}{16 \pi G} r_0^n(1 + nN\sinh^2\alpha) ~,
\end{split}
\end{align}
while the charge density and conjugate electric potential are
\begin{equation} \label{eq:chargedcharges}
	Q_p = \frac{\Omega_{(n+1)}}{16 \pi G} n \sqrt{N} r_0^n \cosh\alpha \sinh\alpha ~, \quad \Phi_p= \sqrt{N} \tanh\alpha ~.
\end{equation}
Alternatively, we can express the stress-energy tensor as \cite{Emparan:2011hg}
\begin{align}
\begin{split}
	T_{ab} =& \; \mathcal{T} s \left( u_au_b - \frac{1}{n} \gamma_{ab} \right) - \gamma_{ab} Q_p \Phi_p ~,
\end{split}
\end{align}
with the temperature and entropy density
\begin{equation} \label{eq:ts}
	\mathcal{T} = \frac{n}{4\pi r_0 (\cosh\alpha)^N} \, , \quad s = \frac{\Omega_{(n+1)}}{4 G} r_0^{n+1} (\cosh\alpha)^N ~.
	\end{equation}

\subsection{Black branes carrying Maxwell charge}
\label{app:charged0}

The asymptotic stress-energy tensor and current of the charged black $p$-brane given with Maxwell charge \eqref{eq:dsmax} take the form
\begin{equation} 
	T_{ab} = \epsilon \, u_a u_b + P P_{ab} ~, \quad J_{1}^{a} = \mathcal{Q} u^{a} ~~,
\end{equation}
where the energy density and pressure are given by
\begin{align}
\begin{split} \label{eq:e0}
	\epsilon &= \frac{\Omega_{(n+1)}}{16 \pi G} r_0^n ( n+1+nN\sinh^2\alpha) ~, \quad
	P = -\frac{\Omega_{(n+1)}}{16 \pi G} r_0^n ~,
\end{split}
\end{align}
while the charge density, chemical potential, temperature and entropy density are the same as in \eqref{eq:chargedcharges} and \eqref{eq:ts}, respectively.

\subsection{D\texorpdfstring{$p$}{p}-F1 bound state}
\label{app:dpf1}

In this appendix we briefly review the effective currents and charges of the D$p$-F1 bound state solution considered in Sec.~\ref{sec:DpF1}. For convenience we will take $\varphi = 0$. The stress-energy tensor can be expressed in the form \cite{Caldarelli:2010xz}
\begin{equation} \label{eq:DpF1stress}
	T_{ab} = \epsilon \, u_a u_b + P v_a v_b + P_{\perp} \perp_{ab} ~,
\end{equation}
where $\perp_{ab} = \eta_{ab} + u_a u_b - v_a v_b$ and the energy density and pressures are
\begin{align}
\begin{split}
	\epsilon &= \frac{\Omega_{(n+1)}}{16 \pi G} r_0^n ( n+1+n\sinh^2\alpha) ~, \\
	P &= -\frac{\Omega_{(n+1)}}{16 \pi G} (1 + n\sinh^2\alpha) ~, \quad
	P_{\perp} = -\frac{\Omega_{(n+1)}}{16 \pi G} (1 + n\sinh^2\alpha \cos^2\xi) ~.
\end{split}
\end{align}
Furthermore, we have the currents
\begin{equation}
	j_2 = \mathcal{Q}_{\text{F1}} \; u \wedge v \, , \quad 
	J_{p-1} = \mathcal{Q}_{p-2} \star_{(p+1)} (u \wedge v) \, , \quad
	J_{p+1} = Q_p \star_{(p+1)} \mathbf{1} ~,
\end{equation}
where the string and top charges are given by
\begin{equation} \label{eqn:DpF1charges0}
		\mathcal{Q}_{\text{F1}} = \sin\xi \mathcal{Q} \, , \quad 
		Q_p = \cos\xi \mathcal{Q} \, , \quad \text{with} \quad
		\mathcal{Q} = \frac{\Omega_{(n+1)}}{16 \pi G} n r_0^n \cosh\alpha \sinh\alpha ~,
\end{equation}
and electric potentials (conjugate to the charges) are
\begin{equation} \label{eqn:DpF1pots0}
		\Phi_{\text{F1}} = \sin\xi \Phi \, , \quad 
		\Phi_p = \cos\xi \Phi \, , \quad \text{with} \quad
		\Phi = \tanh \alpha ~.
\end{equation}
The charge associated with the $(p-1)$-current $J_{p-1}$ can be expressed in terms of the above
\begin{equation}
	\mathcal{Q}_{p-2} = \Phi_p \mathcal{Q}_{\text{F1}} = \Phi_{\text{F1}} Q_{p} ~.
\end{equation}
We note that for $\xi=0$, the effective currents and charges reduces to the one given in App.~\ref{app:charged} (in ten dimensions where $N=1$).
Introducing the worldvolume metric $\gamma_{ab}$ and the projector $h_{ab} = -u_a u_b +v_a v_b$, we can alternatively express the stress-energy tensor as \cite{Emparan:2011hg}
\begin{align}
\begin{split}
	T_{ab} 
	 =& \; \mathcal{T} s \left( u_au_b - \frac{1}{n} \gamma_{ab} \right) - h_{ab} \mathcal{Q}_{\text{F1}} \Phi_{\text{F1}} - \gamma_{ab}Q_p \Phi_p ~,
\end{split}
\end{align}
with the temperature and entropy density
\begin{equation}
	\mathcal{T} = \frac{n}{4\pi r_0 \cosh\alpha} \, , \quad s = \frac{\Omega_{(n+1)}}{4 G} r_0^{n+1} \cosh \alpha ~.
	\end{equation}

\subsection{D\texorpdfstring{$p$}{p}-F1 constraint equations} \label{sec:intrcomb}

Let the components of the Einstein equations be denoted by $\mathcal{E}_{\mu\nu} = G_{\mu\nu} - 8\pi G T_{\mu\nu}$ and let the components of the l.h.s. of each of the Eqs.~\eqref{eq:IIAeom} (Eqs.~\eqref{eq:IIBeom}) for type IIA(B) be denoted by $\mathcal{M}_X^{\mu_1 \mu_2 ... }$ where $X$ denotes the associated field. Furthermore let the components of the Hodge dual of those equations be denoted by $\mathcal{N}_{X}^{\mu_1\mu_2...}$.  Then, the constraint equations \eqref{eqn:DpF1bgrcons} appear in the following linear combinations of the system equations
\begin{align} \label{eq:intsyscom}
\begin{split}
\nabla_a T^{ab} - \mathcal{F}^{b} \quad \quad \propto& \quad 
 \mathcal{E}_r
^{\phantom{r}b} + c_1 \mathcal{M}_{A_{p+1}}^{b \Omega} + 
	c_2 \mathcal{M}_{B}^{b \sigma_2 ... \sigma_{p} \Omega} +
	c_3 \mathcal{M}_{A_{p-1}}^{b \sigma_0 \sigma_1 \Omega} \\
\nabla_a j_{2}^{a b} \quad \propto& \quad \mathcal{N}_{B}^{r b} \, , \\
\nabla_a J_{p+1}^{a a_1 ...a_p} \quad \propto& \quad \mathcal{N}_{A_{p+1}}^{r a_1 ...a_p} \, , \\
\nabla_a J_{p-1}^{a a_1 ... a_{p-2}} - \frac{1}{3!} \mathcal{H}_{abc}  J_{p+1}^{abc a_1 ... a_{p-2}} \quad \propto& \quad \mathcal{N}_{A_{p+1}}^{r \sigma_0 \sigma_1 a_1 ... a_{p-2}} + c_4 \mathcal{N}_{A_{p-1}}^{r a_1 ... a_{p-2} } ~,
\end{split}
\end{align}
where $c_i$ are $r$-dependent functions, $\Omega$ collectively denotes the coordinates on the transverse sphere and the F1-string is aligned along the $\sigma^1$-direction. $\mathcal{F}^b$ is the collection of force terms appearing in Eq.\eqref{eqn:DpF1bgrcons}. Notice that the combinations of the system equations are exactly such that the l.h.s. of Eqs.~\eqref{eq:intsyscom} is $r$-independent.

\section{Entropy current analysis with a dilaton forcing function} \label{app:entropy}
In this appendix we perform an entropy current analysis of the forced fluids considered in Sec.~\ref{sec:hydroqp} and show that in order to have stationary flows one must require the fluid to be aligned with a worldvolume Killing vector field. This analysis follows closely that of \cite{Armas:2013goa}.

The fluids of Sec.~\ref{sec:hydroqp} have only the temperature $\mathcal{T}$ and the fluid velocity $u^{a}$ as degrees of freedom. These fluids are characterized also by a conserved total charge $Q_p$ but this charge is not a degree freedom, instead it only labels different families of such fluids. The thermodynamic properties given in App.~\eqref{app:charged} are those of a neutral perfect fluid. Therefore, the dynamical equations along the worldvolume are just those of \eqref{eq:c12}, which we write explicitly as
\beq \label{eq:eom2}
u^{a}\partial_a\mathcal{T}=-\frac{1}{\mathcal{T}}\frac{\partial \mathcal{T}}{\partial s}\left(\mathcal{T}s\theta+j_\phi \dot\varphi\right)~~,~~{P^{c}}_b\partial^{b}\mathcal{T}=\frac{1}{s}{P^{c}}_b\left(j_\phi\partial^{b}\varphi-\mathcal{T}sa^{b}\right)~~,
\eeq
where we have defined $\dot\varphi=u^{a}\partial_a\varphi$ and introduced the fluid expansion $\theta$ via the decomposition
\beq
\nabla_au_b=-u_a a_{b}+\sigma_{ab}+\omega_{ab}+\frac{\theta}{p}P_{ab}~~,
\eeq
where $\sigma_{ab}$ and $\omega_{ab}$ are the fluid shear and vorticity respectively. Up to first order in derivatives the most general stress-energy tensor, dilaton current and entropy current allowed by symmetries are\footnote{One can also consider elastic corrections due to deformations of the surface where the fluid lives as in \cite{Armas:2013goa} but we are not concerned with these corrections here.}
\begin{align}
\begin{split}
T^{ab}&=P\gamma^{ab}+(\epsilon+P)u^{a}u^{b}-\zeta \sigma^{ab}-\eta \theta P^{ab}~~,\\
j_\phi&=j_{\phi}^{(0)}+\alpha_1\theta+\alpha_2\dot\varphi~~,\\
J_s^{a}&=su^{a}+\beta_1 \theta u^{a}+\beta_2 a^{a}+\alpha_3 u^{a}\dot\varphi +\alpha_4 {P^{a}}_b\partial^{b}\varphi~~,
\end{split}
\end{align}
where all coefficients $\zeta,\eta,\beta_i,\alpha_i$ are functions of the temperature $\mathcal{T}$ and $j_\phi^{(0)}$ is the leading order dilaton current, which in the case of the branes of Sec.~\ref{sec:hydroqp} is given by $a_pQ_p\Phi_p/2$, however, we have left it arbitrary in our analysis in this appendix.\footnote{In particular, the case studied in \cite{Bhattacharyya:2008ji} has $j_\phi^{(0)}=0$.}

We wish to impose positivity of the divergence of the entropy current $\nabla_a J^{a}_s\ge0$, thereby ensuring that the second law of thermodynamics is satisfied. Using \eqref{eq:eom2} we find
\begin{align} \label{eq:entropy}
\begin{split}
\nabla_a J^{a}_s=& \; \zeta \sigma^2+\eta \theta^2+\left(\beta_1-s\beta_1'\frac{\partial \mathcal{T}}{\partial s}+\frac{\beta_2}{p}\right)\theta^2-\beta_2'a^2+\beta_2\left(\sigma^{2}+\omega^2\right) \\
&+(\beta_1+\beta_2)u^{a}\nabla_a \theta+\beta_2u^{a}u^{b}\mathcal{R}_{ab}\\
&-\frac{j_\phi^{(0)}}{\mathcal{T}}\dot \varphi -\frac{\alpha_1}{\mathcal{T}}\theta\dot\varphi-\frac{\alpha_2}{\mathcal{T}}\dot\varphi^2\\
&+\left(\alpha_3+\alpha_4-\frac{\partial \mathcal{T}}{\partial s}s\alpha_3'-\frac{j_\phi^{(0)}}{\mathcal{T}}\frac{\partial \mathcal{T}}{\partial s}\beta_1'\right)\theta\dot\varphi+\left(\alpha_3+\alpha_4-\alpha_4'\mathcal{T}-\frac{j_\phi^{(0)}}{s}\beta_2'\right)a^{b}\partial_b\varphi\\
&+\left(\frac{j_\phi^{(0)}}{s}\alpha_4'P^{ab}\partial_a\varphi\partial_b\varphi+\alpha_4 P^{ab}\partial_a\partial_b\varphi+\alpha_3u^{a}u^{b}\partial_a\partial_b\varphi\right)-\frac{j_\phi^{(0)}}{\mathcal{T}}\frac{\partial \mathcal{T}}{\partial s}\alpha_3'\dot\varphi^2~~,
\end{split}
\end{align}
where we have defined $\sigma^{2}=\sigma_{ab}\sigma^{ab}$ and $\omega^2=\omega_{ab}\omega^{ba}$ and introduced the Ricci tensor $\mathcal{R}_{ab}$ on the worldvolume. The \emph{prime} denotes derivatives with respect to $\mathcal{T}$. The first two lines in \eqref{eq:entropy} are those that are also obtained when no dilaton is present in the background. In particular the second line, being linear in fluid data requires $\beta_1=\beta_2=0$ and hence the first line requires the usual result $\zeta\ge0,\eta\ge0$ which is unaffected in the presence of the dilaton. The first three terms in the last line in \eqref{eq:entropy} are linear in independent data and had been classified in \cite{Bhattacharyya:2008ji}. Therefore, if we wish to require positivity of the divergence of the entropy current for arbitrary background source $\varphi$ we must set $\alpha_3=\alpha_4=0$. We are left with the third line in \eqref{eq:entropy}. The first term in the third line is linear in the fluid data but $j_\phi^{(0)}$ is non-zero. Therefore one must require $-j_\phi^{(0)}\dot \varphi\ge0$.\footnote{This specific condition had already been noticed in \cite{Bhattacharyya:2008ji} using other arguments.} If $j_\phi^{(0)}$ is constant, for example, this condition will impose restrictions on the driving force $\dot\varphi$. The second term in the third line is linear in fluid data and therefore, for arbitrary background sources we must have that $\alpha_1=0$. The third term in the third line is quadratic in the fluid data and therefore we obtain the condition $\alpha_2\le0$.

Consider the forced fluid dynamics case analyzed in \cite{Bhattacharyya:2008ji}. To first order in derivatives the dilaton current found there is given by 
\beq
j_{\phi}=-\frac{1}{16\pi G}(\pi\mathcal{T})^3\dot\varphi~~,
\eeq
and hence we identify $j_\phi^{(0)}=0$ and $\alpha_1=0,\alpha_2=-(\pi\mathcal{T})^3/(16\pi G)<0$ in agreement with the analysis above. The entropy current obtained in \cite{Bhattacharyya:2008ji} contains no first order corrections, also in agreement with the analysis presented here.

Consider now the stationary case for which there is no entropy production $\nabla_a J^{a}_s=0$. Due to the presence of non-zero viscosities $\zeta,\eta$ and leading order dilaton current $j_{\phi}^{(0)}$, stationary configurations must satisfy $\theta=\sigma^{ab}=\dot\varphi=0$.\footnote{We are assuming here that the transport coefficients are non-zero, which is indeed the case for the fluids we are considering \cite{Emparan:2013ila,DiDato:2015dia}.} Indeed, this is only possible if the fluid velocity is aligned with a worldvolume Killing vector field, i.e., $u^{a}=\textbf{k}^{a}/\textbf{k}$.




\addcontentsline{toc}{section}{References}
\footnotesize
\providecommand{\href}[2]{#2}\begingroup\raggedright\endgroup


\begin{thebibliography}{10}

\bibitem{Policastro:2001yc}
G.~Policastro, D.~Son, and A.~Starinets, ``{The Shear viscosity of strongly
  coupled N=4 supersymmetric Yang-Mills plasma},''
  \href{http://dx.doi.org/10.1103/PhysRevLett.87.081601}{{\em Phys.Rev.Lett.}
  {\bf 87} (2001)  081601},
\href{http://arxiv.org/abs/hep-th/0104066}{{\tt arXiv:hep-th/0104066
  [hep-th]}}.

\bibitem{Kovtun:2004de}
P.~Kovtun, D.~Son, and A.~Starinets, ``{Viscosity in strongly interacting
  quantum field theories from black hole physics},''
  \href{http://dx.doi.org/10.1103/PhysRevLett.94.111601}{{\em Phys.Rev.Lett.}
  {\bf 94} (2005)  111601}, \href{http://arxiv.org/abs/hep-th/0405231}{{\tt
  arXiv:hep-th/0405231 [hep-th]}}.
An Essay submitted to 2004 Gravity Research Foundation competition.

\bibitem{Bhattacharyya:2008jc}
S.~Bhattacharyya, V.~E. Hubeny, S.~Minwalla, and M.~Rangamani, ``{Nonlinear
  Fluid Dynamics from Gravity},''
  \href{http://dx.doi.org/10.1088/1126-6708/2008/02/045}{{\em JHEP} {\bf 0802}
  (2008)  045},
\href{http://arxiv.org/abs/0712.2456}{{\tt arXiv:0712.2456 [hep-th]}}.

\bibitem{Baier:2007ix}
R.~Baier, P.~Romatschke, D.~T. Son, A.~O. Starinets, and M.~A. Stephanov,
  ``{Relativistic viscous hydrodynamics, conformal invariance, and
  holography},'' \href{http://dx.doi.org/10.1088/1126-6708/2008/04/100}{{\em
  JHEP} {\bf 04} (2008)  100},
\href{http://arxiv.org/abs/0712.2451}{{\tt arXiv:0712.2451 [hep-th]}}.

\bibitem{Rangamani:2009xk}
M.~Rangamani, ``{Gravity and Hydrodynamics: Lectures on the fluid-gravity
  correspondence},''
  \href{http://dx.doi.org/10.1088/0264-9381/26/22/224003}{{\em Class. Quant.
  Grav.} {\bf 26} (2009)  224003},
\href{http://arxiv.org/abs/0905.4352}{{\tt arXiv:0905.4352 [hep-th]}}.

\bibitem{Emparan:2009cs}
R.~Emparan, T.~Harmark, V.~Niarchos, and N.~A. Obers, ``{World-Volume Effective
  Theory for Higher-Dimensional Black Holes},''
  \href{http://dx.doi.org/10.1103/PhysRevLett.102.191301}{{\em Phys. Rev.
  Lett.} {\bf 102} (2009)  191301},
\href{http://arxiv.org/abs/0902.0427}{{\tt arXiv:0902.0427 [hep-th]}}.

\bibitem{Emparan:2009at}
R.~Emparan, T.~Harmark, V.~Niarchos, and N.~A. Obers, ``{Essentials of
  Blackfold Dynamics},'' \href{http://dx.doi.org/10.1007/JHEP03(2010)063}{{\em
  JHEP} {\bf 03} (2010)  063},
\href{http://arxiv.org/abs/0910.1601}{{\tt arXiv:0910.1601 [hep-th]}}.

\bibitem{Camps:2010br}
J.~Camps, R.~Emparan, and N.~Haddad, ``{Black Brane Viscosity and the
  Gregory-Laflamme Instability},''
  \href{http://dx.doi.org/10.1007/JHEP05(2010)042}{{\em JHEP} {\bf 05} (2010)
  042},
\href{http://arxiv.org/abs/1003.3636}{{\tt arXiv:1003.3636 [hep-th]}}.

\bibitem{Emparan:2013ila}
R.~Emparan, V.~E. Hubeny, and M.~Rangamani, ``{Effective hydrodynamics of black
  D3-branes},'' \href{http://dx.doi.org/10.1007/JHEP06(2013)035}{{\em JHEP}
  {\bf 06} (2013)  035},
\href{http://arxiv.org/abs/1303.3563}{{\tt arXiv:1303.3563 [hep-th]}}.

\bibitem{Bhattacharyya:2008ji}
S.~Bhattacharyya, R.~Loganayagam, S.~Minwalla, S.~Nampuri, S.~P. Trivedi, and
  S.~R. Wadia, ``{Forced Fluid Dynamics from Gravity},''
  \href{http://dx.doi.org/10.1088/1126-6708/2009/02/018}{{\em JHEP} {\bf 02}
  (2009)  018},
\href{http://arxiv.org/abs/0806.0006}{{\tt arXiv:0806.0006 [hep-th]}}.

\bibitem{Gath:2013qya}
J.~Gath and A.~V. Pedersen, ``{Viscous asymptotically flat Reissner-Nordstršm
  black branes},'' \href{http://dx.doi.org/10.1007/JHEP03(2014)059}{{\em JHEP}
  {\bf 03} (2014)  059},
\href{http://arxiv.org/abs/1302.5480}{{\tt arXiv:1302.5480 [hep-th]}}.

\bibitem{DiDato:2015dia}
A.~Di~Dato, J.~Gath, and A.~V. Pedersen, ``{Probing the Hydrodynamic Limit of
  (Super)gravity},'' \href{http://dx.doi.org/10.1007/JHEP04(2015)171}{{\em
  JHEP} {\bf 04} (2015)  171},
\href{http://arxiv.org/abs/1501.05441}{{\tt arXiv:1501.05441 [hep-th]}}.

\bibitem{Emparan:2007wm}
R.~Emparan, T.~Harmark, V.~Niarchos, N.~A. Obers, and M.~J. Rodriguez, ``{The
  Phase Structure of Higher-Dimensional Black Rings and Black Holes},''
  \href{http://dx.doi.org/10.1088/1126-6708/2007/10/110}{{\em JHEP} {\bf 10}
  (2007)  110},
\href{http://arxiv.org/abs/0708.2181}{{\tt arXiv:0708.2181 [hep-th]}}.

\bibitem{Caldarelli:2008pz}
M.~M. Caldarelli, R.~Emparan, and M.~J. Rodriguez, ``{Black Rings in
  (Anti)-de{S}itter space},''
  \href{http://dx.doi.org/10.1088/1126-6708/2008/11/011}{{\em JHEP} {\bf 11}
  (2008)  011},
\href{http://arxiv.org/abs/0806.1954}{{\tt arXiv:0806.1954 [hep-th]}}.

\bibitem{Camps:2008hb}
J.~Camps, R.~Emparan, P.~Figueras, S.~Giusto, and A.~Saxena, ``{Black Rings in
  Taub-NUT and D0-D6 interactions},''
  \href{http://dx.doi.org/10.1088/1126-6708/2009/02/021}{{\em JHEP} {\bf 0902}
  (2009)  021},
\href{http://arxiv.org/abs/0811.2088}{{\tt arXiv:0811.2088 [hep-th]}}.

\bibitem{Emparan:2009vd}
R.~Emparan, T.~Harmark, V.~Niarchos, and N.~A. Obers, ``{New Horizons for Black
  Holes and Branes},'' \href{http://dx.doi.org/10.1007/JHEP04(2010)046}{{\em
  JHEP} {\bf 04} (2010)  046},
\href{http://arxiv.org/abs/0912.2352}{{\tt arXiv:0912.2352 [hep-th]}}.

\bibitem{Grignani:2010xm}
G.~Grignani, T.~Harmark, A.~Marini, N.~A. Obers, and M.~Orselli, ``{Heating up
  the BIon},'' \href{http://dx.doi.org/10.1007/JHEP06(2011)058}{{\em JHEP} {\bf
  1106} (2011)  058},
\href{http://arxiv.org/abs/1012.1494}{{\tt arXiv:1012.1494 [hep-th]}}.

\bibitem{Caldarelli:2010xz}
M.~M. Caldarelli, R.~Emparan, and B.~Van~Pol, ``{Higher-dimensional Rotating
  Charged Black Holes},'' \href{http://dx.doi.org/10.1007/JHEP04(2011)013}{{\em
  JHEP} {\bf 1104} (2011)  013},
\href{http://arxiv.org/abs/1012.4517}{{\tt arXiv:1012.4517 [hep-th]}}.

\bibitem{Armas:2010hz}
J.~Armas and N.~A. Obers, ``{Blackfolds in (Anti)-de Sitter Backgrounds},''
  \href{http://dx.doi.org/10.1103/PhysRevD.83.084039}{{\em Phys.Rev.} {\bf D83}
  (2011)  084039}, \href{http://arxiv.org/abs/1012.5081}{{\tt arXiv:1012.5081
  [hep-th]}}.

\bibitem{Emparan:2011hg}
R.~Emparan, T.~Harmark, V.~Niarchos, and N.~A. Obers, ``{Blackfolds in
  Supergravity and String Theory},''
  \href{http://dx.doi.org/10.1007/JHEP08(2011)154}{{\em JHEP} {\bf 1108} (2011)
   154},
\href{http://arxiv.org/abs/1106.4428}{{\tt arXiv:1106.4428 [hep-th]}}.

\bibitem{Armas:2011uf}
J.~Armas, J.~Camps, T.~Harmark, and N.~A. Obers, ``{The Young Modulus of Black
  Strings and the Fine Structure of Blackfolds},''
  \href{http://dx.doi.org/10.1007/JHEP02(2012)110}{{\em JHEP} {\bf 1202} (2012)
   110},
\href{http://arxiv.org/abs/1110.4835}{{\tt arXiv:1110.4835 [hep-th]}}.

\bibitem{Camps:2012hw}
J.~Camps and R.~Emparan, ``{Derivation of the blackfold effective theory},''
  \href{http://dx.doi.org/10.1007/JHEP03(2012)038}{{\em JHEP} {\bf 1203} (2012)
   038},
\href{http://arxiv.org/abs/1201.3506}{{\tt arXiv:1201.3506 [hep-th]}}.

\bibitem{Niarchos:2012pn}
V.~Niarchos and K.~Siampos, ``{M2-M5 blackfold funnels},''
  \href{http://dx.doi.org/10.1007/JHEP06(2012)175}{{\em JHEP} {\bf 1206} (2012)
   175},
\href{http://arxiv.org/abs/1205.1535}{{\tt arXiv:1205.1535 [hep-th]}}.

\bibitem{Armas:2012ac}
J.~Armas, J.~Gath, and N.~A. Obers, ``{Black Branes as Piezoelectrics},''
  \href{http://dx.doi.org/10.1103/PhysRevLett.109.241101}{{\em Phys. Rev.
  Lett.} {\bf 109} (2012)  241101},
\href{http://arxiv.org/abs/1209.2127}{{\tt arXiv:1209.2127 [hep-th]}}.

\bibitem{Armas:2013aka}
J.~Armas, J.~Gath, and N.~A. Obers, ``{Electroelasticity of Charged Black
  Branes},'' \href{http://dx.doi.org/10.1007/JHEP10(2013)035}{{\em JHEP} {\bf
  1310} (2013)  035},
\href{http://arxiv.org/abs/1307.0504}{{\tt arXiv:1307.0504 [hep-th]}}.

\bibitem{Armas:2015kra}
J.~Armas and M.~Blau, ``{Blackfolds, Plane Waves and Minimal Surfaces},''
  \href{http://dx.doi.org/10.1007/JHEP07(2015)156}{{\em JHEP} {\bf 07} (2015)
  156},
\href{http://arxiv.org/abs/1503.08834}{{\tt arXiv:1503.08834 [hep-th]}}.

\bibitem{Armas:2015nea}
J.~Armas and M.~Blau, ``{New Geometries for Black Hole Horizons},''
  \href{http://dx.doi.org/10.1007/JHEP07(2015)048}{{\em JHEP} {\bf 07} (2015)
  048},
\href{http://arxiv.org/abs/1504.01393}{{\tt arXiv:1504.01393 [hep-th]}}.

\bibitem{Armas:2012bk}
J.~Armas, T.~Harmark, N.~A. Obers, M.~Orselli, and A.~V. Pedersen, ``{Thermal
  Giant Gravitons},'' \href{http://dx.doi.org/10.1007/JHEP11(2012)123}{{\em
  JHEP} {\bf 1211} (2012)  123},
\href{http://arxiv.org/abs/1207.2789}{{\tt arXiv:1207.2789 [hep-th]}}.

\bibitem{Armas:2013ota}
J.~Armas, N.~A. Obers, and A.~V. Pedersen, ``{Null-Wave Giant Gravitons from
  Thermal Spinning Brane Probes},''
  \href{http://dx.doi.org/10.1007/JHEP10(2013)109}{{\em JHEP} {\bf 10} (2013)
  109},
\href{http://arxiv.org/abs/1306.2633}{{\tt arXiv:1306.2633 [hep-th]}}.

\bibitem{Niarchos:2015moa}
V.~Niarchos, ``{Open/closed string duality and relativistic fluids},''
\href{http://arxiv.org/abs/1510.03438}{{\tt arXiv:1510.03438 [hep-th]}}.

\bibitem{Grignani:2016bpq}
G.~Grignani, T.~Harmark, A.~Marini, and M.~Orselli, ``{The Born-Infeld/Gravity
  Correspondence},''
\href{http://arxiv.org/abs/1602.01640}{{\tt arXiv:1602.01640 [hep-th]}}.

\bibitem{Grignani:2011mr}
G.~Grignani, T.~Harmark, A.~Marini, N.~A. Obers, and M.~Orselli,
  ``{Thermodynamics of the hot BIon},''
  \href{http://dx.doi.org/10.1016/j.nuclphysb.2011.06.002}{{\em Nucl.Phys.}
  {\bf B851} (2011)  462--480},
\href{http://arxiv.org/abs/1101.1297}{{\tt arXiv:1101.1297 [hep-th]}}.

\bibitem{Grignani:2013ewa}
G.~Grignani, T.~Harmark, A.~Marini, and M.~Orselli, ``{Thermal DBI action for
  the D3-brane at weak and strong coupling},''
  \href{http://dx.doi.org/10.1007/JHEP03(2014)114}{{\em JHEP} {\bf 1403} (2014)
   114},
\href{http://arxiv.org/abs/1311.3834}{{\tt arXiv:1311.3834 [hep-th]}}.

\bibitem{Pasti:1997gx}
P.~Pasti, D.~P. Sorokin, and M.~Tonin, ``{Covariant action for a D = 11
  five-brane with the chiral field},''
  \href{http://dx.doi.org/10.1016/S0370-2693(97)00188-3}{{\em Phys. Lett.} {\bf
  B398} (1997)  41--46},
\href{http://arxiv.org/abs/hep-th/9701037}{{\tt arXiv:hep-th/9701037
  [hep-th]}}.

\bibitem{Niarchos:2014maa}
V.~Niarchos, ``{Supersymmetric Perturbations of the M5 brane},''
  \href{http://dx.doi.org/10.1007/JHEP05(2014)023}{{\em JHEP} {\bf 05} (2014)
  023},
\href{http://arxiv.org/abs/1402.4132}{{\tt arXiv:1402.4132 [hep-th]}}.

\bibitem{Myers:1986un}
R.~C. Myers and M.~J. Perry, ``{Black Holes in Higher Dimensional
  Space-Times},''
\href{http://dx.doi.org/10.1016/0003-4916(86)90186-7}{{\em Ann. Phys.} {\bf
  172} (1986)  304}.

\bibitem{Harmark:2002tr}
T.~Harmark and N.~A. Obers, ``{Black holes on cylinders},''
  \href{http://dx.doi.org/10.1088/1126-6708/2002/05/032}{{\em JHEP} {\bf 05}
  (2002)  032},
\href{http://arxiv.org/abs/hep-th/0204047}{{\tt arXiv:hep-th/0204047
  [hep-th]}}.

\bibitem{Kol:2003if}
B.~Kol, E.~Sorkin, and T.~Piran, ``{Caged black holes: Black holes in
  compactified space-times. 1. Theory},''
  \href{http://dx.doi.org/10.1103/PhysRevD.69.064031}{{\em Phys. Rev.} {\bf
  D69} (2004)  064031},
\href{http://arxiv.org/abs/hep-th/0309190}{{\tt arXiv:hep-th/0309190
  [hep-th]}}.

\bibitem{rindler2001relativity}
W.~Rindler, {\em Relativity: Special, General, and Cosmological}.
\newblock Relativity: Special, General, and Cosmological. Oxford University
  Press, 2001.

\bibitem{Bhattacharya:2011eea}
J.~Bhattacharya, S.~Bhattacharyya, and S.~Minwalla, ``{Dissipative Superfluid
  dynamics from gravity},''
  \href{http://dx.doi.org/10.1007/JHEP04(2011)125}{{\em JHEP} {\bf 1104} (2011)
   125},
\href{http://arxiv.org/abs/1101.3332}{{\tt arXiv:1101.3332 [hep-th]}}.


\bibitem{Bergshoeff:2001pv}
E.~Bergshoeff, R.~Kallosh, T.~Ortin, D.~Roest, and A.~Van~Proeyen, ``{New
  formulations of D = 10 supersymmetry and D8 - O8 domain walls},''
  \href{http://dx.doi.org/10.1088/0264-9381/18/17/303}{{\em Class.Quant.Grav.}
  {\bf 18} (2001)  3359--3382},
\href{http://arxiv.org/abs/hep-th/0103233}{{\tt arXiv:hep-th/0103233
  [hep-th]}}.

\bibitem{Dall'Agata:1998va}
G.~Dall'Agata, K.~Lechner, and M.~Tonin, ``{D = 10, N = IIB supergravity:
  Lorentz invariant actions and duality},''
  \href{http://dx.doi.org/10.1088/1126-6708/1998/07/017}{{\em JHEP} {\bf 9807}
  (1998)  017},
\href{http://arxiv.org/abs/hep-th/9806140}{{\tt arXiv:hep-th/9806140
  [hep-th]}}.

\bibitem{Bandos:2003et}
I.~A. Bandos, A.~J. Nurmagambetov, and D.~P. Sorokin, ``{Various faces of type
  IIA supergravity},''
  \href{http://dx.doi.org/10.1016/j.nuclphysb.2003.10.036}{{\em Nucl.Phys.}
  {\bf B676} (2004)  189--228},
\href{http://arxiv.org/abs/hep-th/0307153}{{\tt arXiv:hep-th/0307153
  [hep-th]}}.

\bibitem{becker2006string}
K.~Becker, M.~Becker, and J.~Schwarz, {\em String Theory and M-Theory: A Modern
  Introduction}.
\newblock Cambridge University Press, 2006.
\newblock \url{https://books.google.be/books?id=WgUkSTJWQacC}.

\bibitem{Bandos:1997gd}
I.~A. Bandos, N.~Berkovits, and D.~P. Sorokin, ``{Duality symmetric
  eleven-dimensional supergravity and its coupling to M-branes},''
  \href{http://dx.doi.org/10.1016/S0550-3213(98)00102-3}{{\em Nucl.Phys.} {\bf
  B522} (1998)  214--233},
\href{http://arxiv.org/abs/hep-th/9711055}{{\tt arXiv:hep-th/9711055
  [hep-th]}}.

\bibitem{Horowitz:1991cd}
G.~T. Horowitz and A.~Strominger, ``{Black strings and P-branes},''
\href{http://dx.doi.org/10.1016/0550-3213(91)90440-9}{{\em Nucl. Phys.} {\bf
  B360} (1991)  197--209}.

\bibitem{Erdmenger:2008rm}
J.~Erdmenger, M.~Haack, M.~Kaminski, and A.~Yarom, ``{Fluid dynamics of
  R-charged black holes},''
  \href{http://dx.doi.org/10.1088/1126-6708/2009/01/055}{{\em JHEP} {\bf 0901}
  (2009)  055},
\href{http://arxiv.org/abs/0809.2488}{{\tt arXiv:0809.2488 [hep-th]}}.

\bibitem{Banerjee:2008th}
N.~Banerjee, J.~Bhattacharya, S.~Bhattacharyya, S.~Dutta, R.~Loganayagam, {\em
  et al.}, ``{Hydrodynamics from charged black branes},''
  \href{http://dx.doi.org/10.1007/JHEP01(2011)094}{{\em JHEP} {\bf 1101} (2011)
   094},
\href{http://arxiv.org/abs/0809.2596}{{\tt arXiv:0809.2596 [hep-th]}}.

\bibitem{Gorbonos:2004uc}
D.~Gorbonos and B.~Kol, ``{A Dialogue of multipoles: Matched asymptotic
  expansion for caged black holes},''
  \href{http://dx.doi.org/10.1088/1126-6708/2004/06/053}{{\em JHEP} {\bf 0406}
  (2004)  053},
\href{http://arxiv.org/abs/hep-th/0406002}{{\tt arXiv:hep-th/0406002
  [hep-th]}}.

\bibitem{Harmark:2000wv}
T.~Harmark, ``{Supergravity and space-time noncommutative open string
  theory},'' \href{http://dx.doi.org/10.1088/1126-6708/2000/07/043}{{\em JHEP}
  {\bf 07} (2000)  043},
\href{http://arxiv.org/abs/hep-th/0006023}{{\tt arXiv:hep-th/0006023
  [hep-th]}}.

\bibitem{Crossley:2015evo}
M.~Crossley, P.~Glorioso, and H.~Liu, ``{Effective field theory of dissipative
  fluids},''
\href{http://arxiv.org/abs/1511.03646}{{\tt arXiv:1511.03646 [hep-th]}}.

\bibitem{Haehl:2015uoc}
F.~M. Haehl, R.~Loganayagam, and M.~Rangamani, ``{Topological sigma models \&
  dissipative hydrodynamics},''
  \href{http://dx.doi.org/10.1007/JHEP04(2016)039}{{\em JHEP} {\bf 04} (2016)
  039},
\href{http://arxiv.org/abs/1511.07809}{{\tt arXiv:1511.07809 [hep-th]}}.

\bibitem{Gibbons:1976ue}
G.~W. Gibbons and S.~W. Hawking, ``{Action Integrals and Partition Functions in
  Quantum Gravity},''
\href{http://dx.doi.org/10.1103/PhysRevD.15.2752}{{\em Phys. Rev.} {\bf D15}
  (1977)  2752--2756}.

\bibitem{Armas:2013hsa}
J.~Armas, ``{How Fluids Bend: the Elastic Expansion for Higher-Dimensional
  Black Holes},'' \href{http://dx.doi.org/10.1007/JHEP09(2013)073}{{\em JHEP}
  {\bf 1309} (2013)  073},
\href{http://arxiv.org/abs/1304.7773}{{\tt arXiv:1304.7773 [hep-th]}}.

\bibitem{Grignani:2012iw}
G.~Grignani, T.~Harmark, A.~Marini, N.~A. Obers, and M.~Orselli, ``{Thermal
  string probes in AdS and finite temperature Wilson loops},''
  \href{http://dx.doi.org/10.1007/JHEP06(2012)144}{{\em JHEP} {\bf 1206} (2012)
   144},
\href{http://arxiv.org/abs/1201.4862}{{\tt arXiv:1201.4862 [hep-th]}}.

\bibitem{Niarchos:2012cy}
V.~Niarchos and K.~Siampos, ``{Entropy of the self-dual string soliton},''
  \href{http://dx.doi.org/10.1007/JHEP07(2012)134}{{\em JHEP} {\bf 07} (2012)
  134},
\href{http://arxiv.org/abs/1206.2935}{{\tt arXiv:1206.2935 [hep-th]}}.

\bibitem{Niarchos:2013ia}
V.~Niarchos and K.~Siampos, ``{The black M2-M5 ring intersection spins},'' {\em
  PoS} {\bf Corfu2012} (2013)  088,
\href{http://arxiv.org/abs/1302.0854}{{\tt arXiv:1302.0854 [hep-th]}}.

\bibitem{Armas:2014nea}
J.~Armas and M.~Blau, ``{Black probes of Schršdinger spacetimes},''
  \href{http://dx.doi.org/10.1007/JHEP08(2014)140}{{\em JHEP} {\bf 08} (2014)
  140},
\href{http://arxiv.org/abs/1405.1301}{{\tt arXiv:1405.1301 [hep-th]}}.

\bibitem{Giataganas:2014mla}
D.~Giataganas and K.~Goldstein, ``{Tension of Confining Strings at Low
  Temperature},'' \href{http://dx.doi.org/10.1007/JHEP02(2015)123}{{\em JHEP}
  {\bf 02} (2015)  123},
\href{http://arxiv.org/abs/1411.4995}{{\tt arXiv:1411.4995 [hep-th]}}.

\bibitem{Bena:2009xk}
I.~Bena, M.~Grana, and N.~Halmagyi, ``{On the Existence of Meta-stable Vacua in
  Klebanov-Strassler},'' \href{http://dx.doi.org/10.1007/JHEP09(2010)087}{{\em
  JHEP} {\bf 09} (2010)  087},
\href{http://arxiv.org/abs/0912.3519}{{\tt arXiv:0912.3519 [hep-th]}}.

\bibitem{Blaback:2011pn}
J.~Blaback, U.~H. Danielsson, D.~Junghans, T.~Van~Riet, T.~Wrase, and
  M.~Zagermann, ``{(Anti-)Brane backreaction beyond perturbation theory},''
  \href{http://dx.doi.org/10.1007/JHEP02(2012)025}{{\em JHEP} {\bf 02} (2012)
  025},
\href{http://arxiv.org/abs/1111.2605}{{\tt arXiv:1111.2605 [hep-th]}}.

\bibitem{Bena:2016fqp}
  I.~Bena, J.~Bl{\aa}b\"ack and D.~Turton,
  ``{Loop corrections to the antibrane potential},''
   \href{http://dx.doi.org/10.1007/JHEP07(2016)132}{{\em JHEP} {\bf 07} (2016)
  132},
  \href{http://arxiv.org/abs/1602.05959}{{\tt arXiv:1602.05959 [hep-th]}}.



\bibitem{Klebanov:2000hb}
I.~R. Klebanov and M.~J. Strassler, ``{Supergravity and a confining gauge
  theory: Duality cascades and chi SB resolution of naked singularities},''
  \href{http://dx.doi.org/10.1088/1126-6708/2000/08/052}{{\em JHEP} {\bf 08}
  (2000)  052},
\href{http://arxiv.org/abs/hep-th/0007191}{{\tt arXiv:hep-th/0007191
  [hep-th]}}.

\bibitem{Kachru:2002gs}
S.~Kachru, J.~Pearson, and H.~L. Verlinde, ``{Brane / flux annihilation and the
  string dual of a nonsupersymmetric field theory},''
  \href{http://dx.doi.org/10.1088/1126-6708/2002/06/021}{{\em JHEP} {\bf 06}
  (2002)  021},
\href{http://arxiv.org/abs/hep-th/0112197}{{\tt arXiv:hep-th/0112197
  [hep-th]}}.

\bibitem{Bena:2011wh}
I.~Bena, G.~Giecold, M.~Grana, N.~Halmagyi, and S.~Massai, ``{The backreaction
  of anti-D3 branes on the Klebanov-Strassler geometry},''
  \href{http://dx.doi.org/10.1007/JHEP06(2013)060}{{\em JHEP} {\bf 06} (2013)
  060},
\href{http://arxiv.org/abs/1106.6165}{{\tt arXiv:1106.6165 [hep-th]}}.

\bibitem{Klebanov:2010qs}
I.~R. Klebanov and S.~S. Pufu, ``{M-Branes and Metastable States},''
  \href{http://dx.doi.org/10.1007/JHEP08(2011)035}{{\em JHEP} {\bf 08} (2011)
  035},
\href{http://arxiv.org/abs/1006.3587}{{\tt arXiv:1006.3587 [hep-th]}}.

\bibitem{Bena:2010gs}
I.~Bena, G.~Giecold, and N.~Halmagyi, ``{The Backreaction of Anti-M2 Branes on
  a Warped Stenzel Space},''
  \href{http://dx.doi.org/10.1007/JHEP04(2011)120}{{\em JHEP} {\bf 04} (2011)
  120},
\href{http://arxiv.org/abs/1011.2195}{{\tt arXiv:1011.2195 [hep-th]}}.

\bibitem{Massai:2011vi}
S.~Massai, ``{Metastable Vacua and the Backreacted Stenzel Geometry},''
  \href{http://dx.doi.org/10.1007/JHEP06(2012)059}{{\em JHEP} {\bf 06} (2012)
  059},
\href{http://arxiv.org/abs/1110.2513}{{\tt arXiv:1110.2513 [hep-th]}}.

\bibitem{Gutperle:2002ai}
M.~Gutperle and A.~Strominger, ``{Space - like branes},''
  \href{http://dx.doi.org/10.1088/1126-6708/2002/04/018}{{\em JHEP} {\bf 04}
  (2002)  018},
\href{http://arxiv.org/abs/hep-th/0202210}{{\tt arXiv:hep-th/0202210
  [hep-th]}}.

\bibitem{Sen:2002nu}
A.~Sen, ``{Rolling tachyon},''
  \href{http://dx.doi.org/10.1088/1126-6708/2002/04/048}{{\em JHEP} {\bf 04}
  (2002)  048},
\href{http://arxiv.org/abs/hep-th/0203211}{{\tt arXiv:hep-th/0203211
  [hep-th]}}.

\bibitem{Kutasov:2003er}
D.~Kutasov and V.~Niarchos, ``{Tachyon effective actions in open string
  theory},'' \href{http://dx.doi.org/10.1016/S0550-3213(03)00498-X}{{\em Nucl.
  Phys.} {\bf B666} (2003)  56--70},
\href{http://arxiv.org/abs/hep-th/0304045}{{\tt arXiv:hep-th/0304045
  [hep-th]}}.

\bibitem{Armas:2013goa}
J.~Armas, ``{(Non)-Dissipative Hydrodynamics on Embedded Surfaces},''
  \href{http://dx.doi.org/10.1007/JHEP09(2014)047}{{\em JHEP} {\bf 09} (2014)
  047},
\href{http://arxiv.org/abs/1312.0597}{{\tt arXiv:1312.0597 [hep-th]}}.

\bibitem{Armas:2014rva}
J.~Armas and T.~Harmark, ``{Constraints on the effective fluid theory of
  stationary branes},'' \href{http://dx.doi.org/10.1007/JHEP10(2014)063}{{\em
  JHEP} {\bf 10} (2014)  63},
\href{http://arxiv.org/abs/1406.7813}{{\tt arXiv:1406.7813 [hep-th]}}.

\bibitem{Armas:2015ssd}
J.~Armas, J.~Bhattacharya, and N.~Kundu, ``{Surface transport in
  plasma-balls},'' \href{http://dx.doi.org/10.1007/JHEP06(2016)015}{{\em JHEP}
  {\bf 06} (2016)  015},
\href{http://arxiv.org/abs/1512.08514}{{\tt arXiv:1512.08514 [hep-th]}}.

\bibitem{Armas:2015qsv}
J.~Armas, N.~A. Obers, and M.~Sanchioni, ``{Gravitational Tension, Spacetime
  Pressure and Black Hole Volume},''
\href{http://arxiv.org/abs/1512.09106}{{\tt arXiv:1512.09106 [hep-th]}}. To appear in JHEP. 

\end{thebibliography}
\end{document}